    \input amstexl


\catcode`\@=11
\ifx\amstexloaded@\relax\else
 \errmessage{AmS-TeX must be loaded before LamS-TeX}\fi
\ifx\laxread@\undefined\else\catcode`\@=\active \fi
\def\err@#1{\errmessage{LamS-TeX error: #1}}
\def^^L{\par}
\let\+\tabalign
\def\newcount{\alloc@0\count\countdef\insc@unt}
\def\newdimen{\alloc@1\dimen\dimendef\insc@unt}
\def\newskip{\alloc@2\skip\skipdef\insc@unt}
\def\newmuskip{\alloc@3\muskip\muskipdef\@cclvi}
\def\newbox{\alloc@4\box\chardef\insc@unt}
\let\newtoks\relax
\def\newhelp#1#2{\newtoks#1#1\expandafter{\csname#2\endcsname}}
\def\newtoks{\alloc@5\toks\toksdef\@cclvi}
\def\newread{\alloc@6\read\chardef\sixt@@n}
\def\newwrite{\alloc@7\write\chardef\sixt@@n}
\def\newfam{\alloc@8\fam\chardef\sixt@@n}
\def\newlanguage{\alloc@9\language\chardef\@cclvi}
\def\newinsert#1{\global\advance\insc@unt by\m@ne
  \ch@ck0\insc@unt\count
  \ch@ck1\insc@unt\dimen
  \ch@ck2\insc@unt\skip
  \ch@ck4\insc@unt\box
  \allocationnumber=\insc@unt
  \global\chardef#1=\allocationnumber
  \wlog{\string#1=\string\insert\the\allocationnumber}}
\def\newif#1{\count@\escapechar \escapechar\m@ne
  \expandafter\expandafter\expandafter
   \edef\@if#1{true}{\let\noexpand#1=\noexpand\iftrue}%
  \expandafter\expandafter\expandafter
   \edef\@if#1{false}{\let\noexpand#1=\noexpand\iffalse}%
  \@if#1{false}\escapechar\count@}

\def\Err@#1{\errhelp\defaulthelp@\err@{#1}}
{\catcode`\@=\active
 \edef\next{\gdef\noexpand@{\futurelet\noexpand\next
  \csname at\string@\endcsname}}
 \next
}
\def\at@{\ifcat\noexpand\next a\let\next@\at@@\else
 \ifcat\noexpand\next0\let\next@\at@@\else
 \ifcat\noexpand\next\relax\let\next@\at@@\else
 \let\next@\at@@@\fi\fi\fi\next@}
\def\at@@@{\errhelp\athelp@\err@{Invalid use of @}}
\def\at@@#1{\expandafter
 \ifx\csname\string#1@at\endcsname\relax\let\next@\at@@@\else
 \DN@{\csname\string#1@at\endcsname}\fi\next@}
\def\atdef@#1{\expandafter\def\csname\string#1@at\endcsname}
\newif\iftest@
\def\tagin@#1{\tagin@false
 \DN@##1\tag##2##3\next@{\test@true\ifx\tagin@##2\test@false\fi}%
 \next@#1\tag\tagin@\next@\tagin@false\iftest@\tagin@true\fi}
\let\lkerns@\relax
\def\nolinebreak{\RIfM@\mathmodeerr@\nolinebreak\else
 \ifhmode\saveskip@\lastskip\unskip
 \nobreak\ifdim\saveskip@>\z@\hskip\saveskip@\fi\lkerns@
 \else\vmodeerr@\nolinebreak\fi\fi}
\def\allowlinebreak{\RIfM@\mathmodeerr@\allowlinebreak\else
 \ifhmode\saveskip@\lastskip\unskip
 \allowbreak\ifdim\saveskip@>\z@\hskip\saveskip@\fi\lkerns@
 \else\vmodeerr@\allowlinebreak\fi\fi}
\def\linebreak{\RIfM@\mathmodeerr@\linebreak\else
 \ifhmode\unskip\unkern\break\lkerns@
 \else\vmodeerr@\linebreak\fi\fi}
\let\nkerns@\relax
\def\newline{\RIfM@\mathmodeerr@\newline\else
 \ifhmode\unskip\unkern\null\hfill\break\nkerns@
 \else\vmodeerr@\newline\fi\fi}%
\def\newbox@{\alloc@@4\box\chardef\insc@unt}
\def\newcount@{\alloc@@0\count\countdef\insc@unt}
\def\accentedsymbol#1#2{\expandafter\newbox@\csname\exstring@#1@box\endcsname
 \setbox\csname\exstring@#1@box\endcsname\hbox{$\m@th#2$}%
 \define#1{\copy\csname\exstring@#1@box\endcsname{}}}
\def\rightadd@#1\to#2{\toks@{\\#1}\toks@@\expandafter{#2}\xdef#2{\the\toks@@
 \the\toks@}\toks@{}\toks@@{}}
\def\fontlist@{\\\tenrm\\\sevenrm\\\fiverm\\\teni\\\seveni\\\fivei
 \\\tensy\\\sevensy\\\fivesy\\\tenex\\\tenbf\\\sevenbf\\\fivebf
 \\\tensl\\\tenit}
\def\font@#1=#2 {\rightadd@#1\to\fontlist@\font#1=#2 }
\def\ismember@#1#2{\global\let\Next@ F\let\next@= #2%
 {\def\\##1{\let\nextii@##1\ifx\nextii@\next@\global\let\Next@ T\fi}#1}%
 \test@false\ifx\Next@ T\test@true\fi\let\next@\relax}
\def\FNSS@#1{\let\FNSS@@#1\FN@\FNSS@@@}
\def\FNSS@@@{\ifx\next\space@\def\FNSS@@@@. {\FN@\FNSS@@@}\else
 \def\FNSS@@@@.{\FNSS@@}\fi\FNSS@@@@.}
\atdef@"{\unskip
 \DN@{\ifx\next`\DN@`{\FN@\nextii@}%
  \else\ifx\next\lq\DN@\lq{\FN@\nextii@}%
  \else\DN@####1{\FN@\nextiii@}\fi\fi
  \next@}%
 \DNii@{\ifx\next`\DN@`{\sldl@``}%
  \else\ifx\next\lq\DN@\lq{\sldl@``}%
  \else\DN@{\dlsl@`}\fi\fi\next@}%
 \def\nextiii@{\ifx\next'\DN@'{\srdr@''}%
  \else\ifx\next\rq\DN@\rq{\srdr@''}%
  \else\DN@{\drsr@'}\fi\fi\next@}%
 \FNSS@\next@}
\def\root{%
  \DN@{\ifx\next\uproot\let\next@\nextii@\else
   \ifx\next\leftroot\let\next@\nextiii@\else
   \let\next@\plainroot@\fi\fi\next@}%
  \DNii@\uproot##1{\uproot@##1\relax\FNSS@\nextiv@}%
  \def\nextiv@{\ifx\next\leftroot\let\next@\nextv@\else
   \let\next@\plainroot@\fi\next@}%
  \def\nextv@\leftroot##1{\leftroot@##1\relax\plainroot@}%
  \def\nextiii@\leftroot##1{\leftroot@##1\relax\FNSS@\nextvi@}%
  \def\nextvi@{\ifx\next\uproot\let\next@\nextvii@\else
   \let\next@\plainroot@\fi\next@}%
  \def\nextvii@\uproot##1{\uproot@##1\relax\plainroot@}%
  \bgroup\uproot@\z@\leftroot@\z@
 \FNSS@\next@}
\def\loop#1\repeat{\def\iterate{#1\relax\expandafter\iterate\fi}%
 \iterate\let\iterate\relax}
\def\gloop@#1\repeat{\gdef\iterate@{#1\relax\expandafter\iterate@\fi}%
 \iterate@\global\let\iterate@\relax}
\def\printoptions{\W@{Do you want S(yntax check),
  G(alleys) or P(ages)?^^JType S, G or P, follow by <return>: }\loop
 \read\m@ne to\ans@
 \edef\next@{\def\noexpand\Ans@{\ans@}}\uppercase\expandafter{\next@}%
 \ifx\Ans@\S@\test@true\syntax\else
 \ifx\Ans@\G@\test@true\galleys\else
 \ifx\Ans@\P@\test@true\else
 \test@false\fi\fi\fi
 \iftest@\else\W@{Type S, G or P, follow by <return>: }%
 \repeat}
\expandafter\let\csname A@;\endcsname;
\expandafter\let\csname A@:\endcsname:
\expandafter\let\csname A@?\endcsname?
\expandafter\let\csname A@!\endcsname!
\def\APdef#1{\def\next@{\expandafter\let\csname A@\string#1\endcsname#1}%
 \afterassignment\next@\def#1}
\let\fextra@\,
\def\tdots@{\unskip
 \DN@{$\m@th\mathinner{\ldotp\ldotp\ldotp}\,
   \ifx\next,\,$\else\ifx\next.\,$\else
   \ifx\next;\,$\else
   \expandafter\ifx\csname A@\string;\endcsname\next\fextra@$\else
   \ifx\next:\,$\else
   \expandafter\ifx\csname A@\string:\endcsname\next\fextra@$\else
   \ifx\next?\,$\else
   \expandafter\ifx\csname A@\string?\endcsname\next\fextra@$\else
   \ifx\next!\,$\else
   \expandafter\ifx\csname A@\string!\endcsname\next\fextra@$\else
   $ \fi\fi\fi\fi\fi\fi\fi\fi\fi\fi}%
 \ \FN@\next@}
\def\extrap@#1{%
 \ifx\next,\DN@{#1\,}\else
 \ifx\next;\DN@{#1\,}\else
 \expandafter\ifx\csname A@\string;\endcsname\next\DN@{#1\fextra@}\else
 \ifx\next.\DN@{#1\,}\else\extra@
 \ifextra@\DN@{#1\,}\else
 \let\next@#1\fi\fi\fi\fi\fi\next@}
\def\dotsc{\DN@{\ifx\next;\plainldots@\,\else
 \expandafter\ifx\csname A@\string;\endcsname\next\plainldots@\fextra@\else
 \ifx\next.\plainldots@\,\else\extra@\plainldots@
 \ifextra@\,\fi\fi\fi\fi}%
 \FN@\next@}
\def\keybin@{\keybin@true
 \ifx\next+\else\ifx\next=\else\ifx\next<\else\ifx\next>\else\ifx\next-\else
 \ifx\next*\else\ifx\next:\else
 \expandafter\ifx\csname A@\string;\endcsname\next\else
 \keybin@false\fi\fi\fi\fi\fi\fi\fi\fi}
\def\boldkey#1{\ifcat\noexpand#1A%
  \ifcmmibloaded@{\fam\cmmibfam#1}\else
   \Err@{First bold symbol font not loaded}\fi
 \else
 \let\next=#1%
 \ifx#1!\mathchar"5\bffam@21 \else
 \expandafter\ifx\csname A@\string!\endcsname\next\mathchar"5\bffam@21 \else
 \ifx#1(\mathchar"4\bffam@28 \else\ifx#1)\mathchar"5\bffam@29 \else
 \ifx#1+\mathchar"2\bffam@2B \else\ifx#1:\mathchar"3\bffam@3A \else
 \expandafter\ifx\csname A@\string:\endcsname\next\mathchar"3\bffam@3A \else
 \ifx#1;\mathchar"6\bffam@3B \else
 \expandafter\ifx\csname A@\string;\endcsname\next\mathchar"6\bffam@3B \else
 \ifx#1=\mathchar"3\bffam@3D \else
 \ifx#1?\mathchar"5\bffam@3F \else
 \expandafter\ifx\csname A@\string?\endcsname\next\mathchar"5\bffam@3F \else
 \ifx#1[\mathchar"4\bffam@5B \else
 \ifx#1]\mathchar"5\bffam@5D \else
 \ifx#1,\mathchari@63B \else
 \ifx#1-\mathcharii@200 \else
 \ifx#1.\mathchari@03A \else
 \ifx#1/\mathchari@03D \else
 \ifx#1<\mathchari@33C \else
 \ifx#1>\mathchari@33E \else
 \ifx#1*\mathcharii@203 \else
 \ifx#1|\mathcharii@06A \else
 \ifx#10\bold0\else\ifx#11\bold1\else\ifx#12\bold2\else\ifx#13\bold3\else
 \ifx#14\bold4\else\ifx#15\bold5\else\ifx#16\bold6\else\ifx#17\bold7\else
 \ifx#18\bold8\else\ifx#19\bold9\else
  \Err@{\noexpand\boldkey can't be used with #1}%
 \fi\fi\fi\fi\fi\fi\fi\fi\fi\fi\fi\fi\fi\fi\fi
 \fi\fi\fi\fi\fi\fi\fi\fi\fi\fi\fi\fi\fi\fi\fi\fi\fi\fi}
\def\arabic#1{#1}
\def\alph#1{\count@#1\relax\advance\count@96 \ifnum\count@>122
 \Err@{\noexpand\alph invalid for numbers > 26}\else\char\count@\fi}
\def\Alph#1{\count@#1\relax\advance\count@64 \ifnum\count@>90
 \Err@{\noexpand\Alph invalid for numbers > 26}\else\char\count@\fi}

\def\Roman#1{\uppercase\expandafter{\romannumeral#1}}
\def\fnsymbol#1{\count@#1\relax
 \count@@\count@
 \advance\count@\m@ne\divide\count@7
 \count@@@\count@\advance\count@@@\@ne
 \multiply\count@7 \advance\count@@-\count@
 \count@\count@@@
 {\loop
  \ifcase\count@@\or*\or\dag\or\ddag\or\P\or\S\or\text{$\|$}\or\#\fi
  \advance\count@\m@ne\ifnum\count@>\z@\repeat}}
\def\cardnine@#1{\ifcase#1\or one\or two\or three\or four\or five\or
 six\or seven\or eight\or nine\fi}
\let\alloc@\alloc@@
\newcount\ten@
\ten@10
\def\cardinal#1{\count@#1\relax
 \ifnum\count@>99 \number\count@
 \else
  \ifnum\count@=\z@ zero%
  \else
   \ifnum\count@<\ten@\cardnine@\count@
   \else
    \ifnum\count@<20
     \advance\count@-\ten@
     \ifcase\count@ ten\or eleven\or twelve\or thirteen\or fourteen\or
      fifteen\or sixteen\or seventeen\or eighteen\or nineteen\fi
    \else
     \count@@\count@\count@@@\count@@
     \divide\count@\ten@\multiply\count@\ten@
     \advance\count@@@-\count@\divide\count@\ten@
     \ifcase\count@\or\or twenty\or thirty\or forty\or fifty\or sixty\or
      seventy\or eighty\or ninety\fi
     \ifnum\count@@@=\z@\else-\cardnine@\count@@@\fi
    \fi
   \fi
  \fi
 \fi}
\def\ordnine@#1{\ifcase#1\or first\or second\or third\or fourth\or fifth\or
 sixth\or seventh\or eighth\or ninth\fi}
\newcount\count@@@@
\def\ordsuffix@{\count@@@@\count@
 \divide\count@\ten@
 \count@@@\count@\count@@\count@
 \divide\count@@\ten@\multiply\count@@\ten@
 \advance\count@@@-\count@@
 \ifnum\count@@@=\@ne th%
 \else
  \count@@@\count@@@@
  \count@@\count@@@@
  \divide\count@@\ten@\multiply\count@@\ten@
  \advance\count@@@-\count@@
  \ifcase\count@@@ th\or st\or nd\or rd\else th\fi
 \fi}
\def\nordinal#1{\count@#1\relax\number\count@\ordsuffix@}
\def\spordinal#1{\count@#1\relax\number\count@$^{\text{\ordsuffix@}}$}
\def\ordinal#1{\count@#1\relax
 \ifnum\count@>99 \number\count@\ordsuffix@
 \else
   \ifnum\count@=\z@ zeroth%
  \else
    \ifnum\count@<\ten@\ordnine@\count@
    \else
     \ifnum\count@<20 \advance\count@-\ten@
      \ifcase\count@ tenth\or eleventh\or twelfth\or thirteenth\or
       fourteenth\or fifteenth\or sixteenth\or seventeenth\or eighteenth\or
       nineteenth\fi
     \else
      \count@@\count@
      \divide\count@\ten@\multiply\count@\ten@
      \count@@@\count@@\advance\count@@@-\count@
      \divide\count@\ten@
      \ifcase\count@\or\or twent\or thirt\or fort\or fift\or sixt\or sevent\or
       eight\or ninet\fi
      \ifnum\count@@@=\z@ ieth\else y-\ordnine@\count@@@\fi
     \fi
    \fi
  \fi
 \fi}
\font@\tensmc=cmcsc10
\textonlyfont@\smc\tensmc
\newtoks\noexpandtoks@
\noexpandtoks@{\let\arabic\relax\let\alph\relax\let\Alph\relax
 \let\Roman\relax\let\fnsymbol\relax\let\rm\relax
 \let\it\relax\let\bf\relax\let\sl\relax\let\smc\relax
 \let\/\relax\let\null\relax}
\def\noexpands@{\the\noexpandtoks@}
\def\Nonexpanding#1{\global\noexpandtoks@
 \expandafter{\the\noexpandtoks@\let#1\relax}}
\def\prevanish@{\saveskip@\z@\ifhmode\saveskip@\lastskip\unskip\fi}
\def\postvanish@{\ifdim\saveskip@>\z@\hskip\saveskip@\fi\FN@\postvanish@@}
\def\postvanish@@{\DN@.{}%
 \ifx\next\space@\ifdim\saveskip@>\z@\DN@. {}\fi\fi\next@.}
\def\invisible#1{\prevanish@\ignorespaces#1\unskip\postvanish@}
\def\vanishlist@{\\\invisible}
\let\noindent@\noindent
\def\noindent{\par\noindent@\FN@\pretendspace@}
\def\pretendspace@{\ismember@\vanishlist@\next
 \iftest@\nobreak\hskip-\p@\hskip\p@\fi}

\newtoks\everypartoks@
\def\noindent@@{\par\everypartoks@\expandafter{\the\everypar}\everypar{}%
 \noindent@\everypar\expandafter{\the\everypartoks@}}
\def\page{\Err@{\noexpand\page has no meaning by itself}}
\let\page@C\pageno
\let\page@P\empty
\let\page@Q\empty
\def\page@S#1{#1\/}
\def\page@F{\rm}
\def\page@N{\arabic}   
\newif\ifindexing@
\def\indexfile{\ifindexing@\else
 \alloc@@7\write\chardef\sixt@@n\ndx@
 \immediate\openout\ndx@=\jobname.ndx
 \global\indexing@true\fi}
\global\advance\insc@unt\m@ne
\ch@ck0\insc@unt\count
\ch@ck1\insc@unt\dimen
\ch@ck2\insc@unt\skip
\ch@ck4\insc@unt\box
\allocationnumber\insc@unt
\global\chardef\margin@\allocationnumber
\dimen\margin@\maxdimen
\count\margin@\z@
\skip\margin@\z@
\newif\ifindexproofing@
\def\indexproofing{\indexproofing@true}
\def\noindexproofing{\indexproofing@false}
\def\unmacro@#1:#2->#3\unmacro@{\def\macpar@{#2}\def\macdef@{#3}}
\def\starparts@#1{\def\stari@{#1}\def\starii@{#1}\let\stariii@\empty
 \test@false
 \DN@##1*##2##3\next@{\ifx\starparts@##2\test@false\else\test@true\fi}%
 \next@#1*\starparts@\next@
 \iftest@\DN@{\starparts@@#1\starparts@@}\else\let\next@\relax\fi\next@}
\def\starparts@@#1*#2\starparts@@{\def\starii@{#1}\def\stariii@{*#2}}
\def\windex@{\ifindexing@
 \expandafter\unmacro@\meaning\stari@\unmacro@
 \edef\macdef@{\string"\macdef@\string"}%
 \edef\next@{\write\ndx@{\macdef@}}\next@
 \write\ndx@{{\number\pageno}{\page@N}{\page@P}{\page@Q}}%
 \fi
 \ifindexproofing@
  \ifx\stariii@\empty\else
   \expandafter\unmacro@\meaning\stariii@\unmacro@\fi
  \insert\margin@{\hbox{\rm\vrule\height9\p@\depth2\p@\width\z@\starii@
  \ifx\stariii@\empty\else\tt\macdef@\fi}}\fi}
\catcode`\"=\active
\def"{\FN@\quote@}
\def\quote@{\ifx\next"\expandafter\quote@@\else\expandafter\quote@@@\fi}
\def\quote@@@#1"{\starparts@{#1}\starii@\windex@}
\def\quote@@"#1"{\prevanish@\starparts@{#1}\windex@\FN@\quote@@@@}
\def\quote@@@@{\ifx\next"\DN@"{\postvanish@}\else
 \let\next@\postvanish@\fi\next@}
\rightadd@"\to\vanishlist@
\def\idefine#1{\DN@{#1}\DNii@{\noexpand#1}%
 \afterassignment\idefine@\def\nextiii@}
\def\idefine@{\ifindexing@
 \expandafter\let\next@\nextiii@
 \expandafter\unmacro@\meaning\nextiii@\unmacro@
 \immediate\write\ndx@{\noexpand\define\nextii@\macpar@{\macdef@}}\fi}
\def\iabbrev*#1#2{\ifindexing@\toks@{#2}%
 \immediate\write\ndx@{\noexpand\abbrev*\noexpand#1{\the\toks@}}\fi}
\newread\laxread@
\newwrite\laxwrite@
\let\fnpages@\empty
\def\Finit@#1#2\Finit@{\let\nextii@#1\def\nextiii@{#2}}
\catcode`\~=11
\def\getparts@ @#1~#2~#3~#4~#5~#6{\def\nextiv@{#1}%
 \def\nextiii@{#2~#3~#4~#5~}\count@#6\relax}
\newif\ifdocument@
\def\document{\ifdocument@\else\global\document@true
 \let\fontlist@\empty
 \immediate\openin\laxread@=\jobname.lax\relax
 {\endlinechar\m@ne\noexpands@\catcode`\@=11 \catcode`\~=11
  \loop\ifeof\laxread@\else
   \read\laxread@ to\next@
   \ifx\next@\empty
   \else
    \expandafter\Finit@\next@\Finit@
    \if\nextii@ F%
     \expandafter\rightadd@\nextiii@\to\fnpages@
    \else
     \expandafter\getparts@\next@
     \edef\next@{\gdef\csname\nextiv@ @L\endcsname{\nextiii@\number\count@}}%
     \next@
    \fi
   \fi
  \repeat}%
 \immediate\closein\laxread@
 \immediate\openout\laxwrite@=\jobname.lax\relax\fi}
\let\thelabel@\relax
\def\thelabels@{\thelabel@ ~\thelabel@@ ~\thelabel@@@ ~\thelabel@@@@ ~}
\def\label#1{\prevanish@
 \ifx\thelabel@\relax
  \Err@{There's nothing here to be labelled}%
 \else
  {\noexpands@
  \expandafter\ifx\csname#1@L\endcsname\relax
   \expandafter\xdef\csname#1@L\endcsname{\thelabels@0}%
   \immediate\write\laxwrite@{@#1~\thelabels@1}%
  \else
   \edef\next@{@~\csname#1@L\endcsname}%
    \expandafter\getparts@\next@
    \ifodd\count@
    \expandafter\xdef\csname#1@L\endcsname{\thelabels@0}%
    \immediate\write\laxwrite@{@#1~\thelabels@1}%
   \else
    \Err@{Label #1 already used}%
   \fi
  \fi
  }%
 \fi
 \postvanish@}
\rightadd@\label\to\vanishlist@
\def\thepages@{\page@N{\number\page@C}~%
 \page@S{\page@P\page@N{\number\page@C}\page@Q}~%
 \number\page@C ~\page@P\page@N{\number\page@C}\page@Q ~}
\def\pagelabel#1{\prevanish@
 \expandafter\ifx\csname#1@L\endcsname\relax
  {\noexpands@
  \expandafter\xdef\csname#1@L\endcsname{\thepages@2}}%
  \write\laxwrite@{@#1~\thepages@3}%
 \else
  {\noexpands@
  \edef\next@{@~\csname#1@L\endcsname}%
  \expandafter\getparts@\next@
  \ifodd\count@
   \ifnum\count@=\@ne
    \expandafter\xdef\csname#1@L\endcsname{\thelabels@2}%
   \fi
   \write\laxwrite@{@#1~\thepages@3}%
  \else
   \Err@{Label #1 already used}%
  \fi
  }%
 \fi
 \postvanish@}
\rightadd@\pagelabel\to\vanishlist@
\newif\ifreferr@
\referr@true
\def\RefErrors{\global\referr@true}
\def\RefWarnings{\global\referr@false}
\setbox\z@\hbox{\global\count@=`^^30}
\ifnum\count@=48 \let\versionthree@\relax\fi
\def\nolabel@#1#2#3{\expandafter\ifx\csname#2@L\endcsname\relax
 \ifreferr@\Err@{No \noexpand\label found for #2}\else
 \W@{Warning: No \noexpand\label found for #2.}%
 \ifx\versionthree@\relax\W@{l.\number\inputlineno\space ... \string#1{#2}}\fi
 \fi#3\else}
\def\csL@#1{{\noexpands@\xdef\Next@{\csname#1@L\endcsname}}}
\def\ref#1{\nolabel@\ref{#1}\relax
 \DNii@##1~##2\nextii@{##1}%
 \csL@{#1}\expandafter\nextii@\Next@\nextii@\fi}
\def\Ref#1{\nolabel@\Ref{#1}\relax
 \DNii@##1~##2~##3\nextii@{##2}%
 \csL@{#1}\expandafter\nextii@\Next@\nextii@\fi}
\def\nref#1{\nolabel@\nref{#1}\relax
 \DNii@##1~##2~##3~##4\nextii@{##3}%
 \csL@{#1}\expandafter\nextii@\Next@\nextii@\fi}
\def\pref#1{\nolabel@\pref{#1}\relax
 \DNii@##1~##2~##3~##4~##5\nextii@{##4}%
 \csL@{#1}\expandafter\nextii@\Next@\nextii@\fi}
\let\pref@\pref
\def\Evaluatenref#1{\nolabel@\Evaluatenref{#1}{\gdef\Nref{-10000 }}%
 \DNii@##1~##2~##3~##4\nextii@{\DNii@{##3}}%
 \csL@{#1}\expandafter\nextii@\Next@\nextii@
 \xdef\Nref{\nextii@}\fi}
\def\Evaluatepref#1{\nolabel@\Evaluatepref{#1}{\global\let\Pref\empty}%
 \DNii@##1~##2~##3~##4~##5\nextii@{\DNii@{##4}}%
 \csL@{#1}\expandafter\nextii@\Next@\nextii@
 \xdef\Pref{\nextii@}\fi}
\def\readlax#1{\immediate\openin\laxread@=#1.lax\relax
 \ifeof\laxread@\W@{}\W@{File #1.lax not found.}\W@{}\fi
 {\endlinechar\m@ne\noexpands@\catcode`\@=11 \catcode`\~=11
  \loop\ifeof\laxread@\else
   \read\laxread@ to\nextv@
   \ifx\nextv@\empty
   \else
    \expandafter\Finit@\nextv@\Finit@
    \ifx\nextii@ F%
    \else
     \expandafter\getparts@\nextv@
     \expandafter\ifx\csname\nextiv@ @L\endcsname\relax
      \edef\next@{\gdef\csname\nextiv@ @L\endcsname
       {\nextiii@\ifnum\count@=\@ne0\else2\fi}}%
      \next@
     \else
      \Err@{Label \nextiv@\space in #1.lax already used}%
     \fi
    \fi
   \fi
  \repeat}%
 \immediate\closein\laxread@}
\catcode`\~=\active

\def\FNSSP@{\FNSS@\pretendspace@}
\everydisplay{\csname displaymath \endcsname}
\expandafter\def\csname displaymath \endcsname#1$${#1$$\FNSSP@}
\def\locallabel@{\let\thelabel@\Thelabel@\let\thelabel@@\Thelabel@@
 \let\thelabel@@@\Thelabel@@@\let\thelabel@@@@\Thelabel@@@@}
\newcount\tag@C
\tag@C\z@
\let\tag@P\empty
\let\tag@Q\empty
\def\tag@S#1{{\rm(}{#1\/}{\rm)}}
\let\tag@N\arabic
\def\tag@F{\rm}
\def\maketag@{\FN@\maketag@@}
\def\maketag@@{\ifx\next\relax\DN@\relax{\FN@\maketag@@}\else
 \ifx\next"\let\next@\maketag@@@\else
 \let\next@\maketag@@@@\fi\fi\next@}
\def\xdefThelabel@#1{\xdef\Thelabel@{#1{\Thelabel@@@}}}
\def\xdefThelabel@@#1{\xdef\Thelabel@@{#1{\Thelabel@@@@}}}
\def\maketag@@@@#1\maketag@{\global\advance\tag@C\@ne
 {\noexpands@
  \xdef\Thelabel@@@{\number\tag@C}%
  \xdefThelabel@\tag@N
  \xdef\Thelabel@@@@{\ifmathtags@$\tag@P\Thelabel@\tag@Q$\else
   \tag@P\Thelabel@\tag@Q\fi}%
  \xdefThelabel@@\tag@S
  }%
 \locallabel@
 \hbox{\tag@F\thelabel@@}%
 #1}
\def\Qlabel@#1{{\noexpands@\xdef\Thelabel@@{#1}%
 \let\style\empty\xdef\Thelabel@@@@{#1}%
 \let\pre\empty\let\post\empty\xdef\Thelabel@{#1}%
 \let\numstyle\empty\xdef\Thelabel@@@{#1}}}
\def\maketag@@@"#1"#2\maketag@{%
 {\let\pre\tag@P\let\post\tag@Q\let\style\tag@S\let\numstyle\tag@N
  \hbox{\tag@F#1}%
  \noexpands@
  \Qlabel@{#1}%
  }%
 \locallabel@
 #2}
\def\align@{\inalign@true\inany@true
 \vspace@\allowdisplaybreak@\displaybreak@\intertext@
 \def\tag{\global\tag@true\ifnum\and@=\z@
  \DN@{&\omit\global\rwidth@\z@&\relax}\else
  \DN@{&\relax}\fi\next@}%
 \iftagsleft@\DN@{\csname align \endcsname}\else
  \DN@{\csname align \space\endcsname}\fi\next@}
\def\noset@{\def\Offset##1##2{\prevanish@\postvanish@}%
 \def\Reset##1##2{\prevanish@\postvanish@}}
\def\measure@#1\endalign{\global\lwidth@\z@\global\rwidth@\z@
 \global\maxlwidth@\z@\global\maxrwidth@\z@
 \global\and@\z@
 \setbox\z@\vbox
  {\noset@\everycr{\noalign{\global\tag@false\global\and@\z@}}\Let@
  \halign{\setboxz@h{$\m@th\displaystyle{\@lign##}$}%
   \global\lwidth@\wdz@
   \ifdim\lwidth@>\maxlwidth@\global\maxlwidth@\lwidth@\fi
   \global\advance\and@\@ne
   &\setboxz@h{$\m@th\displaystyle{{}\@lign##}$}\global\rwidth@\wdz@
   \ifdim\rwidth@>\maxrwidth@\global\maxrwidth@\rwidth@\fi
   \global\advance\and@\@ne
   &\Tag@\eat@{##}\crcr#1\crcr}}%
 \totwidth@\maxlwidth@\advance\totwidth@\maxrwidth@}
\def\prepost@{\global\let\tag@P@\tag@P\global\let\tag@Q@\tag@Q}
\def\reprepost@{\let\tag@P\tag@P@\let\tag@Q\tag@Q@}
\expandafter\def\csname align \space\endcsname#1\endalign
 {\measure@#1\endalign\global\and@\z@
 \ifingather@\everycr{\noalign{\global\and@\z@}}\else\displ@y@\fi
 \Let@\tabskip\centering@
 \halign to\displaywidth
  {\hfil\strut@\setboxz@h{$\m@th\displaystyle{\@lign##\prepost@}$}%
  \boxz@\global\advance\and@\@ne
  \tabskip\z@skip
  &\setboxz@h{$\m@th\displaystyle{{}\@lign##\prepost@}$}%
  \global\rwidth@\wdz@\boxz@\hfil\global\advance\and@\@ne
  \tabskip\centering@
  &\setboxz@h{\@lign\strut@\reprepost@\maketag@##\maketag@}%
  \dimen@\displaywidth\advance\dimen@-\totwidth@
  \divide\dimen@\tw@\advance\dimen@\maxrwidth@\advance\dimen@-\rwidth@
  \ifdim\dimen@<\tw@\wdz@\llap{\vtop{\normalbaselines\null\boxz@}}%
  \else\llap{\boxz@}\fi
  \tabskip\z@skip
  \crcr#1\crcr
  \black@\totwidth@}}
\expandafter\def\csname align \endcsname#1\endalign{\measure@#1\endalign
 \global\and@\z@
 \ifdim\totwidth@>\displaywidth\let\displaywidth@\totwidth@\else
  \let\displaywidth@\displaywidth\fi
 \ifingather@\everycr{\noalign{\global\and@\z@}}\else\displ@y@\fi
 \Let@\tabskip\centering@\halign to\displaywidth
  {\hfil\strut@\setboxz@h{$\m@th\displaystyle{\@lign##\prepost@}$}%
  \global\lwidth@\wdz@\global\lineht@\ht\z@
  \boxz@\global\advance\and@\@ne
  \tabskip\z@skip&\setboxz@h{$\m@th\displaystyle{{}\@lign##\prepost@}$}%
  \ifdim\ht\z@>\lineht@\global\lineht@\ht\z@\fi
  \boxz@\hfil\global\advance\and@\@ne
  \tabskip\centering@&\kern-\displaywidth@
  \setboxz@h{\@lign\strut@\reprepost@\maketag@##\maketag@}%
  \dimen@\displaywidth\advance\dimen@-\totwidth@
  \divide\dimen@\tw@\advance\dimen@\maxlwidth@\advance\dimen@-\lwidth@
  \ifdim\dimen@<\tw@\wdz@
   \rlap{\vbox{\normalbaselines\boxz@\vbox to\lineht@{}}}\else
   \rlap{\boxz@}\fi
  \tabskip\displaywidth@\crcr#1\crcr\black@\totwidth@}}
\def\attag@#1{\let\Maketag@\maketag@\let\TAG@\Tag@
 \let\Prepost@\prepost@\let\Reprepost@\reprepost@
 \let\Tag@\relax\let\maketag@\relax
 \let\prepost@\relax\let\reprepost@\relax
 \ifmeasuring@
  \def\llap@##1{\setboxz@h{##1}\hbox to\tw@\wdz@{}}%
  \def\rlap@##1{\setboxz@h{##1}\hbox to\tw@\wdz@{}}%
 \else\let\llap@\llap\let\rlap@\rlap\fi
 \toks@{\hfil\strut@
  $\m@th\displaystyle{\@lign\the\hashtoks@\prepost@}$%
  \tabskip\z@skip\global\advance\and@\@ne&
  $\m@th\displaystyle{{}\@lign\the\hashtoks@\prepost@}$\hfil
  \ifxat@\tabskip\centering@\fi\global\advance\and@\@ne}%
 \iftagsleft@
  \toks@@{\tabskip\centering@&\Tag@\kern-\displaywidth
   \rlap@{\@lign\reprepost@\maketag@\the\hashtoks@\maketag@}%
   \global\advance\and@\@ne\tabskip\displaywidth}\else
  \toks@@{\tabskip\centering@&\Tag@\llap@{\@lign\reprepost@\maketag@
   \the\hashtoks@\maketag@}\global\advance\and@\@ne\tabskip\z@skip}\fi
 \atcount@#1\relax\advance\atcount@\m@ne
 \loop\ifnum\atcount@>\z@
  \toks@\expandafter{\the\toks@&\hfil$\m@th\displaystyle{\@lign
  \the\hashtoks@\prepost@}$\global\advance\and@\@ne
  \tabskip\z@skip
  &$\m@th\displaystyle{{}\@lign\the\hashtoks@\prepost@}$\hfil\ifxat@
  \tabskip\centering@\fi\global\advance\and@\@ne}\advance\atcount@\m@ne
 \repeat
 \edef\preamble@{\the\toks@\the\toks@@}%
 \edef\preamble@@{\preamble@}%
 \let\maketag@\Maketag@\let\Tag@\TAG@
 \let\prepost@\Prepost@\let\reprepost@\Reprepost@}
\def\unlabel@{\def\label##1{\prevanish@\postvanish@}%
 \def\pagelabel##1{\prevanish@\postvanish@}}
\newcount\tag@CC
\expandafter\def\csname alignat \endcsname#1#2\endalignat
 {\inany@true\xat@false
 \def\tag{\global\tag@true
  \count@#1\relax\multiply\count@\tw@\advance\count@\m@ne
  \gdef\tag@{&}%
  \loop\ifnum\count@>\and@\xdef\tag@{&\omit\tag@}%
  \advance\count@\m@ne\repeat
  \tag@\relax}%
 \vspace@\allowdisplaybreak@\displaybreak@\intertext@
 \displ@y@\measuring@true\tag@CC\tag@C
 \setbox\savealignat@\hbox{\noset@\unlabel@$\m@th\displaystyle\Let@
  \attag@{#1}\vbox{\halign{\span\preamble@@\crcr#2\crcr}}$}%
 \measuring@false
 \Let@\attag@{#1}\tag@C\tag@CC
 \tabskip\centering@\halign to\displaywidth
  {\span\preamble@@\crcr#2\crcr\black@{\wd\savealignat@}}}
\expandafter\def\csname xalignat \endcsname#1#2\endxalignat
 {\inany@true\xat@true
 \def\tag{\global\tag@true
  \count@#1\relax\multiply\count@\tw@\advance\count@\m@ne
  \gdef\tag@{&}%
  \loop\ifnum\count@>\and@\xdef\tag@{&\omit\tag@}%
  \advance\count@\m@ne\repeat
  \tag@\relax}%
 \vspace@\allowdisplaybreak@\displaybreak@\intertext@
 \displ@y@\measuring@true\tag@CC\tag@C
 \setbox\savealignat@\hbox{\noset@\unlabel@$\m@th\displaystyle\Let@
  \attag@{#1}\vbox{\halign{\span\preamble@@\crcr#2\crcr}}$}%
 \measuring@false\Let@\attag@{#1}\tag@C\tag@CC
 \tabskip\centering@\halign to\displaywidth
 {\span\preamble@@\crcr#2\crcr\black@{\wd\savealignat@}}}
\def\gather{\RIfMIfI@\DN@{\onlydmatherr@\gather}\else
 \ingather@true\inany@true\def\tag{&\relax}%
 \vspace@\allowdisplaybreak@\displaybreak@\intertext@
 \displ@y\Let@
 \iftagsleft@\DN@{\csname gather \endcsname}\else
  \DN@{\csname gather \space\endcsname}\fi\fi
 \else\DN@{\onlydmatherr@\gather}\fi\next@}
\def\exstring@{\expandafter\eat@\string}
\def\newcounter#1{\define#1{}%
 \edef\next@{\def\noexpand#1{\futurelet\noexpand\next
  \csname\exstring@#1@Z\endcsname}}\next@
 \edef\next@{\def\csname\exstring@#1@Z\endcsname
  {\global\advance\csname\exstring@#1@C\endcsname\@ne
  {\csname\exstring@#1@F\endcsname\csname\exstring@#1@S\endcsname
   {\csname\exstring@#1@P\endcsname\csname\exstring@#1@N\endcsname
   {\noexpand\number\csname\exstring@#1@C\endcsname}%
   \csname\exstring@#1@Q\endcsname}}%
  \noexpand\ifx\noexpand\next\noexpand\label
   \def\noexpand\next@\noexpand\label########1{{\noexpand\noexpands@
    \xdef\noexpand\Thelabel@{\csname\exstring@#1@N\endcsname
     {\noexpand\number\csname\exstring@#1@C\endcsname}}%
    \xdef\noexpand\Thelabel@@@{\noexpand\number
     \csname\exstring@#1@C\endcsname}%
    \xdef\noexpand\Thelabel@@{\csname\exstring@#1@S\endcsname
     {\csname\exstring@#1@P\endcsname
     \csname\exstring@#1@N\endcsname
     {\noexpand\number\csname\exstring@#1@C\endcsname}%
     \csname\exstring@#1@Q\endcsname}}%
    \xdef\noexpand\Thelabel@@@@{\csname\exstring@#1@P\endcsname
     \csname\exstring@#1@N\endcsname
     {\noexpand\number\csname\exstring@#1@C\endcsname}%
     \csname\exstring@#1@Q\endcsname}}%
    {\noexpand\locallabel@\noexpand\label{########1}}}%
   \noexpand\else\let\noexpand\next@\relax\noexpand\fi\noexpand\next@}}\next@
 \expandafter\newcount@\csname\exstring@#1@C\endcsname
 \expandafter\let\csname\exstring@#1@N\endcsname\arabic
 \expandafter\def\csname\exstring@#1@S\endcsname##1{##1\/}%
 \expandafter\let\csname\exstring@#1@P\endcsname\empty
 \expandafter\let\csname\exstring@#1@Q\endcsname\empty
 \expandafter\def\csname\exstring@#1@F\endcsname{\rm}%
 }
\def\HASH@#1#2{\ifnum#2=\z@\else
 \edef\next@{\toks@{\the\toks@\the\hashtoks@#2}%
 \toks@@{\the\toks@@{\the\hashtoks@#2}}}\next@\expandafter\HASH@\fi}
\def\HASH@@{\toks@{}\toks@@{}\expandafter\HASH@\macpar@00}
\def\usecounter#1#2{\expandafter\ifx\csname\exstring@#1@Z\endcsname
 \relax\Err@{\noexpand#1not created with \string\newcounter}\fi
 \expandafter\let\csname\exstring@#1@@Z\endcsname\relax
 \expandafter\let\csname\exstring@#1@@Z@\endcsname\relax
 \expandafter\let\csname\exstring@#1@@Z@@\endcsname\relax
 \edef\next@{\def\noexpand#2{\futurelet\noexpand\next
  \csname\exstring@#1@@Z\endcsname}}\next@
 \edef\next@{\def\csname\exstring@#1@@Z\endcsname{\noexpand\ifx
  \noexpand\next\noexpand\label\def\noexpand\next@\noexpand\label
   ########1{\csname\exstring@#1@@Z@\endcsname
   {\noexpand#1\noexpand\label{########1}}}%
   \noexpand\else\noexpand\ifx\noexpand\next
   \noexpand"\def\noexpand\next@\noexpand"########1\noexpand"%
   {\csname\exstring@#1@@Z@\endcsname{{\expandafter\noexpand
   \csname\exstring@#1@F\endcsname
   \let\noexpand\pre\expandafter\noexpand\csname\exstring@#1@P\endcsname
   \let\noexpand\post\expandafter\noexpand\csname\exstring@#1@Q\endcsname
   \let\noexpand\style\expandafter\noexpand\csname\exstring@#1@S\endcsname
   \let\noexpand\numstyle\expandafter\noexpand\csname\exstring@#1@N\endcsname
   ########1}}}\noexpand\else
   \def\noexpand\next@{\csname\exstring@#1@@Z@\endcsname{\noexpand#1}}%
   \noexpand\fi\noexpand\fi\noexpand\next@}}\next@
 \def\next@{\expandafter\expandafter\expandafter\unmacro@\expandafter
  \meaning\csname\exstring@#1@@Z@@\endcsname\unmacro@
  \HASH@@
  \edef\next@{\def\csname\exstring@#1@@Z@\endcsname\the\toks@{%
   \expandafter\noexpand\csname\exstring@#1@@Z@@\endcsname\the\toks@@
   \noexpand\FNSSP@}}\next@}%
 \afterassignment\next@
 \expandafter\def\csname\exstring@#1@@Z@@\endcsname}
\def\listbi@{\penalty50 \medskip}
\def\listbii@{\penalty100 \smallskip}
\let\listbiii@\relax
\let\listbiv@\relax
\let\listbv@\relax
\def\listmi@{\advance\leftskip30\p@\relax}
\let\listmii@\listmi@
\let\listmiii@\listmi@
\let\listmiv@\listmi@
\let\listmv@\listmi@
\def\itemi@#1{\noindent@@\llap{#1\hskip5\p@}}
\let\itemii@\itemi@
\let\itemiii@\itemi@
\let\itemiv@\itemi@
\let\itemv@\itemi@
\def\liste@{\penalty-50 \medskip}
\def\listei@{\penalty-100 \smallskip}
\let\listeii@\relax
\let\listeiii@\relax
\let\listeiv@\relax
\expandafter\newcount\csname list@C1\endcsname
\csname list@C1\endcsname\z@
\expandafter\newcount\csname list@C2\endcsname
\csname list@C2\endcsname\z@
\expandafter\newcount\csname list@C3\endcsname
\csname list@C3\endcsname\z@
\expandafter\newcount\csname list@C4\endcsname
\csname list@C4\endcsname\z@
\expandafter\newcount\csname list@C5\endcsname
\csname list@C5\endcsname\z@
\expandafter\let\csname list@P1\endcsname\empty
\expandafter\let\csname list@P2\endcsname\empty
\expandafter\let\csname list@P3\endcsname\empty
\expandafter\let\csname list@P4\endcsname\empty
\expandafter\let\csname list@P5\endcsname\empty
\expandafter\let\csname list@Q1\endcsname\empty
\expandafter\let\csname list@Q2\endcsname\empty
\expandafter\let\csname list@Q3\endcsname\empty
\expandafter\let\csname list@Q4\endcsname\empty
\expandafter\let\csname list@Q5\endcsname\empty
\expandafter\def\csname list@S1\endcsname#1{{\rm(}{#1\/}{\rm)}}
\expandafter\def\csname list@S2\endcsname#1{{\rm(}{#1\/}{\rm)}}
\expandafter\def\csname list@S3\endcsname#1{{\rm(}{#1\/}{\rm)}}
\expandafter\def\csname list@S4\endcsname#1{{\rm(}{#1\/}{\rm)}}
\expandafter\def\csname list@S5\endcsname#1{{\rm(}{#1\/}{\rm)}}
\expandafter\let\csname list@N1\endcsname\arabic
\expandafter\let\csname list@N2\endcsname\arabic
\expandafter\let\csname list@N3\endcsname\arabic
\expandafter\let\csname list@N4\endcsname\arabic
\expandafter\let\csname list@N5\endcsname\arabic
\expandafter\def\csname list@F1\endcsname{\rm}
\expandafter\def\csname list@F2\endcsname{\rm}
\expandafter\def\csname list@F3\endcsname{\rm}
\expandafter\def\csname list@F4\endcsname{\rm}
\expandafter\def\csname list@F5\endcsname{\rm}
\newcount\listlevel@
\listlevel@\z@
\def\list@@C{\csname list@C\number\listlevel@\endcsname}
\def\list@@P{\csname list@P\number\listlevel@\endcsname}
\def\list@@Q{\csname list@Q\number\listlevel@\endcsname}
\def\list@@S{\csname list@S\number\listlevel@\endcsname}
\def\list@@N{\csname list@N\number\listlevel@\endcsname}
\def\list@@F{\csname list@F\number\listlevel@\endcsname}
\newif\iffirstitemi@
\newif\iffirstitemii@
\newif\iffirstitemiii@
\newif\iffirstitemiv@
\newif\iffirstitemv@
\def\Firstitem@true{\csname firstitem\romannumeral\listlevel@
 @true\endcsname}
\def\Firstitem@false{\csname firstitem\romannumeral\listlevel@
 @false\endcsname}
\def\Listm@{\csname listm\romannumeral\listlevel@ @\endcsname}
\def\Item@{\csname item\romannumeral\listlevel@ @\endcsname}
\def\Liste@{\csname liste\romannumeral\listlevel@ @\endcsname}
\newif\iflistcontinue@
\def\keepitem{\listcontinue@true}
\newcount\list@C@
\def\list{%
 \iflistcontinue@\csname list@C1\endcsname\csname list@C@\endcsname\fi
 \global\csname list@C2\endcsname\z@
 \global\csname list@C3\endcsname\z@
 \global\csname list@C4\endcsname\z@
 \global\csname list@C5\endcsname\z@
 \begingroup
 \firstitemi@true
 \listlevel@\@ne
 \def\item{\FN@\item@}%
 \FN@\list@}
\Invalid@\runinitem
\def\list@{\ifx\next\par
 \DN@\par{\FN@\list@}\else
 \ifx\next\runinitem
  \DN@\runinitem{\FN@\runinitem@}\else
  \DN@{\par\dimen@\parskip\parskip\dimen@}\fi\fi\next@}
\newif\ifoutlevel@
\newif\ifrunin@
\def\item@{%
 \ifoutlevel@\Liste@\outlevel@false\fi
 \ifrunin@\runin@false\par
  \dimen@\parskip\parskip\dimen@
  \Listm@\fi
 \iffirstitemi@\listbi@\listmi@\firstitemi@false\else\par\fi
 \iffirstitemii@\listbii@\listmii@\firstitemii@false\else\par\fi
 \iffirstitemiii@\listbiii@\listmiii@\firstitemiii@false\else\par\fi
 \iffirstitemiv@\listbiv@\listmiv@\firstitemiv@false\else\par\fi
 \iffirstitemv@\listbv@\listmv@\firstitemv@false\else\par\fi
 \DN@"##1"{{\let\pre\list@@P\let\post\list@@Q
  \let\style\list@@S\let\numstyle\list@@N
  \vskip-\parskip
  \Item@{\list@@F##1}%
  \noexpands@
  \Qlabel@{##1}}%
  \locallabel@
  \FNSSP@}%
 \DNii@{\global\advance\list@@C\@ne
  {\noexpands@
   \xdef\Thelabel@@@{\number\list@@C}%
   \xdefThelabel@\list@@N
   \xdef\Thelabel@@@@{\list@@P\Thelabel@\list@@Q}%
   \xdefThelabel@@\list@@S
  }%
  \locallabel@
  \vskip-\parskip
  \Item@{\list@@F\thelabel@@}%
  \FN@\pretendspace@}%
 \ifx\next"\expandafter\next@\else\expandafter\nextii@\fi}
\def\runinitem@{%
  \runin@true
  \Firstitem@false
  \DN@"##1"{{\let\pre\list@@P\let\post\list@@Q
   \let\style\list@@S\let\numstyle\list@@N
   \unskip\space{\list@@F##1} %
   \noexpands@
   \Qlabel@{##1}}%
   \locallabel@
   \ignorespaces}%
  \DNii@{\global\advance\list@@C\@ne
   {\noexpands@
    \xdef\Thelabel@@@{\number\list@@C}%
    \xdefThelabel@\list@@N
    \xdef\Thelabel@@@@{\list@@P\Thelabel@\list@@Q}%
    \xdefThelabel@@\list@@S
   }%
   \locallabel@
   \unskip\space{\list@@F\thelabel@@} }%
  \ifx\next"\expandafter\next@\else\expandafter\nextii@\fi}
\def\inlevel{\ifnum\listlevel@=5
 \DN@{\Err@{Already 5 levels down}}\else
 \DN@{\begingroup\advance\listlevel@\@ne
 \Firstitem@true\FN@\inlevel@}\fi\next@}
\def\inlevel@{\ifx\next\par
 \DN@\par{\FN@\inlevel@}\else
 \ifx\next\runinitem
  \DN@\runinitem{\FN@\runinitem@}\else
  \let\next@\relax\fi\fi\next@}
\def\outlevel{\ifnum\listlevel@=\@ne
 \Err@{At top level}\else
 \par\global\list@@C\z@\endgroup\outlevel@true\fi}
\def\endlist{%
 \expandafter\global\csname list@C@\endcsname\csname list@C1\endcsname
 \par
 \global\toks\@ne{}\count@\listlevel@
 {\loop
  \ifnum\count@>\z@\global\toks\@ne\expandafter{\the\toks\@ne\endgroup}%
  \advance\count@\m@ne
  \repeat}%
 \the\toks\@ne
 \liste@
 \listcontinue@false\global\csname list@C1\endcsname\z@
 \vskip-\parskip
 \noindent@@
 \FN@\pretendspace@}
\newif\iffirstdescribe@
\def\describe{\par
 \begingroup\firstdescribe@true
 \def\item##1{%
  \iffirstdescribe@\penalty50 \medskip\vskip-\parskip
  \firstdescribe@false\else\par\fi
  \noindent@@\hangindent2pc\hangafter\@ne
  {\bf##1}\hskip.5em}}

\Invalid@\pullin
\Invalid@\pullinmore
\newif\iffirstpull@
\def\margins{\par\begingroup\firstpull@true
 \def\pullin##1##2{\par
  \iffirstpull@\firstpull@false\else\endgroup\fi
  \begingroup\DN@{##1}%
  \ifx\next@\empty\leftskip\z@\else\ifx\next@\space\leftskip\z@
  \else\leftskip##1\fi\fi
  \DN@{##2}\ifx\next@\empty\rightskip\z@\else\ifx\next@\space
  \rightskip\z@\else\rightskip##2\fi\fi\ignorespaces}%
 \def\pullinmore##1##2{\par
  \xdef\Next@{\leftskip\the\leftskip\relax\rightskip\the\rightskip\relax}%
  \iffirstpull@\firstpull@false\else\endgroup\fi
  \begingroup\Next@
  \DN@{##1}%
  \ifx\next@\empty\else\ifx\next@\space\else\advance\leftskip##1\fi\fi
  \DN@{##2}\ifx\next@\empty\else\ifx\next@\space\else
  \advance\rightskip##2\fi\fi\ignorespaces}}

\newif\ifnopunct@
\newif\ifnospace@
\newif\ifoverlong@
\let\nofrillslist@\empty
\let\overlonglist@\empty
\def\nopunct{\nopunct@true\FN@\nopunct@}
\def\nospace{\nospace@true\FN@\nospace@}
\def\overlong{\overlong@true\FN@\overlong@}
\def\nopunct@{\ifx\next\nospace
 \DN@\nospace{\nospace@true\FN@\nopnos@}\else\ifx\next\overlong
 \DN@\overlong{\overlong@true\FN@\nopol@}\else
 \let\next@\nopunct@@\fi\fi\next@}
\def\nopunct@@#1{\ismember@\nofrillslist@#1%
 \iftest@\let\next@#1\else
 \DN@{\nopunct@false\Err@{\noexpand\nopunct can't be used with
 \string#1}#1}\fi\next@}
\def\nospace@{\ifx\next\nopunct
 \DN@\nopunct{\nopunct@true\FN@\nopnos@}\else\ifx\next\overlong
 \DN@\overlong{\overlong@true\FN@\nosol@}\else
 \let\next@\nospace@@\fi\fi\next@}
\def\nospace@@#1{\ismember@\nofrillslist@#1%
 \iftest@\let\next@#1\else
 \DN@{\nospace@false\Err@{\noexpand\nospace can't be used with
 \string#1}#1}\fi\next@}
\def\overlong@{\ifx\next\nopunct
 \DN@\nopunct{\nopunct@true\FN@\nopol@}\else\ifx\next\nospace
 \DN@\nospace{\nospace@true\FN@\nosol@}\else
 \let\next@\overlong@@\fi\fi\next@}
\def\overlong@@#1{\ismember@\overlonglist@#1%
 \iftest@\let\next@#1\else
 \DN@{\overlong@false\Err@{\noexpand\overlong can't be used with
 \string#1}#1}\fi\next@}
\def\nopnos@{\ifx\next\overlong
 \DN@\overlong{\overlong@true\nopnosol@}\else
 \let\next@\nopnos@@\fi\next@}
\def\nopol@{\ifx\next\nospace
 \DN@\nospace{\nospace@true\nopnosol@}\else
 \let\next@\nopol@@\fi\next@}
\def\nosol@{\ifx\next\nopunct
 \DN@\nopunct{\nopunct@true\nopnosol@}\else
 \let\next@\nosol@@\fi\next@}
\def\nopnos@@#1{\ismember@\nofrillslist@#1%
 \iftest@\let\next@#1\else
 \DN@{\nopunct@false\nospace@false
  \Err@{\noexpand\nopunct\noexpand\nospace
   can't be used with \string#1}#1}\fi\next@}
\def\testii@#1{\ismember@\nofrillslist@#1%
 \iftest@\let\nextiii@ T\else\let\nextiii@ F\fi
 \ismember@\overlonglist@#1%
 \iftest@\let\nextiv@ T\else\let\nextiv@ F\fi
 \test@false\if\nextiii@ T\if\nextiv@ T\test@true\fi\fi}
\def\nopol@@#1{\testii@{#1}%
 \iftest@\let\next@#1%
 \else\DN@{\if\nextiii@ T\else\nopunct@false\fi
  \if\nextiv@ T\else\overlong@false\fi
  \Err@{\if\nextiii@ T\else\noexpand\nopunct\fi
  \if\nextiv@ T\else\noexpand\overlong\fi can't be used
  with \string#1}#1}\fi\next@}
\def\nosol@@#1{\testii@{#1}%
 \iftest@\let\next@#1%
 \else\DN@{\if\nextiii@ T\else\nospace@false\fi
  \if\nextiv@ T\else\overlong@false\fi
  \Err@{\if\nextiii@ T\else\noexpand\nospace\fi
  \if\nextiv@ T\else\noexpand\overlong\fi can't be used
  with \string#1}#1}\fi\next@}
\def\nopnosol@#1{\testii@{#1}%
 \iftest@\let\next@#1%
 \else\DN@{\if\nextiii@ T\else\nopunct@false\nospace@false\fi
  \if\nextiv@ T\else\overlong@false\fi
  \Err@{\if\nextiii@ T\else\noexpand\nopunct\noexpand\nospace\fi
  \if\nextiv@ T\else\noexpand\overlong\fi can't be used
  with \string#1}#1}\fi\next@}
\def\punct@#1{\ifnopunct@\else#1\fi}
\def\addspace@#1{\ifnospace@\else#1\fi}
\def\hss@{\ifoverlong@\z@ plus\@m\p@ minus\@m\p@
 \else \z@ plus\@m\p@\fi}
\rightadd@\demo\to\nofrillslist@
\newif\ifclaim@
\def\exxx@{\expandafter\expandafter\expandafter\eat@\expandafter\string}
\let\colon@:
\def\demo#1{\ifclaim@
 \Err@{Previous \expandafter\noexpand\claimtype@ has
  no matching \string\end\exxx@\claimtype@}%
 \let\next@\relax
 \else
  \par
  \ifdim\lastskip<\smallskipamount\removelastskip\smallskip\fi
  \begingroup
  \noindent@@{\smc\ignorespaces#1\unskip
   \punct@{\null\colon@}\addspace@\enspace}%
  \nopunct@false\nospace@false
  \rm
  \DN@{\FNSSP@}%
 \fi
 \next@}
\def\enddemo{\par\endgroup\nopunct@false\nospace@false\smallskip}
\rightadd@\claim\to\nofrillslist@
\def\claim@F{\smc}
\def\claim@@@F{\csname\exxx@\claimtype@ @F\endcsname}
\def\claimformat@#1#2#3{%
 \medbreak\noindent@@{\smc#1 {\claim@@@F#2} #3%
 \punct@{\null.}\addspace@\enspace}\sl}
\def\claimformat@@#1#2{\claimformat@{\ignorespaces#1\unskip}%
 {\ifx\thelabel@@\empty\unskip\else\thelabel@@\fi}%
 {\ignorespaces#2\unskip}%
 \let\Claimformat@@\claimformat@@\FNSSP@}
\let\Claimformat@@\claimformat@@
\def\claim@@@P{\csname\exxx@\claimtype@ @P\endcsname}
\def\claim@@@Q{\csname\exxx@\claimtype@ @Q\endcsname}
\def\claim@@@S{\csname\exxx@\claimtype@ @S\endcsname}
\def\claim@@@N{\csname\exxx@\claimtype@ @N\endcsname}
\def\claim@@@C{\csname claim@C\claimclass@\endcsname}
\newcount\claim@C
\claim@C\z@
\let\claim@P\empty
\let\claim@Q\empty
\def\claim@S#1{#1\/}
\let\claim@N\arabic
\def\claim{\claim@true\let\claimclass@\empty
 \def\claimtype@{\claim}\FN@\claim@}
\def\claim@{%
 \ifx\next\c
  \let\next@\claim@c
 \else
  \ifx\next"%
   \let\next@\claim@q
  \else
   \begingroup\global\advance\claim@C\@ne
   {\noexpands@
    \xdef\Thelabel@@@{\number\claim@C}%
    \xdefThelabel@\claim@N
    \xdef\Thelabel@@@@{\claim@P\Thelabel@\claim@Q}%
    \xdefThelabel@@\claim@S
   }%
   \locallabel@
   \let\next@\Claimformat@@
  \fi
 \fi
 \next@}
\def\claim@c\c#1{\claim@true\begingroup
 \expandafter
 \ifx\csname claim@C#1\endcsname\relax
  \expandafter\newcount@\csname claim@C#1\endcsname
  \global\csname claim@C#1\endcsname\@ne
 \else
  \global\advance\csname claim@C#1\endcsname\@ne
 \fi
 \def\claimclass@{#1}%
 {\noexpands@
  \xdef\Thelabel@@@{\number\claim@@@C}%
  \xdefThelabel@\claim@@@N
  \xdef\Thelabel@@@@{\claim@@@P\Thelabel@\claim@@@Q}%
  \xdefThelabel@@\claim@@@S
 }%
 \locallabel@
 \FNSS@\claim@c@}
\def\claim@q"#1"{\begingroup
 {\let\pre\claim@@@P\let\post\claim@@@Q
  \let\style\claim@@@S\let\numstyle\claim@@@N
  \noexpands@
  \Qlabel@{#1}}%
 \locallabel@
 \FNSS@\claim@q@}
\def\claim@c@{\ifx\next"%
 \global\advance\claim@@@C\m@ne\let\next@\claim@cq
 \else\let\next@\Claimformat@@\fi\next@}
\def\claim@cq"#1"{{\let\pre\claim@@@P\let\post\claim@@@Q
 \let\style\claim@@@S\let\numstyle\claim@@@N
 \noexpands@
 \Qlabel@{#1}}%
 \locallabel@
 \FNSS@\Claimformat@@}
\def\claim@q@{\ifx\next\c\expandafter\claim@qc
 \else\expandafter\Claimformat@@\fi}
\def\claim@qc\c#1{\expandafter\ifx\csname claim@C#1\endcsname\relax
 \expandafter\newcount@\csname claim@C#1\endcsname
 \global\csname claim@C#1\endcsname\z@\fi
 \FNSS@\Claimformat@@}
\def\endclaim{\endgroup\claim@false\nopunct@false\nospace@false
 \let\Claimformat@@\claimformat@@\medbreak}
\Invalid@\claimclause
\def\newclaim{\FN@\newclaim@}
\def\newclaim@{\ifx\next\claimclause
 \DN@\claimclause##1{\newclaim@@{##1}}\else
 \DN@{\newclaim@@\relax}\fi\next@}
\def\claimlist@{\\\claim}
\newtoks\claim@i
\newtoks\claim@v
\let\noclaimclause@=F
\def\newclaim@@#1#2#3\c#4#5{\define#2{}%
 \rightadd@#2\to\claimlist@\rightadd@#2\to\nofrillslist@%
 \expandafter\def\csname\exstring@#2@P\endcsname{\claim@P}%
 \expandafter\def\csname\exstring@#2@Q\endcsname{\claim@Q}%
 \expandafter\def\csname\exstring@#2@S\endcsname{\claim@S}%
 \expandafter\def\csname\exstring@#2@N\endcsname{\claim@N}%
 \expandafter\def\csname\exstring@#2@F\endcsname{\claim@F}%
 \expandafter\def\csname end\exstring@#2\endcsname{\endclaim}%
 \expandafter\ifx\csname claim@C#4\endcsname\relax
  \expandafter\newcount@\csname claim@C#4\endcsname
  \global\csname claim@C#4\endcsname\z@\fi
 \edef\next@{\let\csname\exstring@#2@C\endcsname
   \csname claim@C#4\endcsname}\next@
 \def#2{\ifx\noclaimclause@ T\else#1\fi
  \global\claim@i{#1}\gdef\claim@iv{#4}\global\claim@v{#5}%
  \def\claimtype@{#2}\def\Claimformat@@{\claimformat@@{#5}}\claim@c\c{#4}}}
\def\shortenclaim#1#2{\define#2{}%
 \ismember@\claimlist@#1%
 \iftest@
  \rightadd@#2\to\nofrillslist@%
  \expandafter\def\csname\exstring@#2@P\endcsname
   {\csname\exstring@#1@P\endcsname}%
  \expandafter\def\csname\exstring@#2@Q\endcsname
   {\csname\exstring@#1@Q\endcsname}%
  \expandafter\def\csname\exstring@#2@S\endcsname
   {\csname\exstring@#1@S\endcsname}%
  \expandafter\def\csname\exstring@#2@N\endcsname
   {\csname\exstring@#1@N\endcsname}%
  \expandafter\def\csname\exstring@#2@F\endcsname
   {\csname\exstring@#1@F\endcsname}%
  \expandafter\def\csname end\exstring@#2\endcsname{\endclaim}%
  \edef\next@{\let\csname\exstring@#2@C\endcsname
    \csname claim\exstring@#1C\endcsname}\next@
  \setbox\z@\vbox{\let\noclaimclause@ T#1""\relax\endgroup}%
  \edef#2{\the\claim@i
   \def\noexpand\claimtype@{\noexpand#2}%
   \def\noexpand\Claimformat@@{\noexpand\claimformat@@{\the\claim@v}\relax}%
   \noexpand\claim@c\noexpand\c{\claim@iv}}%
 \else
  \Err@{\noexpand#1not yet created by \string\newclaim}%
 \fi}
\def\classtest@#1{\DN@{#1}\ifx\next@\claimclass@
 \test@true\else\test@false\fi}
\def\typetest@#1{\DN@{#1}\ifx\next@\claimtype@\test@true\else
  \test@false\fi}
\newif\iftoc@
\def\tocfile{\iftoc@\else\alloc@@7\write\chardef\sixt@@n\toc@
 \immediate\openout\toc@=\jobname.toc
 \alloc@@7\write\chardef\sixt@@n\tic@
 \immediate\openout\tic@=\jobname.tic
 \global\toc@true\fi}
\rightadd@\hl\to\nofrillslist@
\rightadd@\HL\to\overlonglist@
\def\HL@@C{\csname HL@C\HLlevel@\endcsname}
\def\HL@@P{\csname HL@P\HLlevel@\endcsname}
\def\HL@@Q{\csname HL@Q\HLlevel@\endcsname}
\def\HL@@S{\csname HL@S\HLlevel@\endcsname}
\def\HL@@N{\csname HL@N\HLlevel@\endcsname}
\def\HL@@F{\csname HL@F\HLlevel@\endcsname}
\def\HL@@@C{\csname\exxx@\HLtype@ @C\endcsname}
\def\HL@@@P{\csname\exxx@\HLtype@ @P\endcsname}
\def\HL@@@Q{\csname\exxx@\HLtype@ @Q\endcsname}
\def\HL@@@S{\csname\exxx@\HLtype@ @S\endcsname}
\def\HL@@@N{\csname\exxx@\HLtype@ @N\endcsname}
\def\HL#1{\expandafter
 \ifx\csname HL@C#1\endcsname\relax
  \DN@{\Err@{\string\HL#1 not defined in this style}}%
 \else
  \DN@{\gdef\HLlevel@{#1}\def\HLname@{\HL{#1}}\let\HLtype@\relax\FNSS@\HL@}%
 \fi
 \next@}%
\newif\ifquoted@
\let\aftertoc@\relax
\def\HL@{%
 \DN@"##1"##2\endHL{\def\entry@{##2}\quoted@true
  {\noexpands@
  \ifx\HLtype@\relax
   \let\pre\HL@@P\let\post\HL@@Q\let\style\HL@@S\let\numstyle\HL@@N
  \else
   \let\pre\HL@@@P\let\post\HL@@@Q\let\style\HL@@@S\let\numstyle\HL@@@N
  \fi
  \Qlabel@{##1}\let\style\relax\xdef\Qlabel@@@@{##1}%
  \xdef\Thepref@{\Thelabel@@@@}}%
  \csname HL@\HLlevel@\endcsname##2\endHL
  \let\pref\Thepref@
  \csname HL@I\HLlevel@\endcsname
  \csname HL@J\HLlevel@\endcsname
  \let\pref\pref@
  \HLtoc@	
  \aftertoc@
  \let\aftertoc@\relax\overlong@false}%
 \DNii@##1\endHL{\def\entry@{##1}\quoted@false
  {\noexpands@
  \ifx\HLtype@\relax
   \global\advance\HL@@C\@ne
   \xdef\Thelabel@@@{\number\HL@@C}%
   \xdefThelabel@{\HL@@N}%
   \xdef\Thelabel@@@@{\HL@@P\Thelabel@\HL@@Q}%
   \xdefThelabel@@{\HL@@S}%
  \else
   \global\advance\HL@@@C\@ne
   \xdef\Thelabel@@@{\number\HL@@@C}%
   \xdefThelabel@{\HL@@@N}%
   \xdef\Thelabel@@@@{\HL@@@P\Thelabel@\HL@@@Q}%
   \xdefThelabel@@{\HL@@@S}%
  \fi
  \xdef\Thepref@{\Thelabel@@@@}}%
  \csname HL@\HLlevel@\endcsname##1\endHL
  \let\pref\Thepref@
  \csname HL@I\HLlevel@\endcsname
  \csname HL@J\HLlevel@\endcsname
  \let\pref\pref@
  \HLtoc@
  \aftertoc@
  \let\aftertoc@\relax\overlong@false}%
 \ifx\next"\expandafter\next@\else\expandafter\nextii@\fi}%
\Invalid@\endHL
\def\hl@@C{\csname hl@C\hllevel@\endcsname}
\def\hl@@P{\csname hl@P\hllevel@\endcsname}
\def\hl@@Q{\csname hl@Q\hllevel@\endcsname}
\def\hl@@S{\csname hl@S\hllevel@\endcsname}
\def\hl@@N{\csname hl@N\hllevel@\endcsname}
\def\hl@@F{\csname hl@F\hllevel@\endcsname}
\def\hl@@@C{\csname\exxx@\hltype@ @C\endcsname}
\def\hl@@@P{\csname\exxx@\hltype@ @P\endcsname}
\def\hl@@@Q{\csname\exxx@\hltype@ @Q\endcsname}
\def\hl@@@S{\csname\exxx@\hltype@ @S\endcsname}
\def\hl@@@N{\csname\exxx@\hltype@ @N\endcsname}
\def\hl#1{\expandafter
 \ifx\csname hl@C#1\endcsname\relax
  \DN@{\Err@{\string\hl#1 not defined in this style}}%
 \else
  \DN@{\gdef\hllevel@{#1}\def\hlname@{\hl{#1}}\let\hltype@\relax\FNSS@\hl@}%
 \fi
 \next@}
\def\hl@{%
 \DN@"##1"##2{\def\entry@{##2}\quoted@true
  {\noexpands@
  \ifx\hltype@\relax
   \let\pre\hl@@P\let\post\hl@@Q\let\style\hl@@S\let\numstyle\hl@@N
  \else
   \let\pre\hl@@@P\let\post\hl@@@Q\let\style\hl@@@S\let\numstyle\hl@@@N
  \fi
  \Qlabel@{##1}\let\style\relax\xdef\Qlabel@@@@{##1}%
  \xdef\Thepref@{\Thelabel@@@@}}%
  \csname hl@\hllevel@\endcsname{##2}%
  \let\pref\Thepref@
  \csname hl@I\hllevel@\endcsname
  \csname hl@J\hllevel@\endcsname
  \let\pref\pref@
  \hltoc@
  \aftertoc@
  \let\aftertoc@\relax\nopunct@false\nospace@false\FNSSP@}%
 \DNii@##1{\def\entry@{##1}\quoted@false
  {\noexpands@
  \ifx\hltype@\relax
   \global\advance\hl@@C\@ne
   \xdef\Thelabel@@@{\number\hl@@C}%
   \xdefThelabel@{\hl@@N}%
   \xdef\Thelabel@@@@{\hl@@P\Thelabel@\hl@@Q}%
   \xdefThelabel@@{\hl@@S}%
  \else
   \global\advance\hl@@@C\@ne
   \xdef\Thelabel@@@{\number\hl@@@C}%
   \xdefThelabel@{\hl@@@N}%
   \xdef\Thelabel@@@@{\hl@@@P\Thelabel@\hl@@@Q}%
   \xdefThelabel@@{\hl@@@S}%
  \fi
  \xdef\Thepref@{\Thelabel@@@@}}%
  \csname hl@\hllevel@\endcsname{##1}%
  \let\pref\Thepref@
  \csname hl@I\hllevel@\endcsname
  \csname hl@J\hllevel@\endcsname
  \let\pref\pref@
  \hltoc@
  \aftertoc@
  \let\aftertoc@\relax\nopunct@false\nospace@false\FNSSP@}%
 \ifx\next"\expandafter\next@\else\expandafter\nextii@\fi}%
\def\six@#1#2 #3 #4 #5 #6 #7 {\DN@{#2}\ifx\next@\empty
 \DN@##1\six@{}\else
 \write#1{ #2 #3 #4 #5 #6 #7}\DN@{\six@#1}\fi
 \next@}
\def\Sixtoc@{\ifx\macdef@\empty\else
 \DN@##1##2\next@{\def\macdef@{##1##2}}%
 \expandafter\next@\macdef@\next@
 \edef\next@
  {\noexpand\six@\toc@\macdef@
  \space\space\space\space\space\space\space\space\space\space\space\space
  \noexpand\six@}%
 \next@\let\macdef@\relax\fi}
\def\QorThelabel@@@@{\ifquoted@
 \noexpand\noexpand\noexpand"\Qlabel@@@@\noexpand\noexpand\noexpand"\else
 \Thelabel@@@@\fi}
\def\HLtoc@{%
 \iftoc@
 \expandafter\expandafter\expandafter\unmacro@
  \expandafter\meaning\csname HL@W\HLlevel@\endcsname\unmacro@
  {\noexpands@\let\style\relax
   \edef\next@{\write\toc@{\noexpand\noexpand\expandafter\noexpand\HLname@
   {\macdef@}{\QorThelabel@@@@}}}%
  \next@}%
  \expandafter\unmacro@\meaning\entry@\unmacro@
  \Sixtoc@
  \write\toc@{\noexpand\Page{\number\pageno}{\page@N}%
   {\page@P}{\page@Q}^^J}%
 \fi}
\def\hltoc@{%
 \iftoc@
 \expandafter\expandafter\expandafter\unmacro@
  \expandafter\meaning\csname hl@W\hllevel@\endcsname\unmacro@
  {\noexpands@\let\style\relax
  \edef\next@{\write\toc@{%
   \ifnopunct@\noexpand\noexpand\noexpand\nopunct\fi
   \ifnospace@\noexpand\noexpand\noexpand\nospace\fi
   \noexpand\noexpand\expandafter\noexpand\hlname@
   {\macdef@}{\QorThelabel@@@@}}}%
  \next@}%
  \expandafter\unmacro@\meaning\entry@\unmacro@
  \Sixtoc@
  \write\toc@{\noexpand\Page{\number\pageno}{\page@N}%
   {\page@P}{\page@Q}^^J}%
 \fi}
\def\mainfile#1{\def\mainfile@{#1}}
\def\checkmainfile@{\ifx\mainfile@\undefined
 \Err@{No \noexpand\mainfile specified}\fi}
\expandafter\newcount@\csname HL@C1\endcsname
\csname HL@C1\endcsname\z@
\expandafter\def\csname HL@S1\endcsname#1{#1\null.}
\expandafter\let\csname HL@N1\endcsname\arabic
\expandafter\let\csname HL@P1\endcsname\empty
\expandafter\let\csname HL@Q1\endcsname\empty
\expandafter\def\csname HL@F1\endcsname{\bf}
\expandafter\let\csname HL@W1\endcsname\empty
\expandafter\newcount@\csname hl@C1\endcsname
\csname hl@C1\endcsname\z@
\expandafter\def\csname hl@S1\endcsname#1{#1\/}
\expandafter\let\csname hl@N1\endcsname\arabic
\expandafter\let\csname hl@P1\endcsname\empty
\expandafter\let\csname hl@Q1\endcsname\empty
\expandafter\def\csname hl@F1\endcsname{\bf}
\expandafter\let\csname hl@W1\endcsname\empty
\expandafter\def\csname HL@1\endcsname#1\endHL{\bigbreak
 {\locallabel@
  \global\setbox\@ne\vbox{\Let@\tabskip\hss@
  \halign to\hsize{\bf\hfil\ignorespaces##\unskip\hfil\cr
  \expandafter\ifx\csname HL@W1\endcsname\empty\else
   \csname HL@W1\endcsname\space\fi
  {\HL@@F\ifx\thelabel@@\empty\else\thelabel@@\space\fi}%
  \ignorespaces#1\crcr}}%
  }%
 \unvbox\@ne\nobreak\medskip}
\expandafter\def\csname hl@1\endcsname#1{\medbreak\noindent@@
 {\locallabel@
 \bf{\hl@@F\ifx\thelabel@@\empty\else\thelabel@@\space\fi}%
 \ignorespaces#1\unskip\punct@{\null.}\addspace@\enspace}}
\expandafter\def\csname HL@I1\endcsname{\Reset\hl1{1}%
 \ifx\pref\empty\newpre\hl1{}\else\newpre\hl1{\pref.}\fi}
\def\NameHL#1#2{\define#2{}%
 \expandafter\ifx\csname HL@R#1\endcsname\relax
 \else
  \def\nextiv@{\let\nextiii@}%
  \expandafter\nextiv@\csname HL@R#1\endcsname
  \expandafter\let\nextiii@\undefined
  \expandafter\let\csname\exxx@\nextiii@ @C\endcsname\relax
  \expandafter\let\csname\exxx@\nextiii@ @P\endcsname\relax
  \expandafter\let\csname\exxx@\nextiii@ @Q\endcsname\relax
  \expandafter\let\csname\exxx@\nextiii@ @S\endcsname\relax
  \expandafter\let\csname\exxx@\nextiii@ @N\endcsname\relax
  \expandafter\let\csname\exxx@\nextiii@ @F\endcsname\relax
  \expandafter\let\csname\exxx@\nextiii@ @W\endcsname\relax
  \expandafter\let\csname end\exxx@\nextiii@\endcsname\undefined
 \fi
 \expandafter\gdef\csname HL@R#1\endcsname{#2}%
 \expandafter\gdef\csname\exstring@#2@R\endcsname{{HL}{#1}}%
 \iftoc@\write\toc@{\noexpand\NameHL#1\noexpand#2^^J}\fi
 \rightadd@#2\to\overlonglist@
 \edef\next@{\let\csname\exstring@#2@C\endcsname\expandafter\noexpand
  \csname HL@C#1\endcsname}\next@
 \edef\next@{\let\csname\exstring@#2@P\endcsname\expandafter\noexpand
  \csname HL@P#1\endcsname}\next@
 \edef\next@{\let\csname\exstring@#2@Q\endcsname\expandafter\noexpand
  \csname HL@Q#1\endcsname}\next@
 \edef\next@{\let\csname\exstring@#2@S\endcsname\expandafter\noexpand
  \csname HL@S#1\endcsname}\next@
 \edef\next@{\let\csname\exstring@#2@N\endcsname\expandafter\noexpand
  \csname HL@N#1\endcsname}\next@
 \edef\next@{\let\csname\exstring@#2@F\endcsname\expandafter\noexpand
  \csname HL@F#1\endcsname}\next@
 \edef\next@{\let\csname\exstring@#2@W\endcsname\expandafter\noexpand
  \csname HL@W#1\endcsname}\next@
 \edef\next@{\def\noexpand#2####1\expandafter\noexpand
  \csname end\exstring@#2\endcsname
  {\def\noexpand\HLtype@{\noexpand#2}%
   \def\noexpand\HLname@{\noexpand#2}%
   \gdef\noexpand\HLlevel@{#1}%
   \noexpand\FNSS@\noexpand\HL@####1\noexpand\endHL}}%
  \next@
 \edef\next@{\noexpand\Invalid@\expandafter\noexpand
  \csname end\exstring@#2\endcsname}%
 \next@}
\def\Namehl#1#2{\define#2{}%
 \expandafter\ifx\csname hl@R#1\endcsname\relax
 \else
  \def\nextiv@{\let\nextiii@}%
  \expandafter\nextiv@\csname hl@R#1\endcsname
  \expandafter\let\nextiii@\undefined
  \expandafter\let\csname\exxx@\nextiii@ @C\endcsname\relax
  \expandafter\let\csname\exxx@\nextiii@ @P\endcsname\relax
  \expandafter\let\csname\exxx@\nextiii@ @Q\endcsname\relax
  \expandafter\let\csname\exxx@\nextiii@ @S\endcsname\relax
  \expandafter\let\csname\exxx@\nextiii@ @N\endcsname\relax
  \expandafter\let\csname\exxx@\nextiii@ @F\endcsname\relax
  \expandafter\let\csname\exxx@\nextiii@ @W\endcsname\relax
 \fi
 \expandafter\gdef\csname hl@R#1\endcsname{#2}%
 \expandafter\gdef\csname\exstring@#2@R\endcsname{{hl}{#1}}%
 \iftoc@\write\toc@{\noexpand\Namehl#1\noexpand#2^^J}\fi
 \rightadd@#2\to\nofrillslist@%
 \edef\next@{\let\csname\exstring@#2@C\endcsname\expandafter\noexpand
  \csname hl@C#1\endcsname}\next@
 \edef\next@{\let\csname\exstring@#2@P\endcsname\expandafter\noexpand
  \csname hl@P#1\endcsname}\next@
 \edef\next@{\let\csname\exstring@#2@Q\endcsname\expandafter\noexpand
  \csname hl@Q#1\endcsname}\next@
 \edef\next@{\let\csname\exstring@#2@S\endcsname\expandafter\noexpand
  \csname hl@S#1\endcsname}\next@
 \edef\next@{\let\csname\exstring@#2@N\endcsname\expandafter\noexpand
  \csname hl@N#1\endcsname}\next@
 \edef\next@{\let\csname\exstring@#2@F\endcsname\expandafter\noexpand
  \csname hl@F#1\endcsname}\next@
 \edef\next@{\let\csname\exstring@#2@W\endcsname\expandafter\noexpand
  \csname hl@W#1\endcsname}\next@
 \edef\next@{\def\noexpand#2{%
  \def\noexpand\hltype@{\noexpand#2}%
  \def\noexpand\hlname@{\noexpand#2}%
  \gdef\noexpand\hllevel@{#1}%
  \noexpand\FNSS@\noexpand\hl@}}%
 \next@}%
\def\Initialize{\FN@\Init@}
\def\Init@{\ifx\next\HL\let\next@\InitH@\else\ifx\next\hl\let\next@\InitH@
  \else\let\next@\InitS@\fi\fi\next@}
\def\InitH@#1#2{\expandafter\ifx\csname\exstring@#1@C#2\endcsname\relax
 \DN@{\Err@{\noexpand#1level #2 not defined in this style}}\else
 \DN@{\expandafter\gdef\csname\exstring@#1@J#2\endcsname}\fi\next@}
\def\InitC@#1#2{\edef\nextii@{\expandafter\noexpand\csname#1\endcsname{#2}}}
\def\InitS@#1{\expandafter\ifx\csname\exstring@#1@R\endcsname\relax
 \Err@{\noexpand#1not defined in this style}\let\next@\relax\else
 \DN@{\let\next@}\expandafter\next@\csname\exstring@#1@R\endcsname
 \expandafter\InitC@\next@
 \DN@{\expandafter\InitH@\nextii@}\fi\next@}
\def\value#1{\expandafter
 \ifx\csname\exstring@#1@C\endcsname\relax
  \expandafter\ifx\csname\exstring@#1@C1\endcsname\relax
   \DN@{\Err@{\noexpand\value can't be used with \string#1}}%
  \else
   \DN@{\value@#1}%
  \fi
 \else
  \DN@{\number\csname\exstring@#1@C\endcsname\relax}%
 \fi
 \next@}
\def\value@#1#2{\expandafter
 \ifx\csname\exstring@#1@C#2\endcsname\relax
  \DN@{\Err@{\string\value\string#1 can't be followed by \string#2}}%
 \else
  \DN@{\number\csname\exstring@#1@C#2\endcsname\relax}%
 \fi
 \next@}
\newcount\Value
\def\Evaluate#1{\expandafter
 \ifx\csname\exstring@#1@C\endcsname\relax
  \expandafter\ifx\csname\exstring@#1@C1\endcsname\relax
   \DN@{\Err@{\noexpand\Evaluate can't be used with \string#1}}%
  \else
   \DN@{\Evaluate@#1}%
  \fi
 \else
  \DN@{\global\Value\csname\exstring@#1@C\endcsname}%
 \fi
 \next@}
\def\Evaluate@#1#2{\expandafter
 \ifx\csname\exstring@#1@C#2\endcsname\relax
  \DN@{\Err@{\string\Evaluate\string#1 can't be followed by \string#2}}%
 \else
  \DN@{\global\Value\csname\exstring@#1@C#2\endcsname}%
 \fi\next@}
\def\pre#1{\expandafter
 \ifx\csname\exstring@#1@P\endcsname\relax
  \expandafter\ifx\csname\exstring@#1@P1\endcsname\relax
   \DN@{\Err@{\noexpand\pre can't be used with \string#1}}%
  \else
   \DN@{\pre@#1}%
  \fi
 \else
  \DN@{{\csname\exstring@#1@P\endcsname}}%
 \fi
 \next@}
\def\pre@#1#2{\expandafter
 \ifx\csname\exstring@#1@P#2\endcsname\relax
  \DN@{\Err@{\string\pre\string#1 can't be followed by \string#2}}%
 \else
  \DN@{{\csname\exstring@#1@P#2\endcsname}}%
 \fi
 \next@}
\def\post#1{\expandafter
 \ifx\csname\exstring@#1@Q\endcsname\relax
  \expandafter\ifx\csname\exstring@#1@Q1\endcsname\relax
   \DN@{\Err@{\noexpand\post can't be used with \string#1}}%
  \else
   \DN@{\post@#1}%
  \fi
 \else
  \DN@{{\csname\exstring@#1@Q\endcsname}}%
 \fi
 \next@}
\def\post@#1#2{\expandafter
 \ifx\csname\exstring@#1@Q#2\endcsname\relax
  \DN@{\Err@{\string\post\string#1 can't be followed by \string#2}}%
 \else
  \DN@{{\csname\exstring@#1@Q#2\endcsname}}%
 \fi
 \next@}
\def\style#1{\expandafter
 \ifx\csname\exstring@#1@S\endcsname\relax
  \expandafter\ifx\csname\exstring@#1@S1\endcsname\relax
   \DN@{\Err@{\noexpand\style can't be used with \string#1}}%
  \else
   \DN@{\style@#1}%
  \fi
 \else
  \DN@{\csname\exstring@#1@S\endcsname}%
 \fi
 \next@}
\def\style@#1#2{\expandafter
 \ifx\csname\exstring@#1@S#2\endcsname\relax
  \DN@{\Err@{\string\style\string#1 can't be followed by \string#2}}%
 \else
  \DN@{\csname\exstring@#1@S#2\endcsname}%
 \fi
 \next@}
\def\fontstyle#1{\expandafter
 \ifx\csname\exstring@#1@F\endcsname\relax
  \expandafter\ifx\csname\exstring@#1@F1\endcsname\relax
   \DN@{\Err@{\noexpand\fontstyle can't be used with \string#1}}%
  \else
   \DN@{\fontstyle@#1}%
  \fi
 \else
  \DN@##1{{\csname\exstring@#1@F\endcsname##1}}%
 \fi
 \next@}
\def\fontstyle@#1#2{\expandafter
 \ifx\csname\exstring@#1@F#2\endcsname\relax
  \DN@{\Err@{\string\fontstyle\string#1 can't be followed by \string#2}}%
 \else
  \DN@##1{{\csname\exstring@#1@F#2\endcsname##1}}%
 \fi
 \next@}
\def\Reset#1{\expandafter
 \ifx\csname\exstring@#1@C\endcsname\relax
  \expandafter\ifx\csname\exstring@#1@C1\endcsname\relax
   \DN@{\Err@{\noexpand\Reset can't be used with \string#1}}%
  \else
   \DN@{\Reset@#1}%
  \fi
 \else
  \DN@##1{\count@##1\relax\ifx#1\page\else\advance\count@\m@ne\fi
   \global\csname\exstring@#1@C\endcsname\count@}%
 \fi
 \next@}
\def\Reset@#1#2{\expandafter
 \ifx\csname\exstring@#1@C#2\endcsname\relax
  \DN@{\Err@{\string\Reset\string#1 can't be followed by \string#2}}%
 \else
  \DN@##1{\count@##1\relax\advance\count@\m@ne
   \global\csname\exstring@#1@C#2\endcsname\count@}%
 \fi
 \next@}
\def\Offset#1{\expandafter
 \ifx\csname\exstring@#1@C\endcsname\relax
  \expandafter\ifx\csname\exstring@#1@C1\endcsname\relax
   \DN@{\Err@{\noexpand\Offset can't be used with \string#1}}%
  \else
   \DN@{\Offset@#1}%
  \fi
 \else
  \DN@##1{\count@##1\relax\advance\count@\m@ne\global\advance
   \csname\exstring@#1@C\endcsname\count@}%
 \fi
 \next@}
\def\Offset@#1#2{\expandafter
 \ifx\csname\exstring@#1@C#2\endcsname\relax
  \DN@{\Err@{\string\Offset\string#1 can't be followed by \string#2}}%
 \else
  \DN@##1{\count@##1\relax\advance\count@\m@ne
   \global\advance\csname\exstring@#1@C#2\endcsname\count@}%
 \fi
 \next@}
\def\getR@#1#2{\def\nextiv@{\let\nextiii@}\expandafter\nextiv@
 \csname\exstring@#1@R#2\endcsname}
\def\letR@#1#2#3{\expandafter\let\csname#1@#3#2\endcsname\Next@}
\def\letR@@#1#2{\expandafter\let\csname\exstring@#1@#2\endcsname\Next@}
\def\newpre#1{\expandafter
 \ifx\csname\exstring@#1@P\endcsname\relax
  \expandafter\ifx\csname\exstring@#1@P1\endcsname\relax
   \DN@{\Err@{\noexpand\newpre can't be used with \string#1}}%
  \else
   \DN@{\newpre@#1}%
  \fi
 \else
  \DN@{%
   \DNii@{%
    \endgroup
    \expandafter\let\csname\exstring@#1@P\endcsname\Next@
    \expandafter\ifx\csname\exstring@#1@R\endcsname\relax\else
    \getR@#1{}\expandafter\letR@\nextiii@ P\fi
    }%
   \begingroup\noexpands@\afterassignment\nextii@\xdef\Next@}%
 \fi
 \next@}
\def\newpre@#1#2{\expandafter
 \ifx\csname\exstring@#1@P#2\endcsname\relax
  \DN@{\Err@{\string\newpre\string#1 can't be followed by \string#2}}%
 \else
  \DN@{%
   \DNii@{%
    \endgroup
    \expandafter\let\csname\exstring@#1@P#2\endcsname\Next@
    \expandafter\ifx\csname\exstring@#1@R#2\endcsname\relax\else
    \getR@#1{#2}\expandafter\letR@@\nextiii@ P\fi
    }%
   \begingroup\noexpands@\afterassignment\nextii@\xdef\Next@}%
 \fi
 \next@}
\def\newpost#1{\expandafter
 \ifx\csname\exstring@#1@Q\endcsname\relax
  \expandafter\ifx\csname\exstring@#1@Q1\endcsname\relax
   \DN@{\Err@{\noexpand\newpost can't be used with \string#1}}%
  \else
   \DN@{\newpost@#1}%
  \fi
 \else
  \DN@{%
   \DNii@{%
    \endgroup
    \expandafter\let\csname\exstring@#1@Q\endcsname\Next@
    \expandafter\ifx\csname\exstring@#1@R\endcsname\relax\else
    \getR@#1{}\expandafter\letR@\nextiii@ Q\fi
    }%
   \begingroup\noexpands@\afterassignment\nextii@\xdef\Next@}%
 \fi
 \next@}
\def\newpost@#1#2{\expandafter
 \ifx\csname\exstring@#1@Q#2\endcsname\relax
  \DN@{\Err@{\string\newpost\string#1 can't be followed by \string#2}}%
 \else
  \DN@{%
   \DNii@{%
    \endgroup
    \expandafter\let\csname\exstring@#1@Q#2\endcsname\Next@
    \expandafter\ifx\csname\exstring@#1@R#2\endcsname\relax\else
    \getR@#1{#2}\expandafter\letR@@\nextiii@ Q\fi
    }%
   \begingroup\noexpands@\afterassignment\nextii@\xdef\Next@}%
 \fi
 \next@}
\def\newstyle#1{\expandafter
 \ifx\csname\exstring@#1@S\endcsname\relax
  \expandafter\ifx\csname\exstring@#1@S1\endcsname\relax
   \DN@{\Err@{\noexpand\newstyle can't be used
    with \string#1}}%
  \else
   \DN@{\newstyle@#1}%
  \fi
 \else
  \DN@{%
   \DNii@{%
    \expandafter\let\csname\exstring@#1@S\endcsname\Next@
    \expandafter\ifx\csname\exstring@#1@R\endcsname\relax\else
    \getR@#1{}\expandafter\letR@\nextiii@ S\fi
    }%
   \afterassignment\nextii@\gdef\Next@}%
 \fi
 \next@}
\def\newstyle@#1#2{\expandafter
 \ifx\csname\exstring@#1@S#2\endcsname\relax
  \DN@{\Err@{\string\newstyle\string#1 can't be followed by
   \string#2}}%
 \else
  \DN@{%
   \DNii@{%
    \expandafter\let\csname\exstring@#1@S#2\endcsname\Next@
    \expandafter\ifx\csname\exstring@#1@R#2\endcsname\relax\else
    \getR@#1{#2}\expandafter\letR@@\nextiii@ S\fi
    }%
   \afterassignment\nextii@\gdef\Next@}%
 \fi
 \next@}
\def\newnumstyle#1{\expandafter
 \ifx\csname\exstring@#1@N\endcsname\relax
  \expandafter\ifx\csname\exstring@#1@N1\endcsname\relax
   \DN@{\Err@{\noexpand\newnumstyle can't be used with
    \string#1}}%
  \else
   \DN@{\newnumstyle@#1}%
  \fi
 \else
  \DN@##1{%
   \gdef\Next@{##1}%
    \expandafter\let\csname\exstring@#1@N\endcsname\Next@
    \expandafter\ifx\csname\exstring@#1@R\endcsname\relax\else
    \getR@#1{}\expandafter\letR@\nextiii@ N\fi
    }%
 \fi
 \next@}
\def\newnumstyle@#1#2{\expandafter
 \ifx\csname\exstring@#1@N#2\endcsname\relax
  \DN@{\Err@{\string\newnumstyle\string#1 can't be followed by
   \string#2}}%
 \else
  \DN@##1{%
   \gdef\Next@{##1}%
    \expandafter\let\csname\exstring@#1@N#2\endcsname\Next@
    \expandafter\ifx\csname\exstring@#1@R#2\endcsname\relax\else
    \getR@#1{#2}\expandafter\letR@@\nextiii@ N\fi
    }%
  \fi
 \next@}
\def\newfontstyle#1{\expandafter
 \ifx\csname\exstring@#1@F\endcsname\relax
  \expandafter\ifx\csname\exstring@#1@F1\endcsname\relax
   \DN@{\Err@{\noexpand\newfontstyle can't be used with
    \string#1}}%
  \else
   \DN@{\newfontstyle@#1}%
  \fi
 \else
  \DN@##1{%
   \gdef\Next@{##1}%
    \expandafter\let\csname\exstring@#1@F\endcsname\Next@
    \expandafter\ifx\csname\exstring@#1@R\endcsname\relax\else
    \getR@#1{}\expandafter\letR@\nextiii@ F\fi
    }%
 \fi
 \next@}
\def\newfontstyle@#1#2{\expandafter
 \ifx\csname\exstring@#1@F#2\endcsname\relax
  \DN@{\Err@{\string\newfontstyle\string#1 can't be followed by
   \string#2}}%
 \else
  \DN@##1{%
   \gdef\Next@{##1}%
    \expandafter\let\csname\exstring@#1@F#2\endcsname\Next@
    \expandafter\ifx\csname\exstring@#1@R#2\endcsname\relax\else
    \getR@#1{#2}\expandafter\letR@@\nextiii@ F\fi
    }%
 \fi
 \next@}
\def\word#1{\expandafter
 \ifx\csname\exstring@#1@W\endcsname\relax
  \expandafter\ifx\csname\exstring@#1@W1\endcsname\relax
   \DN@{\Err@{\noexpand\word can't be used with \string#1}}%
  \else
   \DN@{\word@#1}%
  \fi
 \else
  \DN@{{\csname\exstring@#1@W\endcsname}}%
 \fi
 \next@}
\def\word@#1#2{\expandafter
 \ifx\csname\exstring@#1@W#2\endcsname\relax
  \DN@{\Err@{\string\word\noexpand#1can't be followed by \string#2}}%
 \else
  \DN@{{\csname\exstring@#1@W#2\endcsname}}%
 \fi
 \next@}
\def\newword#1{\expandafter
 \ifx\csname\exstring@#1@W\endcsname\relax
  \expandafter\ifx\csname\exstring@#1@W1\endcsname\relax
   \DN@{\Err@{\noexpand\newword can't be used  with \string#1}}%
  \else
   \DN@{\newword@#1}%
  \fi
 \else
  \DN@{%
   \DNii@{%
    \expandafter\let\csname\exstring@#1@W\endcsname\Next@
    \expandafter\ifx\csname\exstring@#1@R\endcsname\relax\else
     \getR@#1{}\expandafter\letR@\nextiii@ W\fi
    }%
   \afterassignment\nextii@\gdef\Next@}%
 \fi
 \next@}
\def\newword@#1#2{\expandafter
 \ifx\csname\exstring@#1@W#2\endcsname\relax
  \DN@{\Err@{\string\newword\noexpand#1can't be followed by \string#2}}%
 \else
  \DN@{%
   \DNii@{%
    \expandafter\let\csname\exstring@#1@W#2\endcsname\Next@
    \expandafter\ifx\csname\exstring@#1@R#2\endcsname\relax\else
     \getR@#1{#2}\expandafter\letR@@\nextiii@ W\fi
    }%
   \afterassignment\nextii@\gdef\Next@}%
 \fi
 \next@}
\newif\iffn@
\newcount\footmark@C
\footmark@C\z@
\def\footmark@S#1{$^{#1}$}
\let\footmark@N\arabic
\def\footmark@F{\rm}
\def\foottext@S#1{$^{#1}$}
\def\foottext@F{\rm}
\let\modifyfootnote@\relax
\def\modifyfootnote#1{\def\modifyfootnote@{#1}}
\def\vfootnote@#1{\insert\footins
 \bgroup
 \floatingpenalty\@MM\interlinepenalty\interfootnotelinepenalty
 \leftskip\z@\rightskip\z@\spaceskip\z@\xspaceskip\z@
 \rm\splittopskip\ht\strutbox\splitmaxdepth\dp\strutbox
 \locallabel@\noindent@@{\foottext@F#1}\modifyfootnote@
 \footstrut\FN@\fo@t}
\def\fo@t{\ifcat\bgroup\noexpand\next\expandafter\f@@t\else
 \expandafter\f@t\fi}
\def\f@t#1{#1\@foot}
\def\f@@t{\bgroup\aftergroup\@foot\afterassignment\FNSSP@\let\next@}
\def\@foot{\unskip\lower\dp\strutbox\vbox to\dp\strutbox{}\egroup
 \iffn@\expandafter\fn@false\else
 \expandafter\postvanish@\fi}
\newif\ifplainfn@
\plainfn@true
\def\fancyfootnotes{\plainfn@false}
\newcount\fancyfootmarkcount@
\fancyfootmarkcount@\z@
\newcount\lastfnpage@
\lastfnpage@-\@M
\let\justfootmarklist@\empty
\def\footmark{\let\@sf\empty
 \ifhmode\edef\@sf{\spacefactor\the\spacefactor}\/\fi
 \DN@{\ifx"\next\expandafter\nextii@\else\expandafter\footmark@\fi}%
 \DNii@"##1"{%
  \iffirstchoice@
   {\let\style\footmark@S\let\numstyle\footmark@N
   \footmark@F##1%
   \noexpands@
   \let\style\foottext@S
   \Qlabel@{##1}%
   }%
   \iffn@\else
    {\noexpands@
    \xdef\Next@{{\Thelabel@}{\Thelabel@@}{\Thelabel@@@}{\Thelabel@@@@}}%
    }%
    \expandafter\rightappend@\Next@\to\justfootmarklist@
   \fi
  \fi
  \@sf\relax}%
 \FN@\next@}
\def\footmark@{%
 \iffirstchoice@
  \global\advance\footmark@C\@ne
  \ifplainfn@
   \xdef\adjustedfootmark@{\number\footmark@C}%
  \else
   {\let\\\or\xdef\Next@{\ifcase\number\footmark@C\fnpages@\else
     -\@M\fi}}%
   \ifnum\Next@=-\@M
    \xdef\adjustedfootmark@{\number\footmark@C}%
   \else
    \ifnum\Next@=\lastfnpage@
     \global\advance\fancyfootmarkcount@\@ne
    \else
     \global\fancyfootmarkcount@\@ne
     \global\lastfnpage@\Next@
    \fi
    \xdef\adjustedfootmark@{\number\fancyfootmarkcount@}%
   \fi
  \fi
  {\noexpands@
  \xdef\Thelabel@@@{\adjustedfootmark@}%
  \xdefThelabel@\footmark@N
  \xdef\Thelabel@@@@{\Thelabel@}%
  \xdefThelabel@@\foottext@S
  }%
  \iffn@\else
   {\noexpands@
   \xdef\Next@{{\Thelabel@}{\Thelabel@@}{\Thelabel@@@}{\Thelabel@@@@}}%
   }%
   \expandafter\rightappend@\Next@\to\justfootmarklist@
  \fi
  \ifplainfn@
  \else
   \edef\next@{\write\laxwrite@{F\noexpand\the\pageno}}\next@
  \fi
 \fi
 \footmark@S{\footmark@N{\adjustedfootmark@}}%
 \@sf\relax}
\def\foottext{\prevanish@
 \ifx\justfootmarklist@\empty
  \Err@{There is no \noexpand\footmark for this \string\foottext}\fi
 \DN@\\##1##2\next@{\DN@{##1}\gdef\justfootmarklist@{##2}}%
 \expandafter\next@\justfootmarklist@\next@
 \expandafter\foottext@\next@}
\def\foottext@#1#2#3#4{{\noexpands@
  \xdef\Thelabel@{#1}\xdef\Thelabel@@{#2}%
  \xdef\Thelabel@@@{#3}\xdef\Thelabel@@@@{#4}}%
  \vfootnote@{\thelabel@@}}
\rightadd@\foottext\to\vanishlist@
\def\footnote{\fn@true
 \let\@sf\empty
 \ifhmode\edef\@sf{\spacefactor\the\spacefactor}\/\fi
 \DN@{\ifx"\next\expandafter\nextii@\else\expandafter\nextiii@\fi}%
 \DNii@"##1"{\footmark"##1"\vfootnote@{\let\style\foottext@S
  \let\numstyle\footmark@N##1}}%
 \def\nextiii@{\footmark\vfootnote@{\foottext@S{\footmark@N
  {\adjustedfootmark@}}}}%
 \FN@\next@}
\newdimen\litindent
\litindent20\p@
\newbox\litbox@
\newbox\Litbox@
\newcount\interlitpenalty@
\interlitpenalty@\@M
\newcount\litlines@
{\obeyspaces\gdef\defspace@{\def {\allowbreak\hskip.5emminus.15em}}}
{\obeylines\gdef\letM@{\let^^M\CtrlM@}}
\def\CtrlM@{\egroup
 \ifcase\litlines@\advance\litlines@\@ne\or
 \box\litbox@\advance\litlines@\@ne\else
 \penalty\interlitpenalty@\box\litbox@\fi
 \Lit@}
\def\Lit@{\setbox\litbox@\hbox\bgroup\litdefs@\hskip\litindent}
\newcount\littab@
\littab@8
\def\littab#1{\littab@#1\relax}
{\catcode`\^^I=\active\gdef\letTAB@{\let^^I\TAB@}}
\def\TAB@{\egroup
 \dimen@\wd\litbox@
 \advance\dimen@-\litindent
 \setboxz@h{\tt0}%
 \dimen@ii\littab@\wdz@
 \divide\dimen@\dimen@ii
 \multiply\dimen@\dimen@ii
 \advance\dimen@\littab@\wdz@
 \advance\dimen@\litindent
 \setbox\litbox@\hbox\bgroup\litdefs@\hbox to\dimen@{\unhbox\litbox@\hfil}}
{\catcode`\`=\active\gdef`{\relax\lq}}
\let\litbs@\relax
\let\litbs@@\relax
\def\litbackslash#1{%
 \edef\litbs@{\catcode`\string#1=\z@
 \def\noexpand\litbs@@{\def\expandafter\noexpand\csname\string#1\endcsname
  {\char`\string#1}}}}
\def\litcodes@{\catcode`\\=12
 \catcode`\{=12 \catcode`\}=12
 \catcode`\$=12 \catcode`\&=12
 \catcode`\#=12
 \catcode`\^=12 \catcode`\_=12
 \catcode`\@=12 \catcode`\~=12 \catcode`\"=12
 \catcode`\;=12 \catcode`\:=12 \catcode`\!=12 \catcode`\?=12
 \catcode`\%=12 \litbs@\catcode`\`=\active\obeyspaces\defspace@}
\def\activate@#1#2{{\lccode`\~=`#2%
 \lowercase{%
  \if0#1%
  \gdef\Next@{\def~{\egroup\endgroup\bigskip\vskip-\parskip
   \def\next@{\noindent@@\FN@\pretendspace@}\FNSS@\next@}}\else
  \gdef\Next@{\def~{\egroup\egroup\endgroup}}\fi
  }%
 }}
\def\litdefs@{\let\0\empty\let\1\litdelim@\def\ {\char32 }\litbs@@}%
\def\litdelimiter#1{%
 \edef\litdelim@{\char`#1}%
 \def\lit#1{\leavevmode\begingroup\litcodes@\litdefs@
  \tt\hyphenchar\tentt\m@ne\lit@}%
 \def\lit@##1#1{##1\endgroup\null}%
 \def\Lit#1{\ifhmode$$\abovedisplayskip\bigskipamount
  \abovedisplayshortskip\bigskipamount
  \belowdisplayskip\z@\belowdisplayshortskip\z@
  \postdisplaypenalty\@M
  $$\vskip-\baselineskip\else\bigskip\fi
  \begingroup\litlines@\z@
  \catcode`#1=\active\activate@0#1\Next@
  \def\displaybreak{\egroup\break\litlines@\z@\Lit@}%
  \def\allowdisplaybreak{\egroup\allowbreak\litlines@\z@\Lit@}%
  \def\allowdisplaybreaks{\egroup\allowbreak\interlitpenalty@\z@
   \litlines@\z@\Lit@}%
  \litcodes@\tt\catcode`\^^I=\active\letTAB@
  \obeylines\letM@\Lit@}%
 \def\Litbox##1=#1{\begingroup\ifodd##1\relax\aftergroup\global\fi
  \aftergroup\setbox\aftergroup##1\aftergroup\box\aftergroup\Litbox@
  \def\allowdisplaybreak{\egroup\allowbreak\litlines@\z@\Lit@}%
  \def\allowdisplaybreaks{\egroup\allowbreak\interlitpenalty@\z@
   \litlines@\z@\Lit@}%
  \catcode`#1=\active\activate@1#1\Next@
  \litcodes@\tt\catcode`\^^I=\active\letTAB@
  \obeylines\letM@\global\setbox\Litbox@\vbox\bgroup\litindent\z@%
  \litlines@\z@\Lit@}%
}
\newbox\titlebox@
\setbox\titlebox@\vbox{}
\rightadd@\title\to\overlonglist@
\def\title{\begingroup\Let@
 \global\setbox\titlebox@\vbox\bgroup\tabskip\hss@
 \halign to\hsize\bgroup\bf\hfil\ignorespaces##\unskip\hfil\cr}
\def\endtitle{\crcr\egroup\egroup\endgroup\overlong@false}
\newbox\authorbox@
\rightadd@\author\to\overlonglist@
\def\author{\begingroup\Let@
 \global\setbox\authorbox@\vbox\bgroup\tabskip\hss@
 \halign to\hsize\bgroup\rm\hfil\ignorespaces##\unskip\hfil\cr}
\def\endauthor{\crcr\egroup\egroup\endgroup\overlong@false}
\newbox\affilbox@
\def\affil{\begingroup\Let@
 \global\setbox\affilbox@\vbox\bgroup\tabskip\hss@
 \halign to\hsize\bgroup\rm\hfil\ignorespaces##\unskip\hfil\cr}%
\def\endaffil{\crcr\egroup\egroup\endgroup\overlong@false}
\let\date@\relax
\def\date#1{\gdef\date@{\ignorespaces#1\unskip}}
\def\today{\ifcase\month\or January\or February\or March\or April\or May\or
 June\or July\or August\or September\or October\or November\or December\fi
 \space\number\day, \number\year}
\def\maketitle{\hrule\height\z@\vskip-\topskip
 \vskip24\p@ plus12\p@ minus12\p@
 \unvbox\titlebox@
 \ifvoid\authorbox@\else\vskip12\p@ plus6\p@ minus3\p@\unvbox\authorbox@\fi
 \ifvoid\affilbox@\else\vskip10\p@ plus5\p@ minus2\p@\unvbox\affilbox@\fi
 \ifx\date@\relax\else\vskip6\p@ plus2\p@ minus\p@\centerline{\rm\date@}\fi
 \vskip18\p@ plus12\p@ minus6\p@}
\def\cite{%
 \DNii@(##1)##2{{\rm[}{##2}, {##1\/}{\rm]}}%
 \def\nextiii@##1{{\rm[}{##1\/}{\rm]}}%
 \DN@{\ifx\next(\expandafter\nextii@\else\expandafter\nextiii@\fi}%
 \FN@\next@}
\def\makebib@W{Bibliography}
\def\makebib{\begingroup\rm\bigbreak\centerline{\smc\makebib@W}%
 \nobreak\medskip
 \sfcode`\.=\@m\everypar{}\parindent\z@
 \def\nopunct{\nopunct@true}\def\nospace{\nospace@true}%
 \nopunct@false\nospace@false
 \def\lkerns@{\null\kern\m@ne sp\kern\@ne sp}%
 \def\nkerns@{\null\kern-\tw@ sp\kern\tw@ sp}%
}

\newif\ifnoprepunct@
\newif\ifnoprespace@
\newif\ifnoquotes@
\def\noprepunct{\noprepunct@true}
\def\noprespace{\noprespace@true}
\def\noquotes{\noquotes@true}
\newbox\nobox@
\newbox\keybox@
\newbox\bybox@
\newbox\paperbox@
\newbox\paperinfobox@
\newbox\jourbox@
\newbox\volbox@
\newbox\issuebox@
\newbox\yrbox@
\newbox\pgbox@
\newbox\ppbox@
\newbox\bookbox@
\newbox\inbookbox@
\newbox\bookinfobox@
\newbox\publbox@
\newbox\publaddrbox@
\newbox\edbox@
\newbox\edsbox@
\newbox\langbox@
\newbox\translbox@
\newbox\finalinfobox@
\def\setbibinfo@#1{\edef\next@{\ifnopunct@1\else0\fi
 \ifnospace@1\else0\fi\ifnoprepunct@1\else0\fi\ifnoprespace@1\else0\fi
 \ifnoquotes@1\else0\fi}%
 \DNii@{00000}%
 \ifx\next@\nextii@\else\xdef\bibinfo@{\bibinfo@\the#1,\next@}%
 \fi}
\def\getbibinfo@#1{\ifx\bibinfo@\empty
 \let\next@0\let\nextii@0\let\nextiii@0\let\nextiv@0\let\nextv@0\else
 \edef\next@{\def
  \noexpand\next@####1\the#1,####2####3####4####5####6####7\noexpand\next@
  {\let\noexpand\next@####2\let\noexpand\nextii@####3%
  \let\noexpand\nextiii@####4\let\noexpand\nextiv@####5%
  \let\noexpand\nextv@####6}%
  \noexpand\next@\bibinfo@\the#1,00000\noexpand\next@}\next@
 \fi}
\newif\ifbookinquotes@
\def\bookinquotes{\bookinquotes@true}
\newif\ifpaperinquotes@
\def\paperinquotes{\paperinquotes@true}
\newif\ifininbook@
\def\ininbook{\ininbook@true}
\newif\ifopenquotes@
\def\closequotes@{\ifopenquotes@''\openquotes@false\fi}
\newif\ifbeginbib@
\newif\ifendbib@
\newif\ifprevjour@
\newif\ifprevbook@
\newdimen\bibindent@
\bibindent@20\p@
\def\bib{\global\let\bibinfo@\empty\global\let\translinfo@\relax\beginbib@true
 \begingroup\noindent@
 \hangindent\bibindent@\hangafter\@ne\bib@}
\def\v@id#1{\setbox#1\box\voidb@x}
\def\bib@{\v@id\nobox@\v@id\keybox@\v@id\bybox@\v@id\paperbox@
 \v@id\paperinfobox@\v@id\jourbox@\v@id\volbox@\v@id\issuebox@
 \v@id\yrbox@\v@id\pgbox@\v@id\ppbox@\v@id\bookbox@\v@id\inbookbox@
 \v@id\bookinfobox@\v@id\publbox@\v@id\publaddrbox@\v@id\edbox@
 \v@id\edsbox@\v@id\langbox@\v@id\translbox@\v@id\finalinfobox@
 \bgroup}
\def\Setnonemptybox@#1#2{\unskip\setbibinfo@#1\egroup#2%
 \def\aftergroup@{\ifdim\wd#1=\z@\setbox#1\box\voidb@x\fi}%
 \setbox#1\vbox\bgroup\aftergroup\aftergroup@\hsize\maxdimen\leftskip\z@
 \rightskip\z@\hbadness\@M\hfuzz\maxdimen\noindent}
\def\setnonemptybox@#1{\Setnonemptybox@#1\relax}
\def\no{\setnonemptybox@\nobox@}
\def\key{\setnonemptybox@\keybox@\bf}
\def\by{\setnonemptybox@\bybox@}
\def\bysame{\setnonemptybox@\bybox@\leaders\hrule\hskip3em\null}
\def\paper{\setnonemptybox@\paperbox@
 \ifpaperinquotes@\getbibinfo@\paperbox@
 \if\nextv@1\else``\fi\else\it\fi}
\def\paperinfo{\setnonemptybox@\paperinfobox@}
\def\jour{\Setnonemptybox@\jourbox@\prevjour@true}
\def\vol{\setnonemptybox@\volbox@\bf}
\def\issue{\setnonemptybox@\issuebox@}
\def\yr{\setnonemptybox@\yrbox@}

\def\pg{\setnonemptybox@\pgbox@}
\def\pp{\setnonemptybox@\ppbox@}
\def\book{\Setnonemptybox@\bookbox@\prevbook@true
 \ifbookinquotes@\getbibinfo@\bookbox@
 \if\nextv@1\else``\fi\else\it\fi}
\def\inbook{\Setnonemptybox@\inbookbox@\prevbook@true
 \ifininbook@ in \fi\ifbookinquotes@\getbibinfo@\inbookbox@
 \if\nextv@1\else``\fi\fi}
\def\bookinfo{\setnonemptybox@\bookinfobox@}
\def\publ{\setnonemptybox@\publbox@}
\def\publaddr{\setnonemptybox@\publaddrbox@}
\def\ed{\setnonemptybox@\edbox@}
\def\eds{\setnonemptybox@\edsbox@}
\def\lang{\setnonemptybox@\langbox@}
\def\finalinfo{\setnonemptybox@\finalinfobox@}
\def\setboxzl@{\setbox\z@\lastbox}
\def\getbox@#1{\setbox\z@\vbox{\vskip-\@M\p@
 \unvbox#1%
 \setboxzl@
 \global\setbox\@ne\hbox{\unhbox\z@\unskip\unskip\unpenalty}%
 \ifdim\lastskip=-\@M\p@\else
 \loop\ifdim\lastskip=-\@M\p@
 \else\unskip\unpenalty\setboxzl@
 \global\setbox\@ne\hbox{\unhbox\z@\unhbox\@ne}%
 \repeat\fi}%
 \unhbox\@ne}
\def\adjustpunct@#1{\count@\lastkern
 \ifnum\count@=\z@#1\closequotes@\else
 \ifnum\count@>\tw@#1\closequotes@\else
 \ifnum\count@<-\tw@#1\closequotes@\else
  \unkern\unkern\setboxzl@
  \skip@\lastskip\unskip
  \count@@\lastpenalty\unpenalty
  \ifnum\count@=\tw@\unskip\setboxzl@\fi
  \ifdim\skip@=\z@\else\hskip\skip@\fi
  #1\closequotes@
  \ifnum\count@=\tw@\null\hfill\fi
  \penalty\count@@
 \fi\fi\fi}
\def\prepunct@#1#2{\getbibinfo@#2%
 \ifnopunct@
 \else
  \if\nextiii@0\adjustpunct@#1\fi
 \fi
 \closequotes@
 \ifnospace@
 \else
  \if\nextiv@0\space\else\fi
 \fi
 \nopunct@false\nospace@false
 \if\next@1\nopunct@true\fi
 \if\nextii@1\nospace@true\fi}
\def\ppunbox@#1#2{\prepunct@{#1}#2%
 \getbox@#2}
\let\semicolon@;
\def\endbib@{%
 \ifbeginbib@
  \ifvoid\nobox@
   \ifvoid\keybox@\else\hbox to\bibindent@{[\getbox@\keybox@]\hss}\fi
  \else\hbox to\bibindent@{\hss\getbox@\nobox@. }\fi
  \ifvoid\bybox@\else\getbox@\bybox@\fi
 \else
  \nopunct@true
  \ifvoid\bybox@\else\ppunbox@\relax\bybox@\fi
 \fi
 \ifvoid\translbox@\else\ppunbox@,\translbox@\fi
 \ifvoid\paperbox@\else\ppunbox@,\paperbox@\ifpaperinquotes@
  \if\nextv@1\else\openquotes@true\fi\fi
 \fi
 \ifvoid\paperinfobox@\else\ppunbox@,\paperinfobox@\fi
 \test@false
 \ifvoid\jourbox@\else\test@true\ppunbox@,\jourbox@\fi
 \ifprevjour@\test@true\fi
 \iftest@
  \ifvoid\volbox@\else\ppunbox@\relax\volbox@\fi
  \ifvoid\issuebox@
   \else\prepunct@\relax\issuebox@ no.~\getbox@\issuebox@\fi
  \ifvoid\yrbox@\else\prepunct@\relax\yrbox@(\getbox@\yrbox@)\fi
  \ifvoid\ppbox@\else\ppunbox@,\ppbox@\fi
  \ifvoid\pgbox@\else\prepunct@,\pgbox@ p.~\getbox@\pgbox@\fi
 \fi
 \test@false
 \ifvoid\bookbox@\else\test@true\ppunbox@,\bookbox@\ifbookinquotes@
  \if\nextv@1\else\openquotes@true\fi\fi\fi
 \ifvoid\inbookbox@\else\test@true\ppunbox@,\inbookbox@\ifbookinquotes@
  \if\nextv@1\else\openquotes@true\fi\fi\fi
 \ifprevbook@\test@true\fi
 \iftest@
  \ifvoid\edbox@\else\prepunct@\relax\edbox@(\getbox@\edbox@, ed.)\fi
  \ifvoid\edsbox@\else\prepunct@\relax\edsbox@(\getbox@\edsbox@, eds.)\fi
  \ifvoid\bookinfobox@\else\ppunbox@,\bookinfobox@\fi
  \ifvoid\publbox@\else\ppunbox@,\publbox@\fi
  \ifvoid\publaddrbox@\else\ppunbox@,\publaddrbox@\fi
  \ifvoid\yrbox@\else\ppunbox@,\yrbox@\fi
  \ifvoid\ppbox@\else\prepunct@,\ppbox@ pp.~\getbox@\ppbox@\fi
  \ifvoid\pgbox@\else\prepunct@,\pgbox@ p.~\getbox@\pgbox@\fi
 \fi
 \ifvoid\finalinfobox@
  \ifendbib@
   \ifnopunct@\else.\closequotes@\fi
  \else
  \ifvoid\langbox@\else\space(\getbox@\langbox@)\fi
   \/\semicolon@\closequotes@
  \fi
 \else
  \ifendbib@
   \ppunbox@{.\spacefactor3000\relax}\finalinfobox@
    \ifnopunct@\else.\fi
  \else
   \ppunbox@,\finalinfobox@\/\semicolon@\fi
 \fi
 \ifvoid\langbox@\else\space(\getbox@\langbox@)\fi
}
\def\endbib{\unskip\egroup\endbib@true\endbib@\par\endgroup}
\def\morebib{\unskip\egroup
 \endbib@false\endbib@
 \global\let\bibinfo@\empty\beginbib@false
 \bib@}
\def\anotherbib{\unskip\egroup
 \endbib@false\endbib@
 \global\let\bibinfo@\empty\beginbib@false
 \prevjour@false\prevbook@false\bib@}
\def\transl{\unskip
 \xdef\translinfo@{\the\translbox@,\ifnopunct@1\else0\fi
 \ifnospace@1\else0\fi\ifnoprepunct@1\else0\fi\ifnoprespace@1\else0\fi0}%
 \egroup\endbib@false\endbib@
 \global\let\bibinfo@\translinfo@\beginbib@false
 \bib@
 \egroup
 \def\aftergroup@{\ifdim\wd\translbox@=\z@\setbox\translbox@\box\voidb@x\fi}%
 \setbox\translbox@\vbox\bgroup\aftergroup\aftergroup@
 \hsize\maxdimen\leftskip\z@\rightskip\z@\hbadness\@M\hfuzz\maxdimen
 \noindent}
\newwrite\auxwrite@
\newread\bbl@
\def\UseBibTeX{\immediate\openout\auxwrite@=\jobname.aux
 \let\cite\BTcite@
 \def\nocite##1{\immediate\write\auxwrite@{\string\citation{##1}}}%
 \def\bibliographystyle##1{\immediate\write\auxwrite@{\string
  \bibstyle{##1}}}%
 \def\bibliography@W{Bibliography}%
 \def\bibliography##1{\immediate\write\auxwrite@{\string\bibdata{##1}}%
  \immediate\openin\bbl@=\jobname.bbl
  \ifeof\bbl@
   \W@{No .bbl file}%
  \else
   \immediate\closein\bbl@
   \begingroup\input bibtex \input\jobname.bbl \endgroup
  \fi}%
 }
\def\BTcite@{%
 \DNii@(##1)##2{{\rm[}\BTcite@@##2,\BTcite@@{\rm, }{##1\/}{\rm]}%
  \immediate\write\auxwrite@{\string\citation{##2}}}%
 \def\nextiii@##1{{\rm[}\BTcite@@##1,\BTcite@@\/{\rm]}%
  \immediate\write\auxwrite@{\string\citation{##1}}}%
 \DN@{\ifx\next(\expandafter\nextii@\else\expandafter\nextiii@\fi}%
 \FN@\next@}%
\def\BTcite@@#1,{\BTcite@@@{#1}\FN@\BTcite@@@@}
\def\BTcite@@@@{\ifx\next\BTcite@@
 \expandafter\eat@\else{\rm, }\expandafter\BTcite@@\fi}
\catcode`\~=11
\def\BTcite@@@#1{\nolabel@\cite{#1}\relax
 \DNii@##1~##2\nextii@{##1}%
 \csL@{#1}\expandafter\nextii@\Next@\nextii@\fi}
\catcode`\~=\active

\def\beginthebibliography@#1{\rm\setboxz@h{#1\ }\bibindent@\wdz@
 \bigbreak\centerline{\smc\bibliography@W}\nobreak\medskip
 \sfcode`\.=\@m\everypar{}\parindent\z@}

\newif\iffigproofing@
\def\Figureproofing{\figproofing@true}
\def\noFigureproofing{\figproofing@false}
\newif\ifHby@
\def\Hbyw#1{\global\Hby@true\hbyw\vsize{#1}}
\def\hbyw#1#2{%
 \hbox{%
  \ifHby@
  \else
   \iffigproofing@
    \setbox\z@\vbox{\hrule\width5\p@}\ht\z@\z@
    \vbox to#1{\hrule\height5\p@\width.4\p@\vfil\hrule\height5\p@\width.4\p@}%
    \kern-.4\p@\rlap{\copy\z@}\raise#1\hbox{\rlap{\copy\z@}}%
   \fi
  \fi
  \vbox to#1{\hbox to#2{}\vfil}%
  \ifHby@
  \else
   \iffigproofing@
    \vbox to#1{\hrule\height5\p@\width.4\p@\vfil\hrule\height5\p@\width.4\p@}%
    \kern-.4\p@\llap{\copy\z@}\raise#1\hbox{\llap{\boxz@}}%
   \fi
  \fi}}
\newcount\island@C
\let\island@P\empty
\let\island@Q\empty
\def\island@S#1{#1\null.}
\let\island@N\arabic
\def\island@F{\rm}
\def\island@@@P{\csname\exxx@\islandtype@ @P\endcsname}
\def\island@@@Q{\csname\exxx@\islandtype@ @Q\endcsname}
\def\island@@@S{\csname\exxx@\islandtype@ @S\endcsname}
\def\island@@@N{\csname\exxx@\islandtype@ @N\endcsname}
\def\island@@@F{\csname\exxx@\islandtype@ @F\endcsname}
\def\island@@@C{\csname island@C\islandclass@\endcsname}
\newif\ifplace@
\newif\ifisland@
\def\island{%
 \ifplace@
  \DN@{\let\islandclass@\empty\def\islandtype@{\island}\FN@\island@}%
 \else
  \long\DN@##1\endisland{\Err@{\noexpand\island must be used after some
   type of \string\...place}}%
 \fi
 \next@}
\def\island@{\ifx\next\c\let\next@\island@c\else
 \DN@{\FN@\island@@}\fi\next@}
\def\island@@{\ifcat\bgroup\noexpand\next\let\next@\island@@@\else
 \DN@{\Err@{\noexpand\island must be followed by a {prefix} for
 \string\caption's}}\fi\next@}
\newbox\islandbox@
\newcount\captioncount@
\def\island@@@#1{\def\captionprefix@{#1}\captioncount@\z@
 \global\setbox\islandbox@\vbox\bgroup}
\def\island@c\c#1{%
 \ifplace@
 \DN@{\def\islandclass@{#1}%
  \expandafter\ifx\csname island@C#1\endcsname\relax
  \expandafter\newcount@\csname island@C#1\endcsname
   \global\csname island@C#1\endcsname\z@\fi
  \FNSS@\island@c@}%
 \else
 \DN@{\edef\next@{\long\def\noexpand\next@########1\expandafter\noexpand
  \csname end\exxx@\islandtype@\endcsname{\noexpand\Err@{\noexpand\noexpand
  \expandafter\noexpand
  \islandtype@ must be used after some type of \noexpand\string
   \noexpand\...place}}}\next@\next@}%
 \fi
 \next@}
\def\island@c@{%
 \ifcat\bgroup\noexpand\next
  \let\next@\island@c@@
 \else
  \DN@{\Err@{\noexpand\island\string\c{\expandafter\string\islandclass@} must
   be followed by a {prefix} for \string\caption's}}%
 \fi\next@}
\def\island@c@@#1{\def\captionprefix@{#1}%
 \captioncount@\z@\global\setbox\islandbox@\vbox\bgroup}
\rightadd@\caption\to\nofrillslist@
\newbox\captionbox@
\newbox\Captionbox@
\def\caption{%
 \ifnum\captioncount@=\z@
  \ifnopunct@
   \DN@{\egroup\nopunct@true}%
  \else
   \let\next@\egroup
  \fi
 \else
  \let\next@\relax
 \fi
 \next@
 \advance\captioncount@\@ne
 \FN@\caption@}
\def\caption@{\ifx\next"\expandafter\caption@q\else\expandafter\caption@@\fi}
\def\caption@q"#1"{\quoted@true
 {\noexpands@
 \let\pre\island@@@P\let\post\island@@@Q
 \let\style\island@@@S\let\numstyle\island@@@N
 \Qlabel@{#1}\let\style\relax\xdef\Qlabel@@@@{#1}}%
 \finishcaption@}
\def\caption@@{\quoted@false
 \global\advance\island@@@C\@ne
 {\noexpands@
 \xdef\Thelabel@@@{\number\island@@@C}%
 \xdefThelabel@\island@@@N
 \xdef\Thelabel@@@@{\island@@@P\Thelabel@\island@@@Q}%
 \xdefThelabel@@\island@@@S
 \xdef\Thepref@{\Thelabel@@@@}}%
 \finishcaption@}
\long\def\captionformat@#1#2#3{\rm\strut#1 {\island@@@F#2} #3%
 \punct@.\strut}
\long\def\widerthanisland@#1#2#3{\test@true\setbox\z@\vbox{\hsize\maxdimen
 \noindent@@\captionformat@{#1}{#2}{#3}\par\setboxzl@}%
 \ifdim\wdz@=\z@
  \global\setbox\captionbox@\hbox{\noset@\unlabel@
   \captionformat@{#1}{#2}{#3}}%
  \ifdim\wd\captionbox@>\wd\islandbox@\else\test@false\fi
 \fi}
\long\def\captionformat@@#1#2#3{\widerthanisland@{#1}{#2}{#3}%
 \iftest@
  \global\setbox\captionbox@\vbox{\hsize\wd\islandbox@
   \vskip-\parskip\noindent@@\noset@\unlabel@
   \captionformat@{#1}{#2}{#3}\par}%
 \else
  \global\setbox\captionbox@
   \hbox to\wd\islandbox@{\hfil\box\captionbox@\hfil}%
 \fi}
\long\def\finishcaption@#1{\def\entry@{#1}%
 {\locallabel@
 \captionformat@@
  {\expandafter\ignorespaces\captionprefix@\unskip}%
  {\ifx\thelabel@@\empty\unskip\else\thelabel@@\fi}%
  {\ignorespaces#1\unskip}%
 \ifnum\captioncount@=\@ne
  \global\setbox\islandbox@\vbox{\ticwrite@\vbox{\box\islandbox@}}%
  \global\setbox\Captionbox@\vbox{\box\captionbox@}%
 \else
  \global\setbox\islandbox@\vbox{\unvbox\islandbox@\setboxzl@
   \ticwrite@\boxz@}%
  \global\setbox\Captionbox@\vbox{\unvbox\Captionbox@
   \smallskip\box\captionbox@}%
 \fi}%
 \nopunct@false\nospace@false\ignorespaces}
\def\Sixtic@{\ifx\macdef@\empty\else
 \DN@##1##2\next@{\def\macdef@{##1##2}}%
 \expandafter\next@\macdef@\next@
 \edef\next@
  {\noexpand\six@\tic@\macdef@
  \space\space\space\space\space\space\space\space\space\space\space\space
  \noexpand\six@}%
 \next@\let\macdef@\relax\fi}
\def\ticwrite@{%
 \iftoc@
  {\noexpands@\let\style\relax
  \DN@{\island}%
  \edef\next@{\write\tic@{%
   \ifnopunct@\noexpand\noexpand\noexpand\nopunct\fi
   \ifx\islandtype@\next@\noexpand\noexpand\noexpand\island
    \noexpand\string\noexpand\c{\islandclass@}{\captionprefix@}%
     {\QorThelabel@@@@}\else\noexpand\noexpand\expandafter\noexpand
     \islandtype@{\QorThelabel@@@@}}\fi}%
  \next@}%
  \expandafter\unmacro@\meaning\entry@\unmacro@
  \Sixtic@
  \write\tic@{\noexpand\Page{\number\pageno}{\page@N}{\page@P}{\page@Q}^^J}%
 \fi}
\def\Htrim@#1{%
 \ifHby@
  \dimen@\vsize
  \ifnum\captioncount@=\z@
  \else
   \advance\dimen@-\ht\Captionbox@
   \advance\dimen@-#1%
  \fi
  \global\Hby@false
  \dimen@ii\wd\islandbox@
  \global\setbox\islandbox@\vbox
   {\unvbox\islandbox@\setboxzl@
   \vbox to\z@{\vss\boxz@}\nointerlineskip\hbyw\dimen@\dimen@ii}%
  \global\Hby@true
 \fi}
\newif\ifdata@
\def\iclasstest@#1{\DN@{#1}\ifx\next@\islandclass@
 \test@true\else\test@false\fi}
\skipdef\skipi@=1
\def\endisland{\ifnum\captioncount@=\z@\expandafter\egroup\fi
 \ifdata@
 \else
  \iclasstest@{T}%
  \iftest@
   {\rm\global\skipi@-\dp\strutbox}\global\advance\skipi@\bigskipamount
   \Htrim@\skipi@
   \global\setbox\islandbox@\vbox
    {\ifnum\captioncount@=\z@\else
     \box\Captionbox@
     \nointerlineskip
     \vskip\skipi@\fi
     \box\islandbox@}%
  \else
   {\rm\global\skipi@\dp\strutbox}\global\advance\skipi@\medskipamount
   \Htrim@\skipi@
   \global\setbox\islandbox@\vbox
    {\box\islandbox@
     \ifnum\captioncount@=\z@\else
     \nointerlineskip
     \vskip\skipi@
     \box\Captionbox@
     \fi}%
  \fi
  \ifHby@
  \else
   \dimen@\ht\islandbox@\advance\dimen@\dp\islandbox@
   \ifdim\dimen@>\vsize
    \DN@{\island}%
    \Err@{%
     \ifx\islandtype@\next@\noexpand\island\else
      \expandafter\noexpand\islandtype@\fi
     \ifnum\captioncount@=\z@\else
       with \noexpand\caption\fi
      is larger than page}%
     \ht\islandbox@=\vsize
   \fi
  \fi
 \fi
 \global\Hby@false\island@true}
\def\newisland#1\c#2#3{\define#1{}%
 \iftoc@\immediate\write\tic@{\noexpand\newisland\noexpand#1%
  \string\c{#2}{#3}^^J}\fi
 \expandafter\def\csname\exstring@#1@S\endcsname{\island@S}%
 \expandafter\def\csname\exstring@#1@N\endcsname{\island@N}%
 \expandafter\def\csname\exstring@#1@P\endcsname{\island@P}%
 \expandafter\def\csname\exstring@#1@Q\endcsname{\island@Q}%
 \expandafter\def\csname\exstring@#1@F\endcsname{\island@F}%
 \expandafter\def\csname end\exstring@#1\endcsname{\endisland}%
 \expandafter
 \ifx\csname island@C#2\endcsname\relax
  \expandafter\newcount@\csname island@C#2\endcsname
  \global\csname island@C#2\endcsname\z@
 \fi
 \edef\next@{\noexpand\expandafter\noexpand\let\noexpand
  \csname\exstring@#1@C\noexpand\endcsname
  \csname island@C#2\endcsname}%
 \next@
 \def#1{\def\islandtype@{#1}\island@c\c{#2}{#3}}}
\newisland\Figure\c{F}{Figure}
\newisland\Table\c{T}{Table}
\newbox\islandboxi
\newbox\islandboxii
\newbox\islandboxiii
\newbox\captionboxi
\newbox\captionboxii
\newbox\captionboxiii
\long\def\islandpairdata#1#2{{\data@true
 \place@true
 #1%
 \global\setbox\islandboxi\box\islandbox@
 \global\setbox\captionboxi\box\Captionbox@
 #2%
 \global\setbox\islandboxii\box\islandbox@
 \global\setbox\captionboxii\box\Captionbox@
 }}
\long\def\islandpairbox#1#2{\islandpairdata{#1}{#2}%
 \dimen@\ht\captionboxi
 \ifdim\ht\captionboxii>\dimen@\dimen@\ht\captionboxii\fi
 \ifdim\dimen@>\z@
  \ifdim\ht\captionboxi<\dimen@
   \global\setbox\captionboxi\vbox to\dimen@{\unvbox\captionboxi\vfil}\fi
  \ifdim\ht\captionboxii<\dimen@
   \global\setbox\captionboxii\vbox to\dimen@{\unvbox\captionboxii\vfil}\fi
 \fi
 \global\setbox\islandbox@\vbox
 {\hbox to\hsize{\hfil\box\islandboxi\hfil\box\islandboxii\hfil}%
 \ifdim\dimen@>\z@\nointerlineskip
 {\rm\global\skipi@\dp\strutbox}\global\advance\skipi@\medskipamount
  \vskip\skipi@
  \hbox to\hsize{\hfil\box\captionboxi\hfil\box\captionboxii\hfil}\fi}}	
\long\def\islandpairboxa#1#2{\islandpairdata{#1}{#2}%
 \dimen@\ht\captionboxi
 \ifdim\ht\captionboxii>\dimen@\dimen@\ht\captionboxii\fi
 \ifdim\dimen@>\z@
  \ifdim\ht\captionboxi<\dimen@
   \global\setbox\captionboxi\vbox to\dimen@{\vfil\unvbox\captionboxi}\fi
  \ifdim\ht\captionboxii<\dimen@
   \global\setbox\captionboxii\vbox to\dimen@{\vfil\unvbox\captionboxii}\fi
 \fi
 \dimen@ii\ht\islandboxi
 \ifdim\ht\islandboxii>\dimen@ii \dimen@ii\ht\islandboxii\fi
 \ifdim\dimen@ii>\z@
  \ifdim\ht\islandboxi<\dimen@ii
   \global\setbox\islandboxi\vbox to\dimen@ii{\box\islandboxi\vfil}\fi
  \ifdim\ht\islandboxii<\dimen@ii
   \global\setbox\islandboxii\vbox to\dimen@ii{\box\islandboxii\vfil}\fi
 \fi
 \global\setbox\islandbox@\vbox{\ifdim\dimen@>\z@
  \hbox to\hsize{\hfil\box\captionboxi\hfil\box\captionboxii\hfil}%
  \nointerlineskip{\rm\global\skipi@-\dp\strutbox}%
  \global\advance\skipi@\bigskipamount\vskip\skipi@\fi
  \hbox to\hsize{\hfil\box\islandboxi\hfil\box\islandboxii\hfil}}}
\long\def\islandtripledata#1#2#3{{\data@true\place@true
 #1%
 \global\setbox\islandboxi\box\islandbox@
 \global\setbox\captionboxi\box\Captionbox@
 #2%
 \global\setbox\islandboxii\box\islandbox@
 \global\setbox\captionboxii\box\Captionbox@
 #3%
 \global\setbox\islandboxiii\box\islandbox@
 \global\setbox\captionboxiii\box\Captionbox@
 }}
\long\def\islandtriplebox#1#2#3{\islandtripledata{#1}{#2}{#3}%
 \dimen@\ht\captionboxi
 \ifdim\ht\captionboxii>\dimen@ \dimen@\ht\captionboxii\fi
 \ifdim\ht\captionboxiii>\dimen@ \dimen@\ht\captionboxiii\fi
 \ifdim\dimen@>\z@
  \ifdim\ht\captionboxi<\dimen@
   \global\setbox\captionboxi\vbox to\dimen@{\unvbox\captionboxi\vfil}\fi
  \ifdim\ht\captionboxii<\dimen@
   \global\setbox\captionboxii\vbox to\dimen@{\unvbox\captionboxii\vfil}\fi
  \ifdim\ht\captionboxiii<\dimen@
   \global\setbox\captionboxiii\vbox to\dimen@{\unvbox\captionboxiii\vfil}\fi
 \fi
 \global\setbox\islandbox@\vbox
  {\hbox to\hsize{\hfil\box\islandboxi\hfil\box\islandboxii\hfil
   \box\islandboxiii\hfil}%
 \ifdim\dimen@>\z@\nointerlineskip
  {\rm\global\skipi@\dp\strutbox}\global\advance\skipi@\medskipamount
  \vskip\skipi@
  \hbox to\hsize{\hfil\box\captionboxi\hfil\box\captionboxii\hfil
   \box\captionboxiii\hfil}\fi}}
\def\islandtripleboxa#1#2#3{\islandtripledata{#1}{#2}{#3}%
 \dimen@\ht\captionboxi
 \ifdim\ht\captionboxii>\dimen@ \dimen@\ht\captionboxii\fi
 \ifdim\ht\captionboxiii>\dimen@ \dimen@\ht\captionboxiii\fi
 \ifdim\dimen@>\z@
  \ifdim\ht\captionboxi<\dimen@
   \global\setbox\captionboxi\vbox to\dimen@{\vfil\unvbox\captionboxi}\fi
  \ifdim\ht\captionboxii<\dimen@
   \global\setbox\captionboxii\vbox to\dimen@{\vfil\unvbox\captionboxii}\fi
  \ifdim\ht\captionboxiii<\dimen@
   \global\setbox\captionboxiii\vbox to\dimen@{\vfil\unvbox\captionboxiii}\fi
 \fi
 \dimen@ii\ht\islandboxi
 \ifdim\ht\islandboxii>\dimen@ii \dimen@ii\ht\islandboxii\fi
 \ifdim\ht\islandboxiii>\dimen@ii \dimen@ii\ht\islandboxiii\fi
 \ifdim\dimen@ii>\z@
  \ifdim\ht\islandboxi<\dimen@ii
   \global\setbox\islandboxi\vbox to\dimen@ii{\box\islandboxi\vfil}\fi
  \ifdim\ht\islandboxii<\dimen@ii
   \global\setbox\islandboxii\vbox to\dimen@ii{\box\islandboxii\vfil}\fi
  \ifdim\ht\islandboxiii<\dimen@ii
   \global\setbox\islandboxiii\vbox to\dimen@ii{\box\islandboxiii\vfil}\fi
 \fi
 \global\setbox\islandbox@\vbox
  {\ifdim\dimen@>\z@
  \hbox to\hsize{\hfil\box\captionboxi\hfil\box\captionboxii\hfil
   \box\captionboxiii\hfil}%
  \nointerlineskip{\rm\global\skipi@-\dp\strutbox}%
  \global\advance\skipi@\bigskipamount\vskip\skipi@\fi
  \hbox to\hsize{\hfil\box\islandboxi\hfil\box\islandboxii\hfil
   \box\islandboxiii\hfil}}}
\def\Figurepair#1\and#2\endFigurepair{\island@true
 \islandpairbox{\Figure#1\endFigure}{\Figure#2\endFigure}}
\def\Figuretriple#1\and#2\and#3\endFiguretriple{\island@true
 \islandtriplebox{\Figure#1\endFigure}{\Figure#2\endFigure}%
  {\Figure#3\endFigure}}
\def\Tablepair#1\and#2\endTablepair{\island@true
 \islandpairboxa{\Table#1\endTable}{\Table#2\endTable}}
\def\Tabletriple#1\and#2\and#3\endTabletriple{\island@true
 \islandtripleboxa{\Table#1\endTable}{\Table#2\endTable}%
 {\Table#3\endTable}}
\def\place#1{\place@true\island@false
 #1%
 \ifisland@
  \box\islandbox@
 \else
  \Err@{Whoa ... there's no \string\Figure, \string\Table,
   etc., here}%
 \fi
 \place@false}
\newskip\belowtopfigskip
\belowtopfigskip 15\p@ plus 5\p@ minus5\p@
\newskip\abovebotfigskip
\abovebotfigskip 18\p@ plus 6\p@ minus6\p@
\newdimen\minpagesize
\minpagesize 5pc
\dimen@\belowtopfigskip
\advance\dimen@-\abovebotfigskip
\skip\topins\dimen@
\dimen\topins\z@
\newcount\topinscount@
\newbox\topinsdims@
\def\storedim@{\global\setbox\topinsdims@
 \vbox{\hbox to\dimen@{}\unvbox\topinsdims@}}
\def\advancedimtopins@{%
 \ifnum\pageno=\@ne
 \else
   \advance\dimen@\dimen\topins
   \global\dimen\topins\dimen@
 \fi}
\newcount\flipcount@
\def\fliptopins@{%
 \global\flipcount@\z@
 \ifvoid\topins\else
 \setbox\z@\vbox
  {\vskip\p@
   \unvbox\topins
   \global\setbox\topins\vbox{}%
   \loop
    \test@false
    \ifdim\lastskip=\z@\unskip
     \ifdim\lastskip=\z@
      \test@true\fi\fi
    \iftest@
    \global\advance\flipcount@\@ne
    \setboxzl@
    \global\setbox\topins\vbox{\unvbox\topins\boxz@}%
    \unpenalty
   \repeat}\fi}
\newif\ifPar@
\newcount\Parcount@
\newbox\Parbox@
\expandafter\newbox\csname Parfigbox1\endcsname
\expandafter\newbox\csname Parfigbox2\endcsname
\expandafter\newbox\csname Parfigbox3\endcsname
\expandafter\newbox\csname Parfigbox4\endcsname
\expandafter\newbox\csname Parfigbox5\endcsname
\expandafter\newdimen\csname Parprev1\endcsname
\expandafter\newdimen\csname Parprev2\endcsname
\expandafter\newdimen\csname Parprev3\endcsname
\expandafter\newdimen\csname Parprev4\endcsname
\expandafter\newdimen\csname Parprev5\endcsname
\expandafter\newdimen\csname Parprev6\endcsname
\def\Par{\par\global\csname Parprev1\endcsname\prevdepth
 \global\Parcount@\@ne
 \global\Par@true\global\let\Parlist@\empty
 \global\setbox\Parbox@\vbox\bgroup\break}
\def\place@#1#2{%
 \ifisland@
  \ifhmode
   \ifPar@
    \ifnum\Parcount@>5
     \Err@{Only 5 \string\place's allowed per
      \string\Par...\noexpand\endPar paragraph}%
    \else
     \expandafter\expandafter\expandafter
      \global\expandafter\setbox
       \csname Parfigbox\number\Parcount@\endcsname\box\islandbox@
     \global\advance\Parcount@\@ne
     \xdef\Parlist@{\Parlist@#1}%
    \fi
   \else
    \vadjust{#2}%
   \fi
  \else
   #2%
  \fi
 \else
  \Err@{Whoa ... there's no \string\Figure,
   \string\Table, etc., here}%
 \fi
 \place@false}
\long\def\Aplace#1{\prevanish@
 \place@true\island@false
 #1%
 \place@ a\Aplace@
 \postvanish@}
\long\def\AAplace#1{\prevanish@\place@true\island@false
 #1%
 \place@ A\AAplace@
 \postvanish@}
\newif\ifAA@
\def\AAplace@{\AA@true\Aplace@\AA@false}
\let\AAlist@\empty
\def\Aplace@{\allowbreak
 \dimen@=\ht\islandbox@
 \advance\dimen@\abovebotfigskip
 \ht\islandbox@\dimen@
 \advance\dimen@\dp\islandbox@
 \storedim@
 \ifAA@
  \xdef\AAlist@{\AAlist@1}%
  \advancedimtopins@
 \else
  \xdef\AAlist@{\AAlist@0}%
  \ifnum\topinscount@>\@ne\else\advancedimtopins@\fi
 \fi
 \insert\topins{\penalty\z@\splittopskip\z@\floatingpenalty\z@
  \box\islandbox@}%
 \global\advance\topinscount@\@ne}
\long\def\Bplace#1{\prevanish@\place@true\island@false
 #1%
 \place@ b\Bplace@
 \postvanish@}
\def\Bplace@{\allowbreak
 \ifnum\topinscount@=\z@
  \setbox\z@\vbox{\vbox to-\belowtopfigskip{}}%
  \dimen@-\skip\topins
  \ht\z@\dimen@
  \storedim@
  \advancedimtopins@
  \insert\topins{\boxz@}%
  \global\advance\topinscount@\@ne
  \xdef\AAlist@{\AAlist@0}%
 \fi
 \dimen@\ht\islandbox@
 \advance\dimen@\abovebotfigskip
 \ht\islandbox@\dimen@
 \advance\dimen@\dp\islandbox@
 \storedim@
 \xdef\AAlist@{\AAlist@0}%
 \ifnum\topinscount@>\@ne\else\advancedimtopins@\fi
 \insert\topins{\penalty\z@\splittopskip\z@
  \floatingpenalty\z@
  \box\islandbox@}%
 \global\advance\topinscount@\@ne}
\def\breakisland@{\global\setbox\@ne\lastbox\global\skipi@\lastskip\unskip
 \global\setbox\thr@@\lastbox}%
\def\printisland@{\centerline{\box\thr@@}\nobreak\nointerlineskip
 \vskip\skipi@
 \ifdim\ht\@ne<\z@\box\@ne\else\centerline{\box\@ne}\fi}
\def\bottomfigs@{%
 \count@\@ne
 \loop
  \ifnum\count@<\flipcount@
  \nointerlineskip
  \vskip\abovebotfigskip
  \global\setbox\topins\vbox{\unvbox\topins\setboxzl@
   \unvbox\z@
   \breakisland@}%
  \printisland@
  \advance\count@\@ne
  \repeat}
\def\resetdimtopins@{%
 \global\advance\topinscount@-\flipcount@
 \global\setbox\topinsdims@\vbox
  {\unvbox\topinsdims@
   \count@\z@
   \DN@##1##2\next@{\gdef\AAlist@{##2}}%
   \loop
    \ifnum\count@<\flipcount@\setboxzl@
    \expandafter\next@\AAlist@\next@
    \advance\count@\@ne
    \repeat
   \dimen@\z@
   \count@\z@
   \setbox\tw@\vbox{}%
   \edef\nextiii@{\AAlist@}%
   \DN@##1##2\next@{\DNii@{##1}\def\nextiii@{##2}}%
   \loop
    \test@false
    \ifnum\count@<\topinscount@
    \expandafter\next@\nextiii@\next@
     \ifnum\count@<\tw@
      \test@true
     \else
      \if\nextii@ 1\test@true\fi
     \fi
    \fi
    \iftest@
     \setboxzl@
     \advance\dimen@\wdz@
     \setbox\tw@\vbox{\boxz@\unvbox\tw@}%
     \advance\count@\@ne
    \repeat
    \unvbox\tw@
    \global\dimen\topins\dimen@}}
\def\Place@#1#2{%
 \ifisland@
  \ifhmode
   \ifPar@
    \ifnum\Parcount@>5
     \Err@{Only 5 \string\place's allowed per
       \string\Par...\noexpand\endPar paragraph}%
    \else
     \expandafter\expandafter\expandafter\global\expandafter\setbox
      \csname Parfigbox\number\Parcount@\endcsname\box\islandbox@
     \global\advance\Parcount@\@ne
     \xdef\Parlist@{\Parlist@#1}%
     \vadjust{\break}%
    \fi
   \else
    \Err@{\noexpand#2allowed only in a \string\Par...\noexpand\endPar
     paragraph}%
   \fi
  \else
   #2%
  \fi
 \else
  \Err@{Who ... there's no \string\Figure, \string\Table,
   etc., here}%
 \fi
 \place@false}
\newif\ifC@
\newdimen\Cdim@
\long\def\Cplace#1{\prevanish@\place@true\island@false
 #1%
 \Place@ c\Cplace@
 \postvanish@}
\def\Cplace@{\allowbreak
 \ifnum\topinscount@>\z@\else
  \global\C@true\global\Cdim@\pagetotal\fi
 \Aplace@}
\long\def\Mplace#1{\prevanish@\place@true\island@false
 #1%
 \Place@ m\Mplace@
 \postvanish@}
\long\def\MXplace#1{\prevanish@\place@true\island@false
 #1%
 \Place@ M\MXplace@
 \postvanish@}
\newif\ifMX@
\def\MXplace@{\MX@true\Mplace@\MX@false}
\def\Mplace@{\allowbreak
 \dimen@\ht\islandbox@\advance\dimen@\dp\islandbox@
 \ifdim\pagetotal=\z@\else
  \ifdim\lastskip<\abovebotfigskip\advance\dimen@\abovebotfigskip
  \advance\dimen@-\lastskip\fi
 \fi
 \advance\dimen@\pagetotal
 \ifdim\dimen@>\pagegoal
  \Aplace@
 \else
  \nointerlineskip
  \ifdim\lastskip<\abovebotfigskip\removelastskip\vskip\abovebotfigskip\fi
  \setbox\z@\vbox{\unvbox\islandbox@
   \breakisland@}%
  \printisland@
  \ifnum\topinscount@=\z@
   \setbox\z@\vbox{\vbox to-\belowtopfigskip{}}%
   \dimen@-\skip\topins
   \ht\z@\dimen@
   \storedim@
   \advancedimtopins@
   \insert\topins{\boxz@}%
   \global\advance\topinscount@\@ne
   \xdef\AAlist@{\AAlist@0}%
  \fi
  \ifMX@
   \ifnum\topinscount@=\@ne
    \setbox\z@\vbox{\vbox to-\abovebotfigskip{}}%
    \ht\z@\z@
    \dimen@\z@
    \storedim@
    \advancedimtopins@
    \insert\topins{\boxz@}%
    \global\advance\topinscount@\@ne
    \xdef\AAlist@{\AAlist@0}%
   \fi
  \fi
  \nointerlineskip
  \vskip\belowtopfigskip
 \fi}
\expandafter\newbox\csname Parbox1\endcsname
\expandafter\newbox\csname Parbox2\endcsname
\expandafter\newbox\csname Parbox3\endcsname
\expandafter\newbox\csname Parbox4\endcsname
\expandafter\newbox\csname Parbox5\endcsname
\def\endPar{\egroup
 \count@\@ne
 {\vbadness\@M\vfuzz\maxdimen\splitmaxdepth\maxdimen\splittopskip\ht\strutbox
 \setbox\z@\vsplit\Parbox@ to\ht\Parbox@
 \loop
  \ifnum\count@<\Parcount@
  \expandafter\expandafter\expandafter\global\expandafter\setbox
   \csname Parbox\number\count@\endcsname\vsplit\Parbox@ to\ht\Parbox@
  \count@@\count@\advance\count@@\@ne
  \global\csname Parprev\number\count@@\endcsname
   \dp\csname Parbox\number\count@\endcsname
  \advance\count@\@ne
  \repeat}%
 \vskip\parskip
 \count@\@ne
 \def\nextv@##1##2\nextv@{\DN@{##1}\gdef\Parlist@{##2}}%
 \loop
  \ifnum\count@<\Parcount@
   \dimen@\csname Parprev\number\count@\endcsname
   \advance\dimen@\ht\strutbox
   \ifdim\dimen@<\baselineskip
    \advance\dimen@-\baselineskip\vskip-\dimen@
   \else
    \vskip\lineskip
   \fi
   \unvbox\csname Parbox\number\count@\endcsname
   \global\setbox\islandbox@\box\csname Parfigbox\number\count@\endcsname
   \expandafter\nextv@\Parlist@\nextv@
   \if a\next@\Aplace@\else
   \if A\next@\AAplace@\else
   \if b\next@\Bplace@\else
   \if c\next@\Cplace@\else
   \if m\next@\Mplace@\else
   \if M\next@\MXplace@\fi\fi\fi\fi\fi\fi
  \advance\count@\@ne
  \repeat
 \global\Par@false
 \ifvoid\Parbox@
  \prevdepth\csname Parprev\number\count@\endcsname
 \else
  \dimen@\csname Parprev\number\count@\endcsname\advance\dimen@\ht\strutbox
  \ifdim\dimen@<\baselineskip
    \advance\dimen@-\baselineskip\vskip-\dimen@
  \else
    \vskip\lineskip
  \fi
  \dimen@\dp\Parbox@
  \unvbox\Parbox@
  \prevdepth\dimen@
 \fi}
\def\folio{{\page@F\page@S{\page@P\page@N{\number\page@C}\page@Q}}}
\def\advancepageno{\global\advance\pageno\@ne}
\newif\ifspecialsplit@
\newbox\outbox@
\let\shipout@\shipout
\def\plainoutput{\specialsplit@false\ifvoid\topins\else\ifdim\ht\topins=\z@
 \specialsplit@true\advance\minpagesize-\skip\topins\fi\fi
 \fliptopins@
 \setbox\outbox@\vbox{\makeheadline\pagebody\makefootline}%
 {\noexpands@\let\style\relax
 \shipout@\box\outbox@}%
 \advancepageno
 \resetdimtopins@
 \ifvoid\@cclv\else\unvbox\@cclv\penalty\outputpenalty\fi
 \ifnum\outputpenalty>-\@MM\else\dosupereject\fi}
\def\pagebody{\vbox to\vsize{\boxmaxdepth\maxdepth
 \ifvoid\margin@\else
 \rlap{\kern\hsize\vbox to\z@{\kern4\p@\box\margin@\vss}}\fi
 \pagecontents}}
\newif\ifonlytop@
\def\pagecontents{%
 \onlytop@false
 \ifdim\ht\@cclv<\minpagesize\ifnum\flipcount@<\tw@\ifvoid\footins
  \onlytop@true\fi\fi\fi
 \test@false
 \ifC@
  \ifnum\flipcount@=\@ne
   \global\multiply\Cdim@\tw@
   \ifdim\Cdim@>\ht\@cclv
    \test@true
   \fi
  \fi
 \fi
 \global\C@false
 \iftest@
  \dimen@\ht\@cclv
  \advance\dimen@\skip\topins
  {\vfuzz\maxdimen\vbadness\@M
  \splitmaxdepth\maxdepth\splittopskip\topskip
  \setbox\z@\vsplit\@cclv to\dimen@
  \unvbox\z@}%
  \global\setbox\topins\vbox{\unvbox\topins
   \global\setbox\@ne\lastbox}%
  \setbox\z@\vbox{\unvbox\@ne
   \breakisland@}%
  \nointerlineskip
  \vskip\abovebotfigskip
  \printisland@
 \else
  \ifnum\flipcount@>\z@
   \global\setbox\topins\vbox{\unvbox\topins\global\setbox\@ne\lastbox}%
   \setbox\z@\vbox{\unvbox\@ne
    \breakisland@}%
   \printisland@
   \ifonlytop@\kern-\prevdepth\vfill\else\vskip\belowtopfigskip\fi
  \fi
 \fi
 \ifdim\ht\@cclv<\minpagesize
  \ifonlytop@\else\vfill\fi
 \else
  \ifspecialsplit@
   {\vfuzz\maxdimen\vbadness\@M
   \splitmaxdepth\maxdepth\splittopskip\topskip
   \dimen@ii\ht\@cclv \advance\dimen@ii\skip\topins
   \setbox\z@\vsplit\@cclv to\dimen@ii
   \unvbox\z@}%
  \else
   \unvbox\@cclv
  \fi
 \fi
 \bottomfigs@
 \ifvoid\footins\else\vskip\skip\footins\footnoterule\unvbox\footins\fi}
\newread\readdata@
\def\readthedata@#1{\expandafter
 \ifx\csname#1@D\endcsname\relax
  \immediate\openin\readdata@=#1.dat
  \ifeof\readdata@
   \Err@{No file #1.dat}%
  \else
   {\endlinechar\m@ne\gdef\Next@{}%
   \DNii@##1 ##2 ##3pt{\global\data@ht##1\global\data@dp##2%
    \global\data@wd##3pt}%
   \loop
    \ifeof\readdata@
    \else
    \read\readdata@ to\next@
    \ifx\next@\empty\else
     \edef\next@{\expandafter\nextii@\next@}%
     \expandafter\rightadd@\next@\to\Next@
    \fi
    \repeat}%
   \immediate\closein\readdata@
   \expandafter\expandafter\expandafter\global\expandafter
    \let\csname#1@D\endcsname\Next@\global\let\Next@\relax
  \fi
 \fi}
\newdimen\data@ht
\newdimen\data@dp
\newdimen\data@wd
\newif\ifgetdata@
\def\getdata@#1#2{\global\getdata@true\count@#2\relax
 {\let\\\or\xdef\Next@{\ifcase\number\count@#1\else
 \global\noexpand\getdata@false\fi}}\Next@}
\def\paste#1#2{\readthedata@{#1}%
 \getdata@{\csname#1@D\endcsname}{#2}%
 \ifgetdata@
 \dimen@\data@ht \advance\dimen@\data@dp
  \hbox{\special{dvipaste: #1 #2}%
   \lower\data@dp\vbox to\dimen@{\hbox to\data@wd{}\vfil}}%
 \else
  {\lccode`\Z=`\#\lccode`\N=`\N\lccode`\F=`\F%
   \lowercase{\Err@{No data for File [#1], Z#2}}}%
 \fi}
\newdimen\httable
\newdimen\dptable
\newdimen\wdtable
\def\measuretable#1#2{\readthedata@{#1}%
 \getdata@{\csname#1@D\endcsname}{#2}%
 \ifgetdata@
  \httable\data@ht \dptable\data@dp \wdtable\data@wd
 \else
  {\lccode`\Z=`\#\lccode`\N=`\N\lccode`\F=`\F%
  \lowercase{\Err@{No data for File [#1], Z#2}}}%
 \fi}
\def\East#1#2{\setboxz@h{$\m@th\ssize\;{#1}\;\;$}%
 \setbox\tw@\hbox{$\m@th\ssize\;{#2}\;\;$}\setbox4=\hbox{$\m@th#2$}%
 \dimen@\minaw@
 \ifdim\wdz@>\dimen@\dimen@\wdz@\fi\ifdim\wd\tw@>\dimen@\dimen@\wd\tw@\fi
 \ifdim\wd4 >\z@
  \mathrel{\mathop{\hbox to\dimen@{\rightarrowfill}}\limits^{#1}_{#2}}%
 \else
  \mathrel{\mathop{\hbox to\dimen@{\rightarrowfill}}\limits^{#1}}%
 \fi}
\def\West#1#2{\setboxz@h{$\m@th\ssize\;\;{#1}\;$}%
 \setbox\tw@\hbox{$\m@th\ssize\;\;{#2}\;$}\setbox4=\hbox{$\m@th#2$}%
 \dimen@\minaw@
 \ifdim\wdz@>\dimen@\dimen@\wdz@\fi\ifdim\wd\tw@>\dimen@\dimen@\wd\tw@\fi
 \ifdim\wd4 >\z@
  \mathrel{\mathop{\hbox to\dimen@{\leftarrowfill}}\limits^{#1}_{#2}}%
 \else
  \mathrel{\mathop{\hbox to\dimen@{\leftarrowfill}}\limits^{#1}}%
 \fi}
\font\arrow@i=lams1
\font\arrow@ii=lams2
\font\arrow@iii=lams3
\font\arrow@iv=lams4
\font\arrow@v=lams5
\newdimen\standardcgap
\standardcgap40\p@
\newdimen\hunit
\hunit\tw@\p@
\newdimen\standardrgap
\standardrgap32\p@
\newdimen\vunit
\vunit1.6\p@
\def\Cgaps#1{\RIfM@
 \standardcgap#1\standardcgap\relax\hunit#1\hunit\relax
 \else\nonmatherr@\Cgaps\fi}
\def\Rgaps#1{\RIfM@
 \standardrgap#1\standardrgap\relax\vunit#1\vunit\relax
 \else\nonmatherr@\Rgaps\fi}
\newdimen\getdim@
\def\getcgap@#1{\ifcase#1\or\getdim@\z@\else\getdim@\standardcgap\fi}
\def\getrgap@#1{\ifcase#1\getdim@\z@\else\getdim@\standardrgap\fi}
\def\cgaps{\RIfM@\expandafter\cgaps@\else\expandafter\nonmatherr@
 \expandafter\cgaps\fi}
\def\cgaps@{\ifnum\catcode`\;=\active\expandafter\cgapsA@\else
 \expandafter\cgapsO@\fi}
\def\cgapsO@#1{\toks@{\ifcase\i@\or\getdim@=\z@}%
 \gaps@@\standardcgap#1;\gaps@@\gaps@@
 \edef\next@{\the\toks@\noexpand\else\noexpand\getdim@\noexpand\standardcgap
  \noexpand\fi}%
 \toks@=\expandafter{\next@}%
 \edef\getcgap@##1{\i@##1\relax\the\toks@}\toks@{}}
{\catcode`\;=\active
 \gdef\cgapsA@#1{\toks@{\ifcase\i@\or\getdim@=\z@}%
 \gaps@@\standardcgap#1;\gaps@@\gaps@@
 \edef\next@{\the\toks@\noexpand\else\noexpand\getdim@\noexpand\standardcgap
  \noexpand\fi}%
 \toks@=\expandafter{\next@}%
 \edef\getcgap@##1{\i@##1\relax\the\toks@}\toks@{}}
}
\def\Gaps@@{\gaps@@}
\def\gaps@@#1#2;#3{\mgaps@#1#2\mgaps@
 \edef\next@{\the\toks@\noexpand\or\noexpand\getdim@
  \noexpand#1\the\mgapstoks@@}%
 \toks@\expandafter{\next@}%
 \DN@{#3}%
 \ifx\next@\Gaps@@\def\next@##1\gaps@@{}\else
  \def\next@{\gaps@@#1#3}\fi\next@}
{\catcode`\;=\active
 \gdef\rgaps#1{\RIfM@{\ifnum\catcode`\;=\active\def;{\string;}\fi
   \xdef\Next@{\noexpand\rgaps@{#1}}}%
  \Next@\edef\getrgap@##1{\i@##1\relax\the\toks@}\toks@{}\else
  \nonmatherr@\rgaps\fi}
}
\def\rgaps@#1{\toks@{\ifcase\i@\getdim@=\z@}%
 \gaps@@\standardrgap#1;\gaps@@\gaps@@
 \edef\next@{\the\toks@\noexpand\else\noexpand\getdim@\noexpand\standardrgap
  \noexpand\fi}%
 \toks@=\expandafter{\next@}}
\newbox\ZER@
\def\mgaps@#1{\let\mgapsnext@#1\FNSS@\mgaps@@}
\def\mgaps@@{\ifx\next\w\expandafter\mgaps@@@\else
 \expandafter\mgaps@@@@\fi}
\newtoks\mgapstoks@@
\def\mgaps@@@@#1\mgaps@{\getdim@\mgapsnext@\getdim@#1\getdim@
 \edef\next@{\noexpand\getdim@\the\getdim@}%
 \mgapstoks@@\expandafter{\next@}}
\def\mgaps@@@\w#1#2\mgaps@{\mgaps@@@@#2\mgaps@
 \setbox\ZER@\hbox{$\m@th\hskip15\p@\tsize@#1$}%
 \dimen@\wd\ZER@
 \ifdim\dimen@>\getdim@\getdim@\dimen@\fi
 \edef\next@{\noexpand\getdim@\the\getdim@}%
 \mgapstoks@@\expandafter{\next@}}
\def\changewidth#1#2{\setbox\ZER@{$\m@th#2}%
 \hbox to\wd\ZER@{\hss$\m@th#1$\hss}}
\atdef@({\FN@\ARROW@}
\def\ARROW@{\ifx\next)\let\next@\OPTIONS@\else
 \DN@{\csname\string @(\endcsname}\fi\next@}
\newif\ifoptions@
\def\OPTIONS@){\ifoptions@\let\next@\relax\else
 \DN@{\global\options@true\begingroup\optioncodes@}\fi\next@}
\newif\ifN@
\newif\ifE@
\newif\ifNESW@
\newif\ifH@
\newif\ifV@
\newif\ifHshort@
\expandafter\def\csname\string @(\endcsname #1,#2){%
 \ifoptions@\expandafter\endgroup\fi
 \N@false\E@false\H@false\V@false\Hshort@false
 \ifnum#1>\z@\E@true\fi
 \ifnum#1=\z@\V@true\global\tX@false\global\tY@false\global\a@false\fi
 \ifnum#2>\z@\N@true\fi
 \ifnum#2=\z@\H@true\global\tX@false\global\tY@false\global\a@false
  \ifshort@\Hshort@true\fi\fi
 \NESW@false
 \ifN@\ifE@\NESW@true\fi\else\ifE@\else\NESW@true\fi\fi
 \arrow@{#1}{#2}%
 \global\options@false
 \global\scount@\z@\global\tcount@\z@\global\arrcount@\z@
 \global\s@false\global\sxdimen@\z@\global\sydimen@\z@
 \global\tX@false\global\tXdimen@i\z@\global\tXdimen@ii\z@
 \global\tY@false\global\tYdimen@i\z@\global\tYdimen@ii\z@
 \global\a@false\global\exacount@\z@
 \global\x@false\global\xdimen@\z@
 \global\X@false\global\Xdimen@\z@
 \global\y@false\global\ydimen@\z@
 \global\Y@false\global\Ydimen@\z@
 \global\p@false\global\pdimen@\z@
 \global\label@ifalse\global\label@iifalse
 \global\dl@ifalse\global\ldimen@i\z@
 \global\dl@iifalse\global\ldimen@ii\z@
 \global\short@false\global\unshort@false}
\newif\iflabel@i
\newif\iflabel@ii
\newcount\scount@
\newcount\tcount@
\newcount\arrcount@
\newif\ifs@
\newdimen\sxdimen@
\newdimen\sydimen@
\newif\iftX@
\newdimen\tXdimen@i
\newdimen\tXdimen@ii
\newif\iftY@
\newdimen\tYdimen@i
\newdimen\tYdimen@ii
\newif\ifa@
\newcount\exacount@
\newif\ifx@
\newdimen\xdimen@
\newif\ifX@
\newdimen\Xdimen@
\newif\ify@
\newdimen\ydimen@
\newif\ifY@
\newdimen\Ydimen@
\newif\ifp@
\newdimen\pdimen@
\newif\ifdl@i
\newif\ifdl@ii
\newdimen\ldimen@i
\newdimen\ldimen@ii
\newif\ifshort@
\newif\ifunshort@
\def\zero@#1{\ifnum\scount@=\z@
 \if#1e\global\scount@\m@ne\else
 \if#1t\global\scount@\tw@\else
 \if#1h\global\scount@\thr@@\else
 \if#1'\global\scount@6 \else
 \if#1`\global\scount@7 \else
 \if#1(\global\scount@8 \else
 \if#1)\global\scount@9 \else
 \if#1s\global\scount@12 \else
 \if#1H\global\scount@13 \else
 \Err@{\Invalid@@ option \string\0}\fi\fi\fi\fi\fi\fi\fi\fi\fi
 \fi}
\def\one@#1{\ifnum\tcount@=\z@
 \if#1e\global\tcount@\m@ne\else
 \if#1h\global\tcount@\tw@\else
 \if#1t\global\tcount@\thr@@\else
 \if#1'\global\tcount@4 \else
 \if#1`\global\tcount@5 \else
 \if#1(\global\tcount@\ten@ \else
 \if#1)\global\tcount@11 \else
 \if#1s\global\tcount@12 \else
 \if#1H\global\tcount@13 \else
 \Err@{\Invalid@@ option \string\1}\fi\fi\fi\fi\fi\fi\fi\fi\fi
 \fi}
\def\a@#1{\ifnum\arrcount@=\z@
 \if#10\global\arrcount@\m@ne\else
 \if#1+\global\arrcount@\@ne\else
 \if#1-\global\arrcount@\tw@\else
 \if#1=\global\arrcount@\thr@@\else
 \Err@{\Invalid@@ option \string\a}\fi\fi\fi\fi
 \fi}
\def\ds@{\ifnum\catcode`\;=\active\expandafter\dsA@\else
 \expandafter\dsO@\fi}
\def\dsO@(#1;#2){\ds@@{#1}{#2}}
\def\ds@@#1#2{\ifs@\else
 \global\s@true
 \global\sxdimen@\hunit\global\sxdimen@#1\sxdimen@\relax
 \global\sydimen@\vunit\global\sydimen@#2\sydimen@\relax
 \fi}
\def\dtX@{\ifnum\catcode`\;=\active\expandafter\dtXA@\else
 \expandafter\dtXO@\fi}
\def\dtXO@(#1;#2){\dtX@@{#1}{#2}}
\def\dtX@@#1#2{\iftX@\else
 \global\tX@true
 \global\tXdimen@i\hunit\global\tXdimen@i#1\tXdimen@i\relax
 \global\tXdimen@ii\vunit\global\tXdimen@ii#2\tXdimen@ii\relax
 \fi}
\def\dtY@{\ifnum\catcode`\;=\active\expandafter\dtYA@\else
 \expandafter\dtYO@\fi}
\def\dtYO@(#1;#2){\dtY@@{#1}{#2}}
\def\dtY@@#1#2{\iftY@\else
 \global\tY@true
 \global\tYdimen@i\hunit\global\tYdimen@i#1\tYdimen@i\relax
 \global\tYdimen@ii\vunit\global\tYdimen@ii#2\tYdimen@ii\relax
 \fi}
{\catcode`\;=\active
 \gdef\dsA@(#1;#2){\ds@@{#1}{#2}}
 \gdef\dtXA@(#1;#2){\dtX@@{#1}{#2}}
 \gdef\dtYA@(#1;#2){\dtY@@{#1}{#2}}
}
\def\da@#1{\ifa@\else\global\a@true\global\exacount@#1\relax\fi}
\def\dx@#1{\ifx@\else
 \global\x@true
 \global\xdimen@\hunit\global\xdimen@#1\xdimen@\relax
 \fi}
\def\dX@#1{\ifX@\else
 \global\X@true
 \global\Xdimen@\hunit\global\Xdimen@#1\Xdimen@\relax
 \fi}
\def\dy@#1{\ify@\else
 \global\y@true
 \global\ydimen@\vunit\global\ydimen@#1\ydimen@\relax
 \fi}
\def\dY@#1{\ifY@\else
 \global\Y@true
 \global\Ydimen@\vunit\global\Ydimen@#1\Ydimen@\relax
 \fi}
\def\p@@#1{\ifp@\else
 \global\p@true
 \global\pdimen@\hunit\global\divide\pdimen@\tw@
 \global\pdimen@#1\pdimen@\relax
 \fi}
\def\L@#1{\iflabel@i\else
 \global\label@itrue\gdef\label@i{#1}%
 \fi}
\def\l@#1{\iflabel@ii\else
 \global\label@iitrue\gdef\label@ii{#1}%
 \fi}
\def\dL@#1{\ifdl@i\else
 \global\dl@itrue\global\ldimen@i\hunit\global\ldimen@i#1\ldimen@i\relax
 \fi}
\def\dl@#1{\ifdl@ii\else
 \global\dl@iitrue\global\ldimen@ii\hunit\global\ldimen@ii#1\ldimen@ii\relax
 \fi}
\def\s@{\ifunshort@\else\global\short@true\fi}
\def\uns@{\ifshort@\else\global\unshort@true\global\short@false\fi}
\def\optioncodes@{\let\0\zero@\let\1\one@\let\a\a@\let\ds\ds@\let\dtX\dtX@
 \let\dtY\dtY@\let\da\da@\let\dx\dx@\let\dX\dX@\let\dY\dY@\let\dy\dy@
 \let\p\p@@\let\L\L@\let\l\l@\let\dL\dL@\let\dl\dl@\let\s\s@\let\uns\uns@}
\def\slopes@{\\161\\152\\143\\134\\255\\126\\357\\238\\349\\45{10}\\56{11}%
 \\11{12}\\65{13}\\54{14}\\43{15}\\32{16}\\53{17}\\21{18}\\52{19}\\31{20}%
 \\41{21}\\51{22}\\61{23}}
\newcount\tan@i
\newcount\tan@ip
\newcount\tan@ii
\newcount\tan@iip
\newdimen\slope@i
\newdimen\slope@ip
\newdimen\slope@ii
\newdimen\slope@iip
\newcount\angcount@
\newcount\extracount@
\def\slope@{{\slope@i\secondy@\advance\slope@i-\firsty@
 \ifN@\else\multiply\slope@i\m@ne\fi
 \slope@ii\secondx@\advance\slope@ii-\firstx@
 \ifE@\else\multiply\slope@ii\m@ne\fi
 \ifdim\slope@ii<\z@
  \global\tan@i6 \global\tan@ii\@ne\global\angcount@23
 \else
  \dimen@\slope@i\multiply\dimen@6
  \ifdim\dimen@<\slope@ii
   \global\tan@i\@ne\global\tan@ii6 \global\angcount@\@ne
  \else
   \dimen@\slope@ii\multiply\dimen@6
   \ifdim\dimen@<\slope@i
    \global\tan@i6 \global\tan@ii\@ne\global\angcount@23
   \else
    \global\tan@ip\z@\global\tan@iip\@ne
    \def\\##1##2##3{\global\angcount@##3\relax
     \slope@ip\slope@i\slope@iip\slope@ii
     \multiply\slope@iip##1\relax\multiply\slope@ip##2\relax
     \ifdim\slope@iip<\slope@ip
      \global\tan@ip##1\relax\global\tan@iip##2\relax
     \else
      \global\tan@i##1\relax\global\tan@ii##2\relax
      \def\\####1####2####3{}%
     \fi}%
    \slopes@
    \slope@i\secondy@\advance\slope@i-\firsty@
    \ifN@\else\multiply\slope@i\m@ne\fi
    \multiply\slope@i\tan@ii\multiply\slope@i\tan@iip\multiply\slope@i\tw@
    \count@\tan@i\multiply\count@\tan@iip
    \extracount@\tan@ip\multiply\extracount@\tan@ii
    \advance\count@\extracount@
    \slope@ii\secondx@\advance\slope@ii-\firstx@
    \ifE@\else\multiply\slope@ii\m@ne\fi
    \multiply\slope@ii\count@
    \ifdim\slope@i<\slope@ii
     \global\tan@i\tan@ip\global\tan@ii\tan@iip
     \global\advance\angcount@\m@ne
    \fi
   \fi
  \fi
 \fi}%
}
\def\slope@a#1{{\def\\##1##2##3{\ifnum##3=#1\global\tan@i##1\relax
 \global\tan@ii##2\relax\fi}\slopes@}}
\newcount\i@
\newcount\j@
\newcount\colcount@
\newcount\Colcount@
\newcount\tcolcount@
\newdimen\rowht@
\newdimen\rowdp@
\newcount\rowcount@
\newcount\Rowcount@
\newcount\maxcolrow@
\newtoks\colwidthtoks@
\newtoks\Rowheighttoks@
\newtoks\Rowdepthtoks@
\newtoks\widthtoks@
\newtoks\Widthtoks@
\newtoks\heighttoks@
\newtoks\Heighttoks@
\newtoks\depthtoks@
\newtoks\Depthtoks@
\newif\iffirstCDcr@
\def\dotoks@i{%
 \global\widthtoks@\expandafter{\the\widthtoks@\else\getdim@\z@\fi}%
 \global\heighttoks@\expandafter{\the\heighttoks@\else\getdim@\z@\fi}%
 \global\depthtoks@\expandafter{\the\depthtoks@\else\getdim@\z@\fi}}
\def\dotoks@ii{%
 \global\widthtoks@{\ifcase\j@}%
 \global\heighttoks@{\ifcase\j@}%
 \global\depthtoks@{\ifcase\j@}}
\def\preCD@#1\endCD{\setbox\ZER@
 \vbox{%
  \def\arrow@##1##2{{}}%
  \global\rowcount@\m@ne\global\colcount@\z@\global\Colcount@\z@
  \global\firstCDcr@true\toks@{}%
  \global\widthtoks@{\ifcase\j@}%
  \global\Widthtoks@{\ifcase\i@}%
  \global\heighttoks@{\ifcase\j@}%
  \global\Heighttoks@{\ifcase\i@}%
  \global\depthtoks@{\ifcase\j@}%
  \global\Depthtoks@{\ifcase\i@}%
  \global\Rowheighttoks@{\ifcase\i@}%
  \global\Rowdepthtoks@{\ifcase\i@}%
  \Let@
  \everycr{%
   \noalign{%
    \global\advance\rowcount@\@ne
    \ifnum\colcount@<\Colcount@
    \else
     \global\Colcount@\colcount@\global\maxcolrow@\rowcount@
    \fi
    \global\colcount@\z@
    \iffirstCDcr@
     \global\firstCDcr@false
    \else
     \edef\next@{\the\Rowheighttoks@\noexpand\or\noexpand\getdim@\the\rowht@}%
      \global\Rowheighttoks@\expandafter{\next@}%
     \edef\next@{\the\Rowdepthtoks@\noexpand\or\noexpand\getdim@\the\rowdp@}%
      \global\Rowdepthtoks@\expandafter{\next@}%
     \global\rowht@\z@\global\rowdp@\z@
     \dotoks@i
     \edef\next@{\the\Widthtoks@\noexpand\or\the\widthtoks@}%
      \global\Widthtoks@\expandafter{\next@}%
     \edef\next@{\the\Heighttoks@\noexpand\or\the\heighttoks@}%
      \global\Heighttoks@\expandafter{\next@}%
     \edef\next@{\the\Depthtoks@\noexpand\or\the\depthtoks@}%
      \global\Depthtoks@\expandafter{\next@}%
     \dotoks@ii
    \fi}}%
  \tabskip\z@
  \halign{&\setbox\ZER@\hbox{\vrule\height\ten@\p@\width\z@\depth\z@     
   $\m@th\displaystyle{##}$}\copy\ZER@
   \ifdim\ht\ZER@>\rowht@\global\rowht@\ht\ZER@\fi
   \ifdim\dp\ZER@>\rowdp@\global\rowdp@\dp\ZER@\fi
   \global\advance\colcount@\@ne
   \edef\next@{\the\widthtoks@\noexpand\or\noexpand\getdim@\the\wd\ZER@}%
    \global\widthtoks@\expandafter{\next@}%
   \edef\next@{\the\heighttoks@\noexpand\or\noexpand\getdim@\the\ht\ZER@}%
    \global\heighttoks@\expandafter{\next@}%
   \edef\next@{\the\depthtoks@\noexpand\or\noexpand\getdim@\the\dp\ZER@}%
    \global\depthtoks@\expandafter{\next@}%
   \cr#1\crcr}}%
 \Rowcount@\rowcount@
 \global\Widthtoks@\expandafter{\the\Widthtoks@\fi\relax}%
 \edef\Width@##1##2{\i@##1\relax\j@##2\relax\the\Widthtoks@}%
 \global\Heighttoks@\expandafter{\the\Heighttoks@\fi\relax}%
 \edef\Height@##1##2{\i@##1\relax\j@##2\relax\the\Heighttoks@}%
 \global\Depthtoks@\expandafter{\the\Depthtoks@\fi\relax}%
 \edef\Depth@##1##2{\i@##1\relax\j@##2\relax\the\Depthtoks@}%
 \edef\next@{\the\Rowheighttoks@\noexpand\fi\relax}%
 \global\Rowheighttoks@\expandafter{\next@}%
 \edef\Rowheight@##1{\i@##1\relax\the\Rowheighttoks@}%
 \edef\next@{\the\Rowdepthtoks@\noexpand\fi\relax}%
 \global\Rowdepthtoks@\expandafter{\next@}%
 \edef\Rowdepth@##1{\i@##1\relax\the\Rowdepthtoks@}%
 \global\colwidthtoks@{\fi}%
 \setbox\ZER@\vbox{%
  \unvbox\ZER@
  \count@\rowcount@
  \loop
   \unskip\unpenalty
   \setbox\ZER@\lastbox
   \ifnum\count@>\maxcolrow@\advance\count@\m@ne
   \repeat
  \hbox{%
   \unhbox\ZER@
   \count@\z@
   \loop
    \unskip
    \setbox\ZER@\lastbox
    \edef\next@{\noexpand\or\noexpand\getdim@\the\wd\ZER@\the\colwidthtoks@}%
     \global\colwidthtoks@\expandafter{\next@}%
    \advance\count@\@ne
    \ifnum\count@<\Colcount@
    \repeat}}%
 \edef\next@{\noexpand\ifcase\noexpand\i@\the\colwidthtoks@}%
  \global\colwidthtoks@\expandafter{\next@}%
 \edef\Colwidth@##1{\i@##1\relax\the\colwidthtoks@}%
 \global\colwidthtoks@{}\global\Rowheighttoks@{}\global\Rowdepthtoks@{}%
 \global\widthtoks@{}\global\Widthtoks@{}\global\heighttoks@{}%
 \global\Heighttoks@{}\global\depthtoks@{}\global\Depthtoks@{}%
}
\newcount\xoff@
\newcount\yoff@
\newcount\endcount@
\newcount\rcount@
\newdimen\firstx@
\newdimen\firsty@
\newdimen\secondx@
\newdimen\secondy@
\newdimen\tocenter@
\newdimen\charht@
\newdimen\charwd@
\def\outside@{\Err@{This arrow points outside the \string\CD}}
\newif\ifsvertex@
\newif\iftvertex@
\def\arrow@#1#2{\global\xoff@#1\relax\global\yoff@#2\relax
 \count@\rowcount@\advance\count@-\yoff@
 \ifnum\count@<\@ne\outside@\else\ifnum\count@>\Rowcount@\outside@\fi\fi
 \count@\colcount@\advance\count@\xoff@
 \ifnum\count@<\@ne\outside@\else\ifnum\count@>\Colcount@\outside@\fi\fi
 \tcolcount@\colcount@\advance\tcolcount@\xoff@
 \Width@\rowcount@\colcount@\divide\getdim@\tw@\tocenter@-\getdim@
 \ifdim\getdim@=\z@
  \firstx@\z@\firsty@\mathaxis@\svertex@true
 \else
  \svertex@false
  \ifHshort@
   \Colwidth@\colcount@\divide\getdim@\tw@
   \ifE@ \firstx@\getdim@ \else \firstx@-\getdim@ \fi
  \else
   \ifE@ \firstx@\getdim@ \else \firstx@-\getdim@ \fi
  \fi
  \ifE@
   \ifH@ \advance\firstx@\thr@@\p@ \else \advance\firstx@-\thr@@\p@ \fi  
  \else
   \ifH@ \advance\firstx@-\thr@@\p@ \else \advance\firstx@\thr@@\p@ \fi  
  \fi
  \ifN@
   \Height@\rowcount@\colcount@ \firsty@\getdim@                         
   \ifV@ \advance\firsty@\thr@@\p@ \fi                                   
  \else
   \ifV@
    \Depth@\rowcount@\colcount@ \firsty@-\getdim@                        
    \advance\firsty@-\thr@@\p@                                           
   \else
    \firsty@\z@                                                          
   \fi
  \fi
 \fi
 \ifV@
 \else
  \Colwidth@\colcount@\divide\getdim@\tw@
  \ifE@\secondx@\getdim@\else\secondx@-\getdim@\fi
  \ifE@\else\getcgap@\colcount@\advance\secondx@-\getdim@\fi
  \endcount@\colcount@\advance\endcount@\xoff@
  \count@\colcount@
  \ifE@
   \advance\count@\@ne
   \loop
    \ifnum\count@<\endcount@
    \Colwidth@\count@\advance\secondx@\getdim@
    \getcgap@\count@\advance\secondx@\getdim@
    \advance\count@\@ne
    \repeat
  \else
   \advance\count@\m@ne
   \loop
    \ifnum\count@>\endcount@
    \Colwidth@\count@\advance\secondx@-\getdim@
    \getcgap@\count@\advance\secondx@-\getdim@
    \advance\count@\m@ne
    \repeat
  \fi
  \Colwidth@\count@\divide\getdim@\tw@
  \ifHshort@
  \else
   \ifE@\advance\secondx@\getdim@\else\advance\secondx@-\getdim@\fi
  \fi
  \ifE@\getcgap@\count@\advance\secondx@\getdim@\fi
  \rcount@\rowcount@\advance\rcount@-\yoff@
  \Width@\rcount@\count@\divide\getdim@\tw@
  \tvertex@false
  \ifH@\ifdim\getdim@=\z@\tvertex@true\Hshort@false\fi\fi
  \ifHshort@
  \else
   \ifE@\advance\secondx@-\getdim@\else\advance\secondx@\getdim@\fi
  \fi
  \iftvertex@
   \advance\secondx@.4\p@
  \else
   \ifE@\advance\secondx@-\thr@@\p@\else\advance\secondx@\thr@@\p@\fi    
  \fi
 \fi
 \ifH@
 \else
  \ifN@
   \Rowheight@\rowcount@\secondy@\getdim@
  \else
   \Rowdepth@\rowcount@\secondy@-\getdim@
   \getrgap@\rowcount@\advance\secondy@-\getdim@
  \fi
  \endcount@\rowcount@\advance\endcount@-\yoff@
  \count@\rowcount@
  \ifN@
   \advance\count@\m@ne
   \loop
    \ifnum\count@>\endcount@
    \Rowheight@\count@\advance\secondy@\getdim@
    \Rowdepth@\count@\advance\secondy@\getdim@
    \getrgap@\count@\advance\secondy@\getdim@
    \advance\count@\m@ne
    \repeat
  \else
   \advance\count@\@ne
   \loop
    \ifnum\count@<\endcount@
    \Rowheight@\count@\advance\secondy@-\getdim@
    \Rowdepth@\count@\advance\secondy@-\getdim@
    \getrgap@\count@\advance\secondy@-\getdim@
    \advance\count@\@ne
    \repeat
  \fi
  \tvertex@false
  \ifV@\Width@\count@\colcount@\ifdim\getdim@=\z@\tvertex@true\fi\fi
  \ifN@
   \getrgap@\count@\advance\secondy@\getdim@
   \Rowdepth@\count@\advance\secondy@\getdim@
   \iftvertex@
    \advance\secondy@\mathaxis@
   \else
    \Depth@\count@\tcolcount@\advance\secondy@-\getdim@
    \advance\secondy@-\thr@@\p@                                          
   \fi
  \else
   \Rowheight@\count@\advance\secondy@-\getdim@
   \iftvertex@
    \advance\secondy@\mathaxis@
   \else
    \Height@\count@\tcolcount@\advance\secondy@\getdim@
    \advance\secondy@\thr@@\p@                                           
   \fi
  \fi
 \fi
 \ifV@\else\advance\firstx@\sxdimen@\fi
 \ifH@\else\advance\firsty@\sydimen@\fi
 \iftX@
  \advance\secondy@\tXdimen@ii
  \advance\secondx@\tXdimen@i
  \slope@
 \else
  \iftY@
   \advance\secondy@\tYdimen@ii
   \advance\secondx@\tYdimen@i
   \slope@
   \secondy@\secondx@\advance\secondy@-\firstx@
   \ifNESW@\else\multiply\secondy@\m@ne\fi
   \multiply\secondy@\tan@i\divide\secondy@\tan@ii\advance\secondy@\firsty@
  \else
   \ifa@
    \slope@
    \ifNESW@\global\advance\angcount@\exacount@\else
     \global\advance\angcount@-\exacount@\fi
    \ifnum\angcount@>23 \global\angcount@23 \fi
    \ifnum\angcount@<\@ne\global\angcount@\@ne\fi
    \slope@a\angcount@
    \ifY@
     \advance\secondy@\Ydimen@
    \else
     \ifX@
      \advance\secondx@\Xdimen@
      \dimen@\secondx@\advance\dimen@-\firstx@
      \ifNESW@\else\multiply\dimen@\m@ne\fi
      \multiply\dimen@\tan@i\divide\dimen@\tan@ii
      \advance\dimen@\firsty@\secondy@\dimen@
     \fi
    \fi
   \else
    \ifH@\else\ifV@\else\slope@\fi\fi
   \fi
  \fi
 \fi
 \ifH@\else\ifV@\else\ifsvertex@\else
  \dimen@6\p@\multiply\dimen@\tan@ii
  \count@\tan@i\advance\count@\tan@ii\divide\dimen@\count@
  \ifE@\advance\firstx@\dimen@\else\advance\firstx@-\dimen@\fi
  \multiply\dimen@\tan@i\divide\dimen@\tan@ii
  \ifN@\advance\firsty@\dimen@\else\advance\firsty@-\dimen@\fi
 \fi\fi\fi
 \ifp@
  \ifH@\else\ifV@\else
   \getcos@\pdimen@\advance\firsty@\dimen@\advance\secondy@\dimen@
   \ifNESW@\advance\firstx@-\dimen@ii\else\advance\firstx@\dimen@ii\fi
  \fi\fi
 \fi
 \ifH@\else\ifV@\else
  \ifnum\tan@i>\tan@ii
   \charht@\ten@\p@\charwd@\ten@\p@
   \multiply\charwd@\tan@ii\divide\charwd@\tan@i
  \else
   \charwd@\ten@\p@\charht@\ten@\p@
   \divide\charht@\tan@ii\multiply\charht@\tan@i
  \fi
  \ifnum\tcount@=\thr@@
   \ifN@\advance\secondy@-.3\charht@\else\advance\secondy@.3\charht@\fi
  \fi
  \ifnum\scount@=\tw@
   \ifE@\advance\firstx@.3\charht@\else\advance\firstx@-.3\charht@\fi
  \fi
  \ifnum\tcount@=12
   \ifN@\advance\secondy@-\charht@\else\advance\secondy@\charht@\fi
  \fi
  \iftY@
  \else
   \ifa@
    \ifX@
    \else
     \secondx@\secondy@\advance\secondx@-\firsty@
     \ifNESW@\else\multiply\secondx@\m@ne\fi
     \multiply\secondx@\tan@ii\divide\secondx@\tan@i
     \advance\secondx@\firstx@
    \fi
   \fi
  \fi
 \fi\fi
 \ifH@\harrow@\else\ifV@\varrow@\else\arrow@@\fi\fi}
\newdimen\mathaxis@
\mathaxis@90\p@\divide\mathaxis@36
\def\harrow@b{\ifE@\hskip\tocenter@\hskip\firstx@\fi}
\def\harrow@bb{\ifE@\hskip\xdimen@\else\hskip\Xdimen@\fi}
\def\harrow@e{\ifE@\else\hskip-\firstx@\hskip-\tocenter@\fi}
\def\harrow@ee{\ifE@\hskip-\Xdimen@\else\hskip-\xdimen@\fi}
\def\harrow@{\dimen@\secondx@\advance\dimen@-\firstx@
 \ifE@\let\next@\rlap\else\multiply\dimen@\m@ne\let\next@\llap\fi
 \next@{%
  \harrow@b
  \smash{\raise\pdimen@\hbox to\dimen@
   {\harrow@bb\arrow@ii
    \ifnum\arrcount@=\m@ne\else\ifnum\arrcount@=\thr@@\else
     \ifE@
      \ifnum\scount@=\m@ne
      \else
       \ifcase\scount@\or\or\char118 \or\char117 \or\or\or\char119 \or
       \char120 \or\char121 \or\char122 \or\or\or\arrow@i\char125 \or
       \char117 \hskip\thr@@\p@\char117 \hskip-\thr@@\p@\fi
      \fi
     \else
      \ifnum\tcount@=\m@ne
      \else
       \ifcase\tcount@\char117 \or\or\char117 \or\char118 \or\char119 \or
       \char120 \or\or\or\or\or\char121 \or\char122 \or\arrow@i\char125
       \or\char117 \hskip\thr@@\p@\char117 \hskip-\thr@@\p@\fi
      \fi
     \fi
    \fi\fi
    \dimen@\mathaxis@\advance\dimen@.2\p@
    \dimen@ii\mathaxis@\advance\dimen@ii-.2\p@
    \ifnum\arrcount@=\m@ne
     \let\leads@\null
    \else
     \ifcase\arrcount@
      \def\leads@{\hrule\height\dimen@\depth-\dimen@ii}\or
      \def\leads@{\hrule\height\dimen@\depth-\dimen@ii}\or
      \def\leads@{\hbox to\ten@\p@{%
       \leaders\hrule\height\dimen@\depth-\dimen@ii\hfil
       \hfil
      \leaders\hrule\height\dimen@\depth-\dimen@ii\hskip\z@ plus2fil\relax
       \hfil
       \leaders\hrule\height\dimen@\depth-\dimen@ii\hfil}}\or
     \def\leads@{\hbox{\hbox to\ten@\p@{\dimen@\mathaxis@\advance\dimen@1.2\p@
       \dimen@ii\dimen@\advance\dimen@ii-.4\p@
       \leaders\hrule\height\dimen@\depth-\dimen@ii\hfil}%
       \kern-\ten@\p@
       \hbox to\ten@\p@{\dimen@\mathaxis@\advance\dimen@-1.2\p@
       \dimen@ii\dimen@\advance\dimen@ii-.4\p@
       \leaders\hrule\height\dimen@\depth-\dimen@ii\hfil}}}\fi
    \fi
    \cleaders\leads@\hfil
    \ifnum\arrcount@=\m@ne\else\ifnum\arrcount@=\thr@@\else
     \arrow@i
     \ifE@
      \ifnum\tcount@=\m@ne
      \else
       \ifcase\tcount@\char119 \or\or\char119 \or\char120 \or\char121 \or
       \char122 \or \or\or\or\or\char123 \or\char124 \or
       \char125 \or\char119 \hskip-\thr@@\p@\char119 \hskip\thr@@\p@\fi
      \fi
     \else
      \ifcase\scount@\or\or\char120 \or\char119 \or\or\or\char121 \or\char122
      \or\char123 \or\char124 \or\or\or\char125 \or
      \char119 \hskip-\thr@@\p@\char119 \hskip\thr@@\p@\fi
     \fi
    \fi\fi
    \harrow@ee}}%
  \harrow@e}%
 \iflabel@i
  \dimen@ii\z@\setbox\ZER@\hbox{$\m@th\tsize@@\label@i$}%
  \ifnum\arrcount@=\m@ne
  \else
   \advance\dimen@ii\mathaxis@
   \advance\dimen@ii\dp\ZER@\advance\dimen@ii\tw@\p@
   \ifnum\arrcount@=\thr@@\advance\dimen@ii\tw@\p@\fi
  \fi
  \advance\dimen@ii\pdimen@
  \next@{\harrow@b\smash{\raise\dimen@ii\hbox to\dimen@
   {\harrow@bb\hskip\tw@\ldimen@i\hfil\box\ZER@\hfil\harrow@ee}}\harrow@e}%
 \fi
 \iflabel@ii
  \ifnum\arrcount@=\m@ne
  \else
   \setbox\ZER@\hbox{$\m@th\tsize@\label@ii$}%
   \dimen@ii-\ht\ZER@\advance\dimen@ii-\tw@\p@
   \ifnum\arrcount@=\thr@@\advance\dimen@ii-\tw@\p@\fi
   \advance\dimen@ii\mathaxis@\advance\dimen@ii\pdimen@
   \next@{\harrow@b\smash{\raise\dimen@ii\hbox to\dimen@
    {\harrow@bb\hskip\tw@\ldimen@ii\hfil\box\ZER@\hfil\harrow@ee}}\harrow@e}%
  \fi
 \fi}
\let\tsize@\tsize
\def\tsizeCDlabels{\let\tsize@\tsize}
\def\ssizeCDlabels{\let\tsize@\ssize}
\def\tsize@@{\ifnum\arrcount@=\m@ne\else\tsize@\fi}
\def\varrow@{\dimen@\secondy@\advance\dimen@-\firsty@
 \ifN@\else\multiply\dimen@\m@ne\fi
 \setbox\ZER@\vbox to\dimen@
  {\ifN@\vskip-\Ydimen@\else\vskip\ydimen@\fi
   \ifnum\arrcount@=\m@ne\else\ifnum\arrcount@=\thr@@\else
    \hbox{\arrow@iii
     \ifN@
      \ifnum\tcount@=\m@ne
      \else
       \ifcase\tcount@\char117 \or\or\char117 \or\char118 \or\char119 \or
       \char120 \or\or\or\or\or\char121 \or\char122 \or\char123 \or
       \vbox{\hbox{\char117}\nointerlineskip\vskip\thr@@\p@
       \hbox{\char117}\vskip-\thr@@\p@}\fi
      \fi
     \else
      \ifcase\scount@\or\or\char118 \or\char117 \or\or\or\char119 \or
      \char120 \or\char121 \or\char122 \or\or\or\char123 \or
      \vbox{\hbox{\char117}\nointerlineskip\vskip\thr@@\p@
      \hbox{\char117}\vskip-\thr@@\p@}\fi
     \fi}%
    \nointerlineskip
   \fi\fi
   \ifnum\arrcount@=\m@ne
    \let\leads@\null
   \else
    \ifcase\arrcount@\let\leads@\vrule\or\let\leads@\vrule\or
    \def\leads@{\vbox to\ten@\p@{%
     \hrule\height1.67\p@\depth\z@\width.4\p@
     \vfil
     \hrule\height3.33\p@\depth\z@\width.4\p@
     \vfil
     \hrule\height1.67\p@\depth\z@\width.4\p@}}\or
    \def\leads@{\hbox{\vrule\height\p@\hskip\tw@\p@\vrule}}\fi
   \fi
  \cleaders\leads@\vfill\nointerlineskip
   \ifnum\arrcount@=\m@ne\else\ifnum\arrcount@=\thr@@\else
    \hbox{\arrow@iv
     \ifN@
      \ifcase\scount@\or\or\char118 \or\char117 \or\or\or\char119 \or
      \char120 \or\char121 \or\char122 \or\or\or\arrow@iii\char123 \or
      \vbox{\hbox{\char117}\nointerlineskip\vskip-\thr@@\p@
      \hbox{\char117}\vskip\thr@@\p@}\fi
     \else
      \ifnum\tcount@=\m@ne
      \else
       \ifcase\tcount@\char117 \or\or\char117 \or\char118 \or\char119 \or
       \char120 \or\or\or\or\or\char121 \or\char122 \or\arrow@iii\char123 \or
       \vbox{\hbox{\char117}\nointerlineskip\vskip-\thr@@\p@
       \hbox{\char117}\vskip\thr@@\p@}\fi
      \fi
     \fi}%
   \fi\fi
   \ifN@\vskip\ydimen@\else\vskip-\Ydimen@\fi}%
 \ifN@
  \dimen@ii\firsty@
 \else
  \dimen@ii-\firsty@\advance\dimen@ii\ht\ZER@\multiply\dimen@ii\m@ne
 \fi
 \rlap{\smash{\hskip\tocenter@\hskip\pdimen@\raise\dimen@ii\box\ZER@}}%
 \iflabel@i
  \setbox\ZER@\vbox to\dimen@{\vfil
   \hbox{$\m@th\tsize@@\label@i$}\vskip\tw@\ldimen@i\vfil}%
  \rlap{\smash{\hskip\tocenter@\hskip\pdimen@
  \ifnum\arrcount@=\m@ne\let\next@\relax\else\let\next@\llap\fi
  \next@{\raise\dimen@ii\hbox{\ifnum\arrcount@=\m@ne\hskip-.5\wd\ZER@\fi
   \box\ZER@\ifnum\arrcount@=\m@ne\else\hskip\tw@\p@\fi}}}}%
 \fi
 \iflabel@ii
  \ifnum\arrcount@=\m@ne
  \else
   \setbox\ZER@\vbox to\dimen@{\vfil
    \hbox{$\m@th\tsize@\label@ii$}\vskip\tw@\ldimen@ii\vfil}%
   \rlap{\smash{\hskip\tocenter@\hskip\pdimen@
   \rlap{\raise\dimen@ii\hbox{\ifnum\arrcount@=\thr@@\hskip4.5\p@\else
    \hskip2.5\p@\fi\box\ZER@}}}}%
  \fi
 \fi
}
\newdimen\goal@
\newdimen\shifted@
\newcount\Tcount@
\newcount\Scount@
\newbox\shaft@
\newcount\slcount@
\def\getcos@#1{%
 \ifnum\tan@i<\tan@ii
  \dimen@#1%
  \ifnum\slcount@<8 \count@9 \else \ifnum\slcount@<12 \count@8 \else
   \count@7 \fi\fi
  \multiply\dimen@\count@\divide\dimen@\ten@
  \dimen@ii\dimen@\multiply\dimen@ii\tan@i\divide\dimen@ii\tan@ii
 \else
  \dimen@ii#1%
  \count@-\slcount@\advance\count@24
  \ifnum\count@<8 \count@9 \else \ifnum\count@<12 \count@8
   \else\count@7 \fi\fi
  \multiply\dimen@ii\count@\divide\dimen@ii\ten@
  \dimen@\dimen@ii\multiply\dimen@\tan@ii\divide\dimen@\tan@i
 \fi}
\newdimen\adjust@
\def\Nnext@{\ifN@\let\next@\raise\else\let\next@\lower\fi}
\def\arrow@@{\slcount@\angcount@
 \ifNESW@
  \ifnum\angcount@<\ten@
   \let\arrowfont@\arrow@i\global\advance\angcount@\m@ne
   \global\multiply\angcount@13
  \else
   \ifnum\angcount@<19
    \let\arrowfont@\arrow@ii\global\advance\angcount@-\ten@
    \global\multiply\angcount@13
   \else
    \let\arrowfont@\arrow@iii\global\advance\angcount@-19
    \global\multiply\angcount@13
  \fi\fi
  \Tcount@\angcount@
 \else
  \ifnum\angcount@<5
   \let\arrowfont@\arrow@iii\global\advance\angcount@\m@ne
   \global\multiply\angcount@13 \global\advance\angcount@65
  \else
   \ifnum\angcount@<14
    \let\arrowfont@\arrow@iv\global\advance\angcount@-5
    \global\multiply\angcount@13
   \else
    \ifnum\angcount@<23
     \let\arrowfont@\arrow@v\global\advance\angcount@-14
     \global\multiply\angcount@13
    \else
     \let\arrowfont@\arrow@i\global\angcount@117
  \fi\fi\fi
  \ifnum\angcount@=117 \Tcount@115 \else\Tcount@\angcount@\fi
 \fi
 \Scount@\Tcount@
 \ifE@
  \ifnum\tcount@=\z@\advance\Tcount@\tw@\else\ifnum\tcount@=13
   \advance\Tcount@\tw@\else\advance\Tcount@\tcount@\fi\fi
  \ifnum\scount@=\z@\else\ifnum\scount@=13 \advance\Scount@\thr@@\else
   \advance\Scount@\scount@\fi\fi
 \else
  \ifcase\tcount@\advance\Tcount@\thr@@\or\or\advance\Tcount@\thr@@\or
  \advance\Tcount@\tw@\or\advance\Tcount@6 \or\advance\Tcount@7
  \or\or\or\or\or\advance\Tcount@8 \or\advance\Tcount@9 \or
  \advance\Tcount@12 \or\advance\Tcount@\thr@@\fi
  \ifcase\scount@\or\or\advance\Scount@\thr@@\or\advance\Scount@\tw@\or
  \or\or\advance\Scount@4 \or\advance\Scount@5 \or\advance\Scount@\ten@
  \or\advance\Scount@11 \or\or\or\advance\Scount@12 \or\advance
  \Scount@\tw@\fi
 \fi
 \ifcase\arrcount@\or\or\global\advance\angcount@\@ne\else\fi
 \ifN@\shifted@\firsty@\else\shifted@-\firsty@\fi
 \ifE@\else\advance\shifted@\charht@\fi
 \goal@\secondy@\advance\goal@-\firsty@
 \ifN@\else\multiply\goal@\m@ne\fi
 \setbox\shaft@\hbox{\arrowfont@\char\angcount@}%
 \ifnum\arrcount@=\thr@@
  \getcos@{1.5\p@}%
  \setbox\shaft@\hbox to\wd\shaft@{\arrowfont@
   \rlap{\hskip\dimen@ii
    \smash{\ifNESW@\let\next@\lower\else\let\next@\raise\fi
     \next@\dimen@\hbox{\arrowfont@\char\angcount@}}}%
   \rlap{\hskip-\dimen@ii
    \smash{\ifNESW@\let\next@\raise\else\let\next@\lower\fi
      \next@\dimen@\hbox{\arrowfont@\char\angcount@}}}\hfil}%
 \fi
 \rlap{\smash{\hskip\tocenter@\hskip\firstx@
  \ifnum\arrcount@=\m@ne
  \else
   \ifnum\arrcount@=\thr@@
   \else
    \ifnum\scount@=\m@ne
    \else
     \ifnum\scount@=\z@
     \else
      \setbox\ZER@\hbox{\ifnum\angcount@=117 \arrow@v\else\arrowfont@\fi
       \char\Scount@}%
      \ifNESW@
       \ifnum\scount@=\tw@
        \dimen@\shifted@\advance\dimen@-\charht@
        \ifN@\hskip-\wd\ZER@\fi
        \Nnext@
        \next@\dimen@\copy\ZER@
        \ifN@\else\hskip-\wd\ZER@\fi
       \else
        \Nnext@
        \ifN@\else\hskip-\wd\ZER@\fi
        \next@\shifted@\copy\ZER@
        \ifN@\hskip-\wd\ZER@\fi
       \fi
       \ifnum\scount@=12
        \advance\shifted@\charht@\advance\goal@-\charht@
        \ifN@\hskip\wd\ZER@\else\hskip-\wd\ZER@\fi
       \fi
       \ifnum\scount@=13
        \getcos@{\thr@@\p@}%
        \ifN@\hskip\dimen@\else\hskip-\wd\ZER@\hskip-\dimen@\fi
        \adjust@\shifted@\advance\adjust@\dimen@ii
        \Nnext@
        \next@\adjust@\copy\ZER@
        \ifN@\hskip-\dimen@\hskip-\wd\ZER@\else\hskip\dimen@\fi
       \fi
      \else
       \ifN@\hskip-\wd\ZER@\fi
       \ifnum\scount@=\tw@
        \ifN@\hskip\wd\ZER@\else\hskip-\wd\ZER@\fi
        \dimen@\shifted@\advance\dimen@-\charht@
        \Nnext@
        \next@\dimen@\copy\ZER@
        \ifN@\hskip-\wd\ZER@\fi
       \else
        \Nnext@
        \next@\shifted@\copy\ZER@
        \ifN@\else\hskip-\wd\ZER@\fi
       \fi
       \ifnum\scount@=12
        \advance\shifted@\charht@\advance\goal@-\charht@
        \ifN@\hskip-\wd\ZER@\else\hskip\wd\ZER@\fi
       \fi
       \ifnum\scount@=13
        \getcos@{\thr@@\p@}%
        \ifN@\hskip-\wd\ZER@\hskip-\dimen@\else\hskip\dimen@\fi
        \adjust@\shifted@\advance\adjust@\dimen@ii
        \Nnext@
        \next@\adjust@\copy\ZER@
        \ifN@\hskip\dimen@\else\hskip-\dimen@\hskip-\wd\ZER@\fi
       \fi	
      \fi
  \fi\fi\fi\fi
  \ifnum\arrcount@=\m@ne
  \else
   \loop
    \ifdim\goal@>\charht@
    \ifE@\else\hskip-\charwd@\fi
    \Nnext@
    \next@\shifted@\copy\shaft@
    \ifE@\else\hskip-\charwd@\fi
    \advance\shifted@\charht@\advance\goal@-\charht@
    \repeat
   \ifdim\goal@>\z@
    \dimen@\charht@\advance\dimen@-\goal@
    \divide\dimen@\tan@i\multiply\dimen@\tan@ii
    \ifE@\hskip-\dimen@\else\hskip-\charwd@\hskip\dimen@\fi
    \adjust@\shifted@\advance\adjust@-\charht@\advance\adjust@\goal@
    \Nnext@
    \next@\adjust@\copy\shaft@
    \ifE@\else\hskip-\charwd@\fi
   \else
    \adjust@\shifted@\advance\adjust@-\charht@
   \fi
  \fi
  \ifnum\arrcount@=\m@ne
  \else
   \ifnum\arrcount@=\thr@@
   \else
    \ifnum\tcount@=\m@ne
    \else
     \setbox\ZER@
      \hbox{\ifnum\angcount@=117 \arrow@v\else\arrowfont@\fi\char\Tcount@}%
     \ifnum\tcount@=\thr@@
      \advance\adjust@\charht@
      \ifE@\else\ifN@\hskip-\charwd@\else\hskip-\wd\ZER@\fi\fi
     \else
      \ifnum\tcount@=12
       \advance\adjust@\charht@
       \ifE@\else\ifN@\hskip-\charwd@\else\hskip-\wd\ZER@\fi\fi
      \else
       \ifE@\hskip-\wd\ZER@\fi
     \fi\fi
     \Nnext@
     \next@\adjust@\copy\ZER@
     \ifnum\tcount@=13
      \hskip-\wd\ZER@
      \getcos@{\thr@@\p@}%
      \ifE@\hskip-\dimen@\else\hskip\dimen@\fi
      \advance\adjust@-\dimen@ii
      \Nnext@
      \next@\adjust@\box\ZER@
     \fi
  \fi\fi\fi}}%
 \iflabel@i
  \rlap{\hskip\tocenter@
  \dimen@\firstx@\advance\dimen@\secondx@\divide\dimen@\tw@
  \advance\dimen@\ldimen@i
  \dimen@ii\firsty@\advance\dimen@ii\secondy@\divide\dimen@ii\tw@
  \global\multiply\ldimen@i\tan@i\global\divide\ldimen@i\tan@ii
  \ifNESW@\advance\dimen@ii\ldimen@i\else\advance\dimen@ii-\ldimen@i\fi
  \setbox\ZER@\hbox{\ifNESW@\else\ifnum\arrcount@=\thr@@\hskip4\p@\else
   \hskip\tw@\p@\fi\fi
   $\m@th\tsize@@\label@i$\ifNESW@\ifnum\arrcount@=\thr@@\hskip4\p@\else
   \hskip\tw@\p@\fi\fi}%
  \ifnum\arrcount@=\m@ne
   \ifNESW@\advance\dimen@.5\wd\ZER@\advance\dimen@\p@\else
    \advance\dimen@-.5\wd\ZER@\advance\dimen@-\p@\fi
   \advance\dimen@ii-.5\ht\ZER@
  \else
   \advance\dimen@ii\dp\ZER@
   \ifnum\slcount@<6 \advance\dimen@ii\tw@\p@\fi
  \fi
  \hskip\dimen@
  \ifNESW@\let\next@\llap\else\let\next@\rlap\fi
  \next@{\smash{\raise\dimen@ii\box\ZER@}}}%
 \fi
 \iflabel@ii
  \ifnum\arrcount@=\m@ne
  \else
   \rlap{\hskip\tocenter@
   \dimen@\firstx@\advance\dimen@\secondx@\divide\dimen@\tw@
   \ifNESW@\advance\dimen@\ldimen@ii\else\advance\dimen@-\ldimen@ii\fi
   \dimen@ii\firsty@\advance\dimen@ii\secondy@\divide\dimen@ii\tw@
   \global\multiply\ldimen@ii\tan@i\global\divide\ldimen@ii\tan@ii
   \advance\dimen@ii\ldimen@ii
   \setbox\ZER@\hbox{\ifNESW@\ifnum\arrcount@=\thr@@\hskip4\p@\else
    \hskip\tw@\p@\fi\fi
    $\m@th\tsize@\label@ii$\ifNESW@\else\ifnum\arrcount@=\thr@@\hskip4\p@
    \else\hskip\tw@\p@\fi\fi}%
   \advance\dimen@ii-\ht\ZER@
   \ifnum\slcount@<9 \advance\dimen@ii-\thr@@\p@\fi
   \ifNESW@\let\next@\rlap\else\let\next@\llap\fi
   \hskip\dimen@\next@{\smash{\raise\dimen@ii\box\ZER@}}}%
  \fi
 \fi
}
\def\outCD@#1{\def#1{\Err@{\noexpand#1must not be used within \string\CD}}}
\newskip\preCDskip@
\newskip\postCDskip@
\preCDskip@\z@
\postCDskip@\z@
\def\preCDspace#1{\RIfMIfI@
 \onlydmatherr@\preCDspace\else\advance\preCDskip@#1\relax\fi\else
 \onlydmatherr@\preCDspace\fi}
\def\postCDspace#1{\RIfMIfI@
 \onlydmatherr@\postCDspace\else\advance\postCDskip@#1\relax\fi\else
 \onlydmatherr@\postCDspace\fi}
\def\predisplayspace#1{\RIfMIfI@
 \onlydmatherr@\predisplayspace\else
 \advance\abovedisplayskip#1\relax
 \advance\abovedisplayshortskip#1\relax\fi
 \else\onlydmatherr@\preCDspace\fi}
\def\postdisplayspace#1{\RIfMIfI@
 \onlydmatherr@\postdisplayspace\else
 \advance\belowdisplayskip#1\relax
 \advance\belowdisplayshortskip#1\relax\fi
 \else\onlydmatherr@\postdisplayspace\fi}
\def\PreCDSpace#1{\global\preCDskip@#1\relax}
\def\PostCDSpace#1{\global\postCDskip@#1\relax}
\def\CD#1\endCD{%
 \outCD@\cgaps\outCD@\rgaps\outCD@\Cgaps\outCD@\Rgaps
 \preCD@#1\endCD
 \advance\abovedisplayskip\preCDskip@
 \advance\abovedisplayshortskip\preCDskip@
 \advance\belowdisplayskip\postCDskip@
 \advance\belowdisplayshortskip\postCDskip@
 \vcenter{\offinterlineskip
  \vskip\preCDskip@\Let@\global\colcount@\@ne\global\rowcount@\z@
  \everycr{%
   \noalign{%
    \ifnum\rowcount@=\Rowcount@
    \else
     \getrgap@\rowcount@\vskip\getdim@
     \global\advance\rowcount@\@ne\global\colcount@\@ne
    \fi}}%
  \tabskip\z@
  \halign{&\global\xoff@\z@\global\yoff@\z@
   \getcgap@\colcount@\hskip\getdim@
   \hfil\vrule\height\ten@\p@\width\z@\depth\z@
   $\m@th\displaystyle{##}$\hfil
   \global\advance\colcount@\@ne\cr
   #1\crcr}\vskip\postCDskip@}%
 \preCDskip@\z@\postCDskip@\z@
 \def\getcgap@##1{\ifcase##1\or\getdim@\z@\else\getdim@\standardcgap\fi}%
 \def\getrgap@##1{\ifcase##1\getdim@\z@\else\getdim@\standardrgap\fi}%
 \let\Width@\relax\let\Height@\relax\let\Depth@\relax\let\Rowheight@\relax
 \let\Rowdepth@\relax\let\Colwidth@\relax
}

\def\alloc@#1#2#3#4#5{\global\advance\count1#1by\@ne
  \ch@ck#1#4#2%
  \allocationnumber=\count1#1%
  \global#3#5=\allocationnumber
  \wlog{\string#5=\string#2\the\allocationnumber}}
\catcode`\@=\active

\catcode`\"=12
\font\black=cmbx10
\font\sblack=cmbx7
\font\ssblack=cmbx5
\font\blackital=cmmib10  \skewchar\blackital='177
\font\sblackital=cmmib7  \skewchar\sblackital='177
\font\ssblackital=cmmib5  \skewchar\ssblackital='177
\font\sanss=txss
\font\ssanss=txss scaled 650
\font\sssanss=txss scaled 500
\font\blackboard=msbm10
\font\sblackboard=msbm7
\font\ssblackboard=msbm5
\font\caligr=zplmr7y
\font\scaligr=zplmr7y scaled 650
\font\sscaligr=zplmr7y scaled 500

\font\fraktur=eufm10
\font\sfraktur=eufm7
\font\ssfraktur=eufm5

\font\bsymb=cmsy10 scaled\magstep2
\def\all#1{\setbox0=\hbox{\lower1.5pt\hbox{\bsymb
       \char"38}}\setbox1=\hbox{$_{#1}$} \box0\lower2pt\box1\;}
\def\exi#1{\setbox0=\hbox{\lower1.5pt\hbox{\bsymb \char"39}}
       \setbox1=\hbox{$_{#1}$} \box0\lower2pt\box1\;}

\def\tx#1{{\fam0\relax#1}}

\newfam\bifam
\textfont\bifam=\blackital
\scriptfont\bifam=\sblackital
\scriptscriptfont\bifam=\ssblackital
\def\bi#1{{\fam\bifam\relax#1}}

\newfam\blfam
\textfont\blfam=\black
\scriptfont\blfam=\sblack
\scriptscriptfont\blfam=\ssblack

\newfam\bbfam
\textfont\bbfam=\blackboard
\scriptfont\bbfam=\sblackboard
\scriptscriptfont\bbfam=\ssblackboard
\def\bb#1{{\fam\bbfam\relax#1}}

\newfam\ssfam
\textfont\ssfam=\sanss
\scriptfont\ssfam=\ssanss
\scriptscriptfont\ssfam=\sssanss
\def\ss#1{{\fam\ssfam\relax#1}}

\newfam\clfam
\textfont\clfam=\caligr
\scriptfont\clfam=\scaligr
\scriptscriptfont\clfam=\sscaligr

\newfam\frfam
\textfont\frfam=\fraktur
\scriptfont\frfam=\sfraktur
\scriptscriptfont\frfam=\ssfraktur

\font\syy=cmsy10 scaled 1150

\def\WE{{\setbox0=\hbox{\syy\char"5E}\box0}}

\def\hpb#1{\setbox0=\hbox{${#1}$}
    \copy0 \kern-\wd0 \kern.2pt \box0}
\def\vpb#1{\setbox0=\hbox{${#1}$}
    \copy0 \kern-\wd0 \raise.08pt \box0}

\def\pmb#1{\setbox0\hbox{${#1}$} \copy0 \kern-\wd0 \kern.2pt \box0}
\def\pmbb#1{\setbox0\hbox{${#1}$} \copy0 \kern-\wd0
      \kern.2pt \copy0 \kern-\wd0 \kern.2pt \box0}
\def\pmbbb#1{\setbox0\hbox{${#1}$} \copy0 \kern-\wd0
      \kern.2pt \copy0 \kern-\wd0 \kern.2pt
    \copy0 \kern-\wd0 \kern.2pt \box0}
\def\pmxb#1{\setbox0\hbox{${#1}$} \copy0 \kern-\wd0
      \kern.2pt \copy0 \kern-\wd0 \kern.2pt
      \copy0 \kern-\wd0 \kern.2pt \copy0 \kern-\wd0 \kern.2pt \box0}
\def\pmxbb#1{\setbox0\hbox{${#1}$} \copy0 \kern-\wd0 \kern.2pt
      \copy0 \kern-\wd0 \kern.2pt
      \copy0 \kern-\wd0 \kern.2pt \copy0 \kern-\wd0 \kern.2pt
      \copy0 \kern-\wd0 \kern.2pt \box0}

\mathchardef\za="710B  
\mathchardef\zb="710C  
\mathchardef\zg="710D  
\mathchardef\zd="710E  
\mathchardef\zve="710F 
\mathchardef\zz="7110  
\mathchardef\zh="7111  
\mathchardef\zvy="7112 
\mathchardef\zi="7113  
\mathchardef\zk="7114  
\mathchardef\zl="7115  
\mathchardef\zm="7116  
\mathchardef\zn="7117  
\mathchardef\zx="7118  
\mathchardef\zp="7119  
\mathchardef\zr="711A  
\mathchardef\zs="711B  
\mathchardef\zt="711C  
\mathchardef\zu="711D  
\mathchardef\zvf="711E 
\mathchardef\zq="711F  
\mathchardef\zc="7120  
\mathchardef\zw="7121  
\mathchardef\ze="7122  
\mathchardef\zy="7123  
\mathchardef\zvp="7124 
\mathchardef\zvr="7125 
\mathchardef\zvs="7126 
\mathchardef\zf="7127  
\mathchardef\zG="7000  
\mathchardef\zD="7001  
\mathchardef\zY="7002  
\mathchardef\zL="7003  
\mathchardef\zX="7004  
\mathchardef\zP="7005  
\mathchardef\zS="7006  
\mathchardef\zU="7007  
\mathchardef\zF="7008  
\mathchardef\zC="7009  
\mathchardef\zW="700A  

\catcode`\"=\active


\loadmsam
\loadmsbm
\newsymbol\blacksquare 1004
\newsymbol\blacklozenge 1007
\newsymbol\leqslant 1336
\newsymbol\geqslant 133E
\newsymbol\centerdot 1205
\newsymbol\shortparallel 2371

\newcount\secnum  \secnum=0 
\newcount\subsecnum
\newcount\subsubsecnum 

\newcount\Asecnum  \secnum=0 
\newcount\Asubsecnum

\def\sect#1{ 
            \global\advance \secnum by 1 \subsecnum=0 \subsubsecnum=0 
			\vskip2pt
            \noindent{\bf \the \secnum.\ #1. \vskip-4.5mm}
			\nobreak
			\leftline{}}

\def\ssca#1{
	    	\global\advance\subsecnum by 1 \subsubsecnum=0
			\vskip1pt
            \noindent{\bf \the \secnum.\the\subsecnum\  #1. \vskip-4.5mm}
		    \nobreak
			\leftline{}}

\def\sssa#1{\global\advance\subsubsecnum by 1 
            \vskip1pt
			\noindent{\bf \the \secnum .\the\subsecnum.\the\subsubsecnum\ #1. \vskip-4.5mm} 
	        \nobreak
		    \leftline{}}

\def\Asect#1{ 
            \global\advance \Asecnum by 1 \Asubsecnum=0 
			\vskip2pt
            \noindent{\bf \the \Asecnum.\ #1. \vskip-4.5mm}
			\nobreak
			\leftline{}}

\def\Assect#1{
	    	\global\advance\Asubsecnum by 1 
			\vskip1pt
            \noindent{\bf A\the \Asecnum .\the\Asubsecnum\  #1. \vskip-4.5mm}
		    \nobreak
			\leftline{}}

\def\sssa#1{\global\advance\subsubsecnum by 1 
            \vskip1pt
			\noindent{\bf \the \secnum .\the\subsecnum.\the\subsubsecnum\ #1. \vskip-4.5mm} 
	        \nobreak
		    \leftline{}}

\font\tfont=cmb10 

\define\Title#1{\vskip2mm\centerline{\tfont #1}\vskip1.5mm}

		\catcode`\"=12
	\font\kropa=lcircle10 scaled 1700
	\def\ybl{\setbox0=\hbox{\kropa \char"70} \kern1.5pt \raise.35pt \box0}
	\def\zbl{\setbox0=\hbox{\kropa \char"70} \kern1.5pt \raise2.9pt \box0}
		\catcode`\"=\active

	\def\lpr{{\setbox0=\hbox{\vrule height .15pt width 3.5pt depth 0pt}
\setbox1=\hbox{\vrule height 5.8pt width .3pt depth 0pt}\kern2pt\box0\box1\kern3pt}}

\def\*{{\textstyle *}}

\def\polar{{\textstyle\circ}}
\newsymbol\blacktriangle 104E
\newsymbol\blacktriangleleft 134A
\newsymbol\blacktriangledown 1048

\def\sP{{\scriptscriptstyle\P}}

\def\N{{\bb N}}
\def\R{{\bb R}}

\def\Z{{\bb Z}}

\def\*{{\textstyle *}}
\def\s*{{\scriptstyle *}}

\def\by{{\bi y}}

\def\bc{{\bi c}}
\def\bd{{\bi d}}

\def\sA{{\ss A}}

\def\sC{{\ss C}}

\def\sG{{\ss G}}
\def\sH{{\ss H}}

\def\sK{{\ss K}}

\def\sO{{\ss O}}
\def\sP{{\ss P}}

\def\sT{{\ss T}}

\def\sF{{\ss F}}

\def\rd{\tx{d}}
\def\xi{\tx{i}}
\def\xD{\tx{D}}

\def\sgn{\operatorname{sgn}}

\def\det{\operatorname{det}}

\def\dim{\operatorname{dim}}


	\def\wA{{\widetilde A}}

	\def\of{{\overline f}}

	\def\of{{\overline f}}

	\def\ozf{{\overline \zf}}

	\def\wA{{\widetilde A}}

	\def\wE{{\widetilde E}}
	\def\wF{{\widetilde F}}
	\def\wG{{\widetilde G}}
	\def\wJ{{\widetilde J}}

	\def\-{{-}}
	\def\+{{+}}

    \input paper.st\relax
    \hsize=42pc
    \hoffset=-10pt
    \vsize=52pc
    \voffset=6pt
    \TagsOnRight
    \document
    \input xy
    \xyoption{all}


		\catcode`\"=12


		\catcode`\"=\active

	\define\compose#1#2#3#4#5#6{{\setbox0=\hbox{\raise#2\hbox{\kern#3\hbox{${#1}$}}}
\setbox1=\hbox{\raise#5\hbox{\kern#6\hbox{${#4}$}}}\box0\box1}}

	\define\position#1#2#3{{\setbox0=\hbox{\raise#2\hbox{\kern#3\hbox{${#1}$}}}}}

	\def\lpr{{\setbox0=\hbox{\vrule height .15pt width 3.5pt depth 0pt}
\setbox1=\hbox{\vrule height 5.8pt width .3pt depth 0pt}\kern2pt\box0\box1\kern3pt}}

	\def\rpr{{\setbox0=\hbox{\vrule height .15pt width 3.5pt depth 0pt}
\setbox1=\hbox{\vrule height 5.8pt width .3pt depth 0pt}\kern3pt\box0\kern-3.5pt\box1\kern6.5pt}}

	\font\F=cmr12
	\def\star{\setbox0=\hbox{\lower3pt\hbox{\text{\F *}}}\box0}

	\def\idx{\operatorname{idx}}

	\def\iA{{\bi A}}
	\def\iU{{\bi U}}
	\def\iE{{\bi E}}
	\def\iD{{\bi D}}
	\def\iB{{\bi B}}
	\def\iH{{\bi H}}
	\def\iQ{{\bi Q}}
	\def\iJ{{\bi J}}
	\def\iC{{\bi C}}

	\def\oviD{\overline{\bi D}}
	\def\oviH{\overline{\bi H}}
	\def\oviQ{\overline{\bi Q}}
	\def\oviJ{\overline{\bi J}}

	\def\wiA{\widetilde{\bi A}}
	\def\wiU{\widetilde{\bi U}}
	\def\wiE{\widetilde{\bi E}}
	\def\wiB{\widetilde{\bi B}}

	\def\ovA{{\overline A}}
	\def\ovG{{\overline G}}

	\def\ozf{{\overline \zf}}

	\def\wA{{\widetilde A}}
	\def\wE{{\widetilde E}}
	\def\wF{{\widetilde F}}
	\def\wG{{\widetilde G}}
	\def\wJ{{\widetilde J}}

	\def\We{{\operatorname{We}}}
	\def\Div{{\operatorname{Div}}}

	\def\-{{-}}
	\def\+{{+}}

	\title
	Space-time orientations and Maxwell's equations
	\endtitle

	\author
        Giuseppe Marmo \\
        Dipartimento di Scienze Fisiche \\
        Universit\`a Federico II di Napoli \\
        Complesso Universitario di Monte Sant'Angelo \\
        Via Cinthia, 80126 Napoli, Italy \\
        Istituto Nazionale di Fisica Nucleare, 
        Sezione di Napoli, Italy \\
        {\tt Giuseppe.Marmo\@na.infn.it} \\
            \\
        Emanuele Parasecoli \\
		Borgo Panni Vallone, 56 \\
		60019 Senigallia (AN) Italy \\
			\\
        W\l odzimierz M. Tulczyjew \\
        Dipartimento di Fisica \\
        Universit\`a di Camerino \\
        62032 Camerino, Italy \\
	    Associated with \\
        Istituto Nazionale di Fisica Nucleare,
        Sezione di Napoli, Italy \\
		{\tt tulczy\@libero.it} \\
	\endauthor

		\thanks{Supported by PRIN SINTESI}

	\maketitle

\leftline{\bf Abstract.}

	An analysis of the concept of orientation used in electrodynamics is presented.  At least two different versions are
encountered in the literature.  Both are clearly identified and comparisons are made.

\leftline{\bf Keywords.}

	Excitations, field strengths, orientation, odd and even forms, extended Lorentz group.

		\sect{Introduction.}
	While studying transformation properties of electromagnetic fields under time reflection we encountered disagreements
between different formulations of electrodynamics in physics literature.  Rigorous formulations identify electromagnetic
objects as even and odd differential forms known to Schouten [1] [2] under different names and reintroduced by de Rham [3].
These identifications imply certain response of electromagnetic fields to time reflection.  Standard formulations of
electrodynamics used by theoretical physicists are usually presented in a frame dependent form and are not in agreement
with the rigorous space-time formulations.  We are presenting both versions of electrodynamics formulated in intrinsic,
frame independent fashion in the affine Minkowski space-time.

	Interpretation of electromagnetic quantities as differential forms was introduced by Cartan [4].  We have consulted a
number of texts using differential forms.  These include {\it Classical Electrodynamics} by R. S. Ingarden and A. Jamio\l
kowski  [5], {\it Formal Structure of Electromagnetics} by E. J. Post [6], {\it Applied differential geometry} by William
L. Burke [7], {\it Relativistic Electrodynamics and Differential geometry}, by S. Parrott [8], and {\it Gravitation}, by C.
W. Misner, K. S. Thorn, and J. A. Wheeler [9].  Geometric objects necessary for correct interpretation of physical
quantities were studied by Schouten.  Schouten classified geometric objects according to their transformation properties
including their response to reflections.  Two types of geometric objects appear in electrodynamics.  These are the even and
odd differential forms according to de Rham's terminology.  Some authors use even and odd differential forms in their
formulations of electrodynamics.  Others use the ``star operator'' derived from Hodge theory.  Transformation properties
relative to reflections can be correctly described in terms of Schouten's classification and in terms of de Rham theory.
The star operator uses a fixed orientation and excludes discussion of reflections.  Standard texts on electrodynamics such
as {\it Classical Electrodynamics} by John David Jackson [10] and {\it Field Theory} by L. D. Landau and E. M. Lifshitz
[11] list reflection symmetries of electromagnetic fields in three dimensions.  These symmetries relative to a frame of
reference are accepted by most physicists.  They are not based on a clear concept of orientation and are in disagreement
with the more rigorous formulations.  We list other books [12--14] dealing with similar arguments.

	The present note starts with definition of orientation of vector spaces and their subspaces.  This followed by
definitions of multivectors and differential forms in affine spaces.  Integral and differential relations of
electrodynamics are then stated.  These space-time formulations are translated in reference frame dependent formulations
with with space and time separated.  Eventually, a formulation in terms of traditional vector analysis is reached.  At this
point transformation properties of electromagnetic fields in three dimensions derived from the space-time parities of
electromagnetic fields are compared with transformation properties assumed by Jackson and Landau and Lifshitz  and
differences are discovered.

	The second part of the paper presents an attempt to establish a rigorous space-time geometric basis for transformation
properties accepted by physicists.  The required space-time parities are defined in terms of non standard orientations
related to the Minkowski geometry of space-time.  A set of four orientations is introduced in correspondence to the four
connected components of the Lorentz group.  A choice of parities results in transformation properties assumed by
physicists.  This result holds in all reference frames.

		\Title{A. Electrodynamics with standard orientations.}

		\sect{Orientation of vector spaces.}
	Let $V$ be a vector space of dimension $m \neq 0$.  We denote by $\sF(V)$ the space of linear isomorphisms from $V$ to
$\R^m$ called {\it frames}.  Let $\sG(V)$ be the group of linear automorphisms of $V$.  There is a natural group action
		$$\sG(V) \times \sF(V) \rightarrow \sF(V) \colon (\zr,\zx) \mapsto \zx \circ \zr^{-1}
																										\tag \label{For1}$$
	and $\sF(V)$ is a homogeneous space with respect to this action.

	The sets
		$$\sC^E(V) = \left\{\zr \in \sG(V);\; \det(\zr) > 0 \right\}
																										\tag \label{For2}$$
	and
		$$\sC^P(V) = \left\{\zr \in \sG(V);\; \det(\zr) < 0 \right\}
																										\tag \label{For3}$$
	are the two connected components of the group $\sG(V)$.  The set $\sG^E(V) = \sC^E(V)$ is the component of the unit
element.  It is a normal subgroup.

	The set of {\it orientations} $\sO(V) = \sF(V) \big/ \sG^E(V)$ has two elements.  This set is a homogeneous space for
the quotient group $\sH(V) = \sG(V) \big/ \sG^E(V)$.  The sets $\sC^E(V)$ and $\sC^P(V)$ are the elements of the quotient
group.  Symbols $E$ and $P$ will be used to denote these elements.  The structure of the group $\sH(V)$ is simple.  The
element $E = \sC^E(V)$ is the unit and the element $P = \sC^P(V)$ is an involution.

	There is an ordered base $(e_1,e_2,\ldots ,e_m)$ of $V$ associated with each frame $\zx$.  If
		$$\zx(v) = \pmatrix\mathstrut v^1 \\ \vdots \\ \mathstrut v^m \endpmatrix,
																										\tag \label{For4}$$
	then $v = e_\zk v^\zk$.  For each $\zr \in \sG(V)$ the base $(\zr(e_1),\zr(e_2),\ldots ,\zr(e_m))$ is associated with the frame
$\zx \circ \zr^{-1}$ if $(e_1,e_2,\ldots ,e_m)$ is the base associated with $\zx$.

	The set of frames $\sF(V)$ and the set of orientations $\sO(V)$ of a space $V$ of dimension 0 are empty.

		\sect{Orientation of subspaces.}
	Let $W \subset V$ be a subspace of a vector space $V$.  The subspace has the set $\sO(W)$ of orientations called {\it
inner orientations} of $W$.  Orientations of the quotient space $V\big/W$ are called {\it outer orientations} of $W$.  An
outer orientation $o''$ of $W$ can be determined by specifying an inner orientation $o$ of $W$ together with an orientation
$o'$ of $V$.  Let $(e_1,\ldots,e_n)$ be the base of $W$ associated with a frame $\zx \in o$.  This base can be completed
to a base $(e'_1,\ldots,e'_m)$ of $V$ with $(e'_1,\ldots,e'_n) = (e_1,\ldots,e_n)$.  The extended base can be chosen
to be associated with a frame $\zx' \in o'$.  Let
		$$\zp \colon V \rightarrow V\big/W
																										\tag \label{For5}$$
be the canonical projection.  The sequence
		$$(e''_1,\ldots,e''_{m-n}) = (\zp(e'_{n+1}),\ldots,\zp(e'_m))
																										\tag \label{For6}$$
	is a base of $V\big/W$.  It determines an orientation $o''$ of $V\big/W$.  Hence an outer orientation of $W$.  The
outer orientation $o''$ of $W$ constructed from $o \in \sO(W)$ and $o' \in \sO(V)$ is the same as the orientation
constructed from $Po$ and $Po'$.

	The subspace $W = \{0\}$ has no inner orientations.  Its outer orientations are the orientations of $V$.  The
specification of an outer orientation as a pair of orientations can not be applied.

	In the case of the subspace $W = V$ the quotient space $V\big/W$ is of dimension 0.  The subspace $W$ has no outer
orientation defined as an orientation of $V\big/W$.

		\sect{Multicovectors.}
	A {\it $q$\,-\,covector} in a vector space $V$ is a mapping
		$$a\, \colon V^q \times \sO(V) \rightarrow \R.
																										\tag \label{For7}$$
	This mapping is $q$\,-\,linear and totally antisymmetric in its vector arguments.  A $q$\,-\,covector $a$ is said to be {\it
even}, if
		$$a(v_1, v_2, \ldots ,v_q,Po) = a(v_1, v_2, \ldots ,v_q,o).
																										\tag \label{For8}$$
	It is said to be {\it odd}, if
		$$a(v_1, v_2, \ldots ,v_q,Po) = - a(v_1, v_2, \ldots ,v_q,o).
																										\tag \label{For9}$$
	The vector space of even $q$\,-\,covectors will be denoted by $\WE^q_e V^\*$ and the space of odd $q$\,-\,covectors
will be denoted by $\WE^q_o V^\*$.  The symbol $\WE^q_p V^\*$ will be used to denote either of the two spaces when
the parity need not be specified.

	The {\it exterior product} of a $q$\,-\,covector $a$ with a $q'$\,-\,covector $a'$ is the $(q\+q')$\,-\,covector
		$$\align
	a \wedge a' &\colon V^{q+q'} \times \sO(V) \rightarrow \R \colon (v_1,\ldots,v_{q+q'},o) \\
		&\hskip6mm\mapsto {\sum_{\zs \in S(q+q')}}\dsize{\frac{\text{sgn}(\zs)}{q!q'!}}\,
a(v_{\zs(1)},\ldots,v_{\zs(q)},o)\, a'(v_{\zs(q+1)},\ldots,v_{\zs(q+q')},o),
																										\tag \label{For10}\endalign$$
	If both multicovectors $a$ and $a'$ are even or both are odd, the product $a \wedge a'$ is even.  In other cases the
product is odd.

	The exterior product is commutative in the graded sense.  If $a$ is a $q$\,-\,covector and $a'$ is a $q'$\,-\,covector, then
		$$a' \wedge a = (-1)^{qq'} a \wedge a'.
																										\tag \label{For11}$$

	The exterior product is associative.  The relation
		$$a \wedge (a' \wedge a'') = (a \wedge a') \wedge a''
																										\tag \label{For12}$$
	holds for any three multicovectors $a$, $a'$ and $a''$.

	If the dimension of $V$ is $m$, then
		$$\dim(\wedge_p^q V^\*) = \binom m q.
																										\tag \label{For13}$$

	The group $\sG(V)$ has natural representations in the spaces $\wedge_p^q V^\*$.  A linear automorphism $\zr \in \sG(V)$
applied to a $q$\,-\,covector $a$ produces the $q$\,-\,covector
		$$\align
		(\zr^{-1})^{\*}a\, &\colon \times^q V \times \sO(V) \rightarrow \R \\
			&\colon (v_1, v_2, \ldots ,v_q,o) \mapsto a(\zr^{-1}(v_1), \zr^{-1}(v_2), \ldots ,
\zr^{-1}(v_q),[\zr](o)),
																										\tag \label{For14}\endalign$$
	where $[\zr]$ is the class of $\zr$ in the quotient group $\sH(V) = \sG(V) \big/ \sG^E(V)$.  For a covector $a \in
\wedge_p^q V^{\*}$ we have
		$$(\zr^{-1})^{\*}a(v_1, v_2, \ldots ,v_q,o) = \idx_p a(\zr^{-1}(v_1), \zr^{-1}(v_2), \ldots, \zr^{-1}(v_q),o).
																										\tag \label{For15}$$
	The values of the {\it index} $\idx_p$ are listed in the following table.

\vskip4mm

\setbox1=\hbox{\vrule height-2.5pt width50mm depth3pt}

\setbox3=\hbox{\vrule height-2.5pt width.7pt depth19.5mm}

\setbox11=\hbox{$\hphantom{-}1$}
\setbox12=\hbox{$-1$}

\setbox40=\hbox{$\zr \in P$}
\setbox41=\hbox{$\zr \in E$}

\setbox60=\hbox{$\idx_e(\zr)$}
\setbox61=\hbox{$\idx_o(\zr)$}

\setbox0=\hbox{
\hskip076mm\lower-01mm\copy41\hskip-\wd41\hskip-076mm
\hskip093mm\lower-01mm\copy40\hskip-\wd40\hskip-093mm

\hskip055mm \copy01 \hskip-105mm
\hskip70mm\lower-05mm\copy03\hskip-\wd03\hskip-70mm

\hskip057mm\lower06mm\copy60\hskip-\wd60\hskip-057mm
\hskip077mm\lower06mm\copy11\hskip-\wd11\hskip-077mm
\hskip094mm\lower06mm\copy11\hskip-\wd11\hskip-094mm

\hskip057mm\lower12mm\copy61\hskip-\wd61\hskip-057mm
\hskip077mm\lower12mm\copy11\hskip-\wd11\hskip-077mm
\hskip094mm\lower12mm\copy12\hskip-\wd12\hskip-094mm

}\box0
\vskip4mm

	Even $m$\,-\,covectors follow the transformation rule
		$$(\zr^{-1})^{\*}a = \det(\zr)^{-1}a
																										\tag \label{For16}$$
	and
		$$(\zr^{-1})^{\*}a = |\det(\zr)|^{-1}a
																										\tag \label{For17}$$
	is the transformation rule of odd $m$-covectors.  These rules are derived in the section on multivectors.

		\sect{Multivectors.}
	We denote by $\sK(V^q \times \sO(V))$ the vector space of formal linear combinations of sequences
		$$(v_1, v_2, \ldots ,v_q,o) \in V^q \times \sO(V).
																										\tag \label{For18}$$
	In the space $\sK(V^q \times \sO(V))$ we introduce subspaces
		$$\align
		\sA^{\,p}_q(V) &= \left\{\tsize{\sum_{\,i=1}^{\,n}}\, \zl_i(v^i_1, v^i_2, \ldots ,v^i_q,o^i) \in \sK(V^q \times
\sO(V)) ;\; \tsize{\sum_{\,i=1}^{\,n}}\, \zl_i a(v^i_1, v^i_2, \ldots ,v^i_q,o^i) = 0 \right. \\
	&\hskip50mm \left. \vphantom{\tsize{\sum_{\,i=1}}}\text{ for each }\;\; a \in \WE^q_p V^\* \right\}.
																										\tag \label{For19}\endalign$$
	Subsequently we define quotient spaces
		$$\WE^q_p V = \sK(V^q \times \sO(V)) \big/ \sA^{\,p}_q(V).
																										\tag \label{For20}$$
	Elements of spaces $\WE^q_e V$ and $\WE^q_o V$ are called {\it even $q$\,-\,vectors} and {\it odd $q$\,-\,vectors}
respectively.  We will denote by
		$$\left[\tsize{\sum_{\,i=1}^{\,n}}\, \zl_i(v^i_1, v^i_2, \ldots ,v^i_q,o^i)\right]_p
																										\tag \label{For21}$$
	the equivalence class of the combination
		$$\tsize{\sum_{\,i=1}^{\,n}}\, \zl_i(v^i_1, v^i_2, \ldots ,v^i_q,o^i)
																										\tag \label{For22}$$
	in $\WE^q_p V$.  A multivector is said to be {\it simple} if it is represented by a single element of the space $V^q
\times \sO(V)$ interpreted as a subspace of $\sK(V^q \times \sO(V))$.

	Evaluation of $q$\,-\,covectors on sequences $(v_1, v_2, \ldots ,v_q,o) \in V^q \times \sO(V)$ extends to linear
combinations and their equivalence classes.  If $w$ is a $q$\,-\,vector represented by the linear combination
		$$\tsize{\sum_{\,i=1}^{\,n}}\, \zl_i(v^i_1, v^i_2, \ldots ,v^i_q,o^i)
																										\tag \label{For23}$$
	and $a$ is a $q$\,-\,covector of the same parity as $w$, then
		$$\langle a, w\rangle = \tsize{\sum_{\,i=1}^{\,n}}\, \zl_i a(v^i_1, v^i_2, \ldots ,v^i_q,o^i)
																										\tag \label{For24}$$
	is the evaluation of $a$ on $w$.  We have constructed pairings
		$$\langle \;,\,\rangle \colon \WE^q_p V^\* \times \WE^q_p V \rightarrow \R.
																										\tag \label{For25}$$

	The dimension of the spaces $\WE^q_p V$ is
		$$\dim(\wedge_p^q V) = \binom m q.
																										\tag \label{For26}$$

	The {\it exterior product} of multivectors
		$$w_1 = \left[\tsize{\sum_{\,i=1}^{\,{n_1}}}\, \zl^1_i(v^{1\,i}_1, v^{1\,i}_2, \ldots,
v^{1\,i}_{q_1},o)\right]_{p_1}
																										\tag \label{For27}$$
	and
		$$w_2 = \left[\tsize{\sum_{\,j=1}^{\,{n_2}}}\, \zl^2_j(v^{2\,j}_1, v^{2\,j}_2, \ldots,
v^{2\,j}_{q_2},o)\right]_{p_2}
																										\tag \label{For28}$$
	is the multivector
		$$w_1 \wedge w_2 = \left[\tsize{\sum_{\,i=1}^{\,{n_1}}\sum_{\,j=1}^{\,{n_2}}}\, \zl^1_i\zl^2_j(v^{1\,i}_1,
v^{1\,i}_2, \ldots ,v^{1\,i}_{q_1},v^{2\,i}_1, v^{2\,i}_2, \ldots ,v^{2\,i}_{q_2},o)\right]_p.
																										\tag \label{For29}$$
	Note that we are using a fixed orientation $o$ in all representatives.  The parity of the exterior product is odd if
the parity of one of the factors is odd.  It is even otherwise.

	The image of the multivector
		$$w = \left[\tsize{\sum_{\,i=1}^{\,n}}\, \zl_i(v^i_1, v^i_2, \ldots ,v^i_q,o^i)\right]_p
																										\tag \label{For30}$$
	by an automorphism $\zr \in \sG(V)$ is the multivector
		$$\align
			\zr_\* w &= \left[\tsize{\sum_{\,i=1}^{\,n}}\, \zl_i(\zr(v^i_1), \zr(v^i_2), \ldots
,\zr(v^i_q),[\zr](o^i))\right]_p \\
				&= \idx_p\left[\tsize{\sum_{\,i=1}^{\,n}}\, \zl_i(\zr(v^i_1), \zr(v^i_2), \ldots
,\zr(v^i_q),o^i)\right]_p \\
																										\tag \label{For31}\endalign$$
	where $[\zr]$ is again the class of $\zr$ in the quotient group $\sH(V) = \sG(V) \big/ \sG^E(V)$ and $\idx_p$ is the
index defined in the preceding section.

	The equality
		$$\zr_{\*}(w_1 \wedge w_2) = \zr_{\*}(w_1) \wedge \zr_{\*}(w_2)
																										\tag \label{For32}$$
	holds for multivectors $w_1$ and $w_2$ of any parity.

	An even 0-vector or a {\it scalar} is transformed identically:
		$$\zr_{\*}w = w.
																										\tag \label{For33}$$
	Odd 0-vectors or {\it pseudoscalars} follow the transformation rule
		$$\zr_{\*}w = \frac{\det(\zr)}{|\det(\zr)|}\,w.
																										\tag \label{For34}$$

	The spaces $\WE^m_p V$ are of dimension 1.  It follows that an automorphism $\zr$ multiplies an $m$\,-\,vector by a
number.  In the case of an even $m$\,-\,vector the number is the determinant of $\zr$.  The equality
		$$\zr_{\*}w = \det(\zr)w
																										\tag \label{For35}$$
	is an abstract definition of the determinant valid for endomorphisms as well as automorphisms.  This definition
translates in the elementary definition if a base in $V$ is chosen.

	An odd $m$\,-\,vector $w$ can be presented as the exterior product $u \wedge v$ of a pseudoscalar $u$ and an even
$m$\,-\,vector $v$.  This implies the transformation rule
		$$\zr_{\*}w = |\det(\zr)|w
																										\tag \label{For36}$$
	for odd $m$\,-\,vectors.

	It follows from the definition \Ref{For24} that the pairings \Ref{For25} are invariant in the sense that
		$$\langle (\zr^{-1})^{\*}a, \zr_\* w\rangle = \langle a, w\rangle.
																										\tag \label{For37}$$
	Let $w$ be an arbitrary $m$\,-\,vector.  The transformation rules \Ref{For15} and \Ref{For16} follow from
		$$\det(\zr)\langle (\zr^{-1})^{\*}a, w\rangle = \langle (\zr^{-1})^{\*}a, \zr_\* w\rangle =  \langle a, w\rangle
																										\tag \label{For38}$$
	and
		$$|\det(\zr)|\langle (\zr^{-1})^{\*}a, w\rangle = \langle (\zr^{-1})^{\*}a, \zr_\* w\rangle = \langle a, w\rangle
																										\tag \label{For39}$$
	respectively.

		\sect{The Weyl isomorphism.}
	The space $\wedge_o^m V^\*$ is one-dimensional.  This makes it possible to define the tensor product $\wedge_e^qV
\otimes \wedge_o^m V^\*$ as the set of equivalence classes of pairs $(w,e) \in \wedge_e^qV \times \wedge_o^m V^\*$.  Pairs
$(w,e)$ and $(w',e')$ are equivalent if there is a number $\zl$ such that $w' = \zl w$ and $e = \zl e'$ or $w = \zl w'$ and
$e' = \zl e$.  The equivalence class of a pair $(w,e)$ will be denoted by $w \otimes e$.  A tensor $\overline a \in
\wedge_e^qV \otimes \wedge_o^m V^\*$ will always be presented as a product $w \otimes e$.  The set $\wedge_e^qV \otimes
\wedge_o^m V^\*$ is a vector space with operations
        $$\cdot\, \colon \R \times \left(\wedge_e^qV \otimes \wedge_o^m V^\*\right) \rightarrow \wedge_e^qV \otimes
\wedge_o^m V^\* \colon (\zl,w \otimes e) \mapsto \zl w \otimes e
																										\tag \label{For40}$$
	and
        $$+ \colon \left(\wedge_e^qV \otimes \wedge_o^m V^\*\right) \times \left(\wedge_e^qV \otimes \wedge_o^m V^\*\right)
\rightarrow \wedge_e^qV \otimes \wedge_o^m V^\* \colon (w_1 \otimes e,w_2 \otimes e) \mapsto (w_1 + w_2) \otimes e.
																										\tag \label{For41}$$
	Note that in the definition of the sum the representatives $(w_1,e)$ and $(w_2,e)$ of the elements being added have
been chosen to have the same component $e \in \wedge_o^m V^\*$.  This is always possible since the space $\wedge_o^m V^\*$
is one-dimensional.

	The tensor
		$$\zr_\*(w \otimes e) = \zr_\*w \otimes (\zr^{-1})^{\*}e = |\det(\zr)|^{-1}(\zr_\*w \otimes e)
																										\tag \label{For42}$$
	is obtained by applying	an automorphism $\zr \in \sG(V)$ to a tensor $w \otimes e \in \wedge_e^qV \otimes \wedge_o^m
V^\*$.

	We show that the linear mapping
		$$\We_q \colon \wedge_e^qV \otimes \wedge_o^m V^\* \rightarrow \wedge_o^{m-q}V^\* \colon w \otimes e \mapsto w \lpr e
																										\tag \label{For43}$$
	is an isomorphism.  If $\We_q(w \otimes e) = 0$, then either $e = 0$ hence,  $w \otimes e = 0$ or $e \neq 0$.  In the
latter case we have
		$$\langle e, w \wedge v\rangle = \langle w \lpr e, v\rangle = \langle \We_q(w \otimes e), v\rangle = 0
																										\tag \label{For44}$$
	for an arbitrary multivector $v \in \wedge_o^{m-q}V$.  Since $a \neq 0$ and $\dim(\wedge_o^m V^\*) = 1$ it follows that
$w \wedge v = 0$.  Hence $w =0$ and $w \otimes e = 0$.  We have shown that $\We_q$ is injective.  The spaces $\wedge_e^qV
\otimes \wedge_o^m V^\*$ and $\wedge_o^{m-q}V^\*$ are of the same dimension since
		$$\dim\left(\wedge_e^qV \otimes \wedge_o^m V^\*\right) = \binom m q
																										\tag \label{For45}$$
	and
		$$\dim\left(\wedge_o^{m-q}V^\*\right) = \binom m {m-q} = \binom m q.
																										\tag \label{For46}$$
	It follows that $\We_q$ is bijective.  The mapping $\We_q$ is called the {\it Weyl isomorphism}.

	For an automorphism $\zr$ and an arbitrary multivector $v \in \wedge_o^{m-q}V$ we have
		$$\align
			\langle \We_q(\zr_\*(w \otimes e)),v \rangle &= \langle \We_q(\zr_\*w \otimes (\zr^{-1})^{\*}e)),v \rangle \\
					&= \langle (\zr^{-1})^{\*}e, \zr_\*w \wedge v\rangle \\
					&= \langle e, w \wedge (\zr^{-1})_\*v\rangle \\
					&= \langle w \lpr e, (\zr^{-1})_\*v\rangle \\
					&= \langle (\zr^{-1})^{\*}(w \lpr e), v\rangle \\
					&= \langle (\zr^{-1})^{\*}\We_q(w \otimes e),v \rangle.
																										\tag \label{For47}\endalign$$
	It follows that the Weyl isomorphism is invariant in the sense that
		$$\We_q(\zr_\*(w \otimes e)) = (\zr^{-1})^{\*}\We_q(w \otimes e).
																										\tag \label{For48}$$

		\sect{Differential forms in affine spaces.}
	Let $M$ be an affine space modelled on a vector space $V$.  A {\it differential $q$\,-\,form} on $M$ is a differentiable
function
		$$A \colon M \times V^q \times \sO(V) \rightarrow \R
																										\tag \label{For49}$$
	depending on a point, $q$ vectors and an orientation.  It is $q$\,-\,linear and totally antisymmetric in its vector
arguments.  A differential form $A$ is said to be {\it even}, if
		$$A(x,v_1, v_2, \ldots ,v_q,Po) = A(x,v_1, v_2, \ldots ,v_q,o).
																										\tag \label{For50}$$
	It is said to be {\it odd}, if
		$$A(x,v_1, v_2, \ldots ,v_q,Po) = - A(x,v_1, v_2, \ldots ,v_q,o).
																										\tag \label{For51}$$

	The vector space of even differential $q$\,-\,forms will be denoted by $\zF^q_e(M)$ and space of odd differential
$q$\,-\,forms will be denoted by $\zF^q_o(M)$.  We will use the symbol $\zF^q_p(M)$ to denote either $\zF^q_e(M)$ or
$\zF^q_o(M)$.

	The {\it exterior product} of a $q$\,-\,form $A$ with a $q'$\,-\,form $A'$ is the $(q\+q')$\,-\,form
		$$\align
	A \wedge A' &\colon M \times V^{q+q'} \times \sO(V) \rightarrow \R \colon (x,v_1,\ldots,v_{q+q'},o) \\
		&\hskip6mm\mapsto  {\sum_{\zs \in S(q+q')}} \dsize{\frac{\sgn(\zs)}{q!q'!}}
A(x,v_{\zs(1)},\ldots,v_{\zs(q)},o) A'(x,v_{\zs(q+1)},\ldots,v_{\zs(q+q')},o),
																										\tag \label{For52}\endalign$$
	where $S(q\+q')$ denotes the group of permutations of the set $\{1,\ldots,q\+q'\}$ of integers.  If both forms $A$ and
$A'$ are even or both are odd, the product $A \wedge A'$ is even.  In other cases the product is odd.

	The exterior product is commutative in the graded sense.  If $A$ is a $q$\,-\,form and $A'$ is a $q'$\,-\,form, then
		$$A' \wedge A = (-1)^{qq'} A \wedge A'.
																										\tag \label{For53}$$
	The exterior product is associative.  The relation
		$$A \wedge (A' \wedge A'') = (A \wedge A') \wedge A''
																										\tag \label{For54}$$
	holds for any three forms $A$, $A'$ and $A''$.

	The {\it exterior differential} of a $q$\,-\,form $A$ is the $(q\+1)$\,-\,form
		$$\align
	\rd A \colon M \times &V^{q+1} \times \sO(V) \rightarrow \R \colon (x,v_1,v_2,\ldots,v_{q+1},o) \\
		&\mapsto - {\sum_{i=1}^{q+1}(-1)^i} \dsize{\frac{\rd}{\rd s}} A(x +
sv_i,v_1,v_2,\ldots,\widehat{v_i},\ldots,v_{q+1},o)\big|_{s=0}.
																										\tag \label{For55}\endalign$$
	The parity of the differential $\rd A$ is the same as the parity of the original form $A$.  The operator $\rd$ is a
{\it differential} in the sense that $\rd\rd A = 0$ for each form $A$.

	A form $A$ is said to be {\it closed} if $\rd A = 0$.  It is said to be {\it exact} if there is a form $B$ such that $a
= \rd B$.  The {\it Poincar\'e lemma} states that in an affine space each closed form is exact.

	Given a vector field $X \colon M \rightarrow V$ and a form $A \in \zF^q_p(M)$ we construct forms
		$$\xi_X A \colon M \times V^{q-1} \times \sO(V) \rightarrow \R \colon (x,v_1,\ldots,v_{q-1},o) \mapsto
A(x,X(x)v_1,\ldots,v_{q-1},o)
																										\tag \label{For56}$$
	and
		$$\rd_X A = \xi_X\rd A + \rd\xi_X A.
																										\tag \label{For57}$$
	The operator $\rd_X$ is the {\it Lie derivative} and can be defined in terms of the one parameter group of
diffeomorphisms generated by $X$.

	An $q$\,-\,form $A$ can be interpreted as a mapping
		$$\wA \colon M \rightarrow \WE^q_p V^\*.
																										\tag \label{For58}$$
	The relation between the form $A$ and the mapping $\wA$ is expressed by
		$$\wA(x)(v_1,\ldots,v_q,o) = A(x,v_1,\ldots,v_q,o)
																										\tag \label{For59}$$
	The exterior product and the exterior differential are extended to this alternative interpretation of forms.  Notation
		$$\wA \wedge \wA' = \widetilde{A \wedge A'}
																										\tag \label{For60}$$
	and
		$$\rd \wA = \widetilde{\rd A}
																										\tag \label{For61}$$
	will be used.

		\sect{The metric volume form.}
	Let $g \colon V \rightarrow V^\*$ be a metric tensor of signature $(m\- r,r)$ in the model space $V$ of an affine space
$M$ of dimension $m$.  We define an odd m-form
		$$\sqrt{|g|} \,\colon M \times V^m \times \sO(V) \rightarrow \R
																										\tag \label{For62}$$
	by the formula
		$$\sqrt{|g|}(x,v_1,v_2,\ldots ,v_m,o) \mapsto \pm \sqrt{|\det(\langle g(v_\zk), v_\zl\rangle)|}.
																										\tag \label{For63}$$
	If vectors $(v_1,v_2,\ldots ,v_m)$ are dependent, then $\det(\langle g(v_\zk), v_\zl\rangle) = 0$.  If the vectors are
independent, then they determine an orientation $o' \in \sO(V)$.  The sign $+$ in the formula is chosen if the orientations
$o$ and $o'$ agree.  Otherwise the sign $-$ is chosen.  It follows from elementary properties of determinants that the
formula defines an odd $m$\,-\,form.

	There are two important examples of metric spaces.  One is the Minkowski affine space-time of special relativity.  The
dimension of this space is 4 and the signature of the metric tensor $g$ is $(1,3)$.  The second example is the
affine physical space of dimension 3 with a Euclidean metric tensor $h$.

		\sect{Integration of forms, chains}
	A {\it cell} of dimension $q$ or a $q$\,-\,{\it cell }in $M$ is a pair $(\zq,o)$, where $\zq$ is a differentiable
mapping $\zq \colon \R^q \rightarrow M$ and $o$ is an orientation of $V$.  The {\it integral} of a $q$\,-\,form $A$ on a
cell $\zq$ is the Riemann integral
		$$\int_{(\zq,o)} A = \int_0^1\cdots\int_0^1
A(\zq(s_1,\ldots,s_q),\xD_1\zq(s_1,\ldots,s_q),\ldots,\xD_q\zq(s_1,\ldots,s_q),o) \rd s_1 \cdots \rd s_q.
																										\tag \label{For64}$$
	For each $q$ we introduce the space $C_q$ of formal linear combinations of $q$\,-\,cells.  The formal linear
combinations turn into real linear combinations if cells are recognized as elements of $C_q$.  Integration of forms is
extended to linear combinations by linearity.  The integral of a form $A$ on a combination
		$$\bc = \tsize{\sum_{i=1}^n} \zl_i\zq^i
																										\tag \label{For65}$$
	is the combination
		$$\int_\bc A = \tsize{\sum_{i=1}^n}\, \zl_i\int_{\zq^i}A
																										\tag \label{For66}$$
	of integrals.  Subspaces $X^p_q \subset C_q$ are defined as the sets
		$$X^p_q = \left\{\bc \in C_q ;\; \int_\bc A = 0\;\;\text{ for each }\;\;A \in \zF^q_p(M) \right\}.
																										\tag \label{For67}$$
	Elements of the quotient spaces $\zX^p_q(M) = C_q\big/X^p_q$ are called {\it even chains} or {\it odd chains} of
dimension $q$.  A chain is said to be simple if it has a single cell as one of its representatives.  Integrals of
$q$\,-\,forms on $q$\,-\,chains are well defined.  The integral of a $q$\,-\,form $A$ on the class $\bc$ of $c \in C_q$ is
the integral of $A$ on $c$.

	The {\it boundary operator} $\partial$ assigns to a chain $\bc \in \zX^{\,p}_q(M)$ its {\it boundary} $\partial\bc \in
\zX^{\,p}_{q-1}(M)$.  The boundary of a simple chain represented by a $q$\,-\,cell $(\zq,o)$ is the chain represented by
the combination
		$$\tsize{\sum_{i=1}^q} (-1)^{i-1} ((\zq^{(i,1)},o) - (\zq^{(i,0)},o))
																										\tag \label{For68}$$
	where the $(q\-1)$\,-\,cells $(\zq^{(i,1)},o)$ and $(\zq^{(i,0)},o)$ defined by
		$$\zq^{(i,1)} \colon \R^{q-1} \rightarrow M \colon (s_1,\ldots,\widehat{s_i},\ldots,s_q) \mapsto
\zq(s_1,\ldots,s_{i-1},1,s_{i+1}\ldots,s_q)
																										\tag \label{For69}$$
	and
		$$\zq^{(i,0)} \colon \R^{q-1} \rightarrow M \colon (s_1,\ldots,\widehat{s_i},\ldots,s_q) \mapsto
\zq(s_1,\ldots,s_{i-1},0,s_{i+1}\ldots,s_q)
																										\tag \label{For70}$$
	represent the {\it faces} of the simple chain.  This construction of the boundary is extended to generic chains by
linearity.  The boundary of a boundary is the zero chain.

	Stokes theorem states that the relation
		$$\int_\bc \rd A = \int_{\partial\bc} A
																										\tag \label{For71}$$
	holds for a chain $\bc \in \zX^{\,p}_q(M)$ and a form $A \in \zF^{q-1}_p(M)$.  This means that the boundary operator is
dual to the exterior differential.

		\sect{Currents.}
	An {\it even} or {\it odd de Rham current} of dimension $q$ is a linear function
		$$\iC \colon \zF^q_p(M) \rightarrow \R \colon A \mapsto \int_\iC A.
																										\tag \label{For72}$$
	The spaces of forms are given certain topologies and continuity is required.  Chains will be treated as currents.  They
form a dense subspace in the space of currents.  We will consider only very simple examples of currents other than chains.
Topological considerations are of little importance for these examples.  The boundary of a current is defined by assuming
that Stokes theorem holds for currents.  Thus if $\iC$ is a current of dimension $q$, then the boundary of $\iC$ is the
mapping
		$$\partial\iC \colon \zF_p^{q-1}(M) \rightarrow \R \colon A \mapsto \int_{\partial\iC} A = \int_\iC \rd A.
																										\tag \label{For73}$$

	Let $X \colon M\rightarrow \sT M$ be a vector field and let $\bc$ be a chain of dimension $q$.  The mapping
		$$X \wedge \bc \colon \zF_p^{q+1}(M) \rightarrow \R \colon A \mapsto \int_{X \wedge \bc} A = \int_\bc \xi_X A
																										\tag \label{For74}$$
	is a current of dimension $q \+ 1$ of the same parity as $\bc$.  For the boundary $\partial(X \wedge \bc)$ we have

		$$\int_{\partial(X \wedge \bc)} A = \int_{X \wedge \bc} \rd A = \int_\bc \xi_X\rd A = \int_\bc(\rd_X A - \rd\,\xi_X
A) = \int_\bc\rd_X A - \int_{\partial\bc}\,\xi_X A = \int_\bc\rd_X A - \int_{X \wedge \partial\bc} A.
																										\tag \label{For75}$$

		\sect{Vector densities and Weyl duality.}
	A differentiable mapping
		$$\ovA \colon M \rightarrow \wedge_e^qV \otimes \wedge_o^m V^\*
																										\tag \label{For76}$$
	is called a {\it $q$\,-\,vector density}.  A 0-vector density is called a {\it scalar density}.  We identify the tensor
product $\wedge_e^0V \otimes \wedge_o^m V^\*$ with $\wedge_o^m V^\*$ by taking advantage of the fact that each element of
$\wedge_e^0V \otimes \wedge_o^m V^\*$ has a representative of the form $(1,e) \in \R \times \wedge_o^m V^\*$.  The tensor
$1 \otimes e$ is identified with $e$.  It follows that a scalar density and an odd $m$\,-\,form are essentially the same
object.

	The Weyl isomorphism is used to associate with an $(m\- q)$\,-\,form $A$ a {\it $q$\,-\,vector density}
		$$\ovA \colon M \rightarrow \wedge_e^qV \otimes \wedge_o^m V^\*
																										\tag \label{For77}$$
	defined by
		$$\ovA = \We^{-1}_q \circ \wA
																										\tag \label{For78}$$
	and called the {\it Weyl dual} of $A$.  If $\ovA$ is the Weyl dual of an $(m\- q)$\,-\,form $A$, then the Weyl dual of the
$(m\- q\+ 1)$\,-\,form $\rd A$ is a $(q\- 1)$\,-\,vector density denoted by $\Div\ovA$.  Thus,
		$$\Div\ovA = \We^{-1}_{q-1} \circ \rd(\We_q \circ \ovA).
																										\tag \label{For79}$$

		\sect{Transformation properties of forms and densities.}
	Let $M$ be an affine space modelled on a vector space $V$.  The {\it derivative} of a differentiable mapping $\zf \colon
M \rightarrow M$ is the mapping
		$$\xD\zf \colon M \times V \rightarrow V \colon (x,v) \mapsto \frac{\rd}{\rd s} \zf(x + sv)\big|_{s=0}.
																										\tag \label{For80}$$
	linear in its vector argument.
	A {\it diffeomorphism} of an affine space $M$ is an invertible differentiable mapping $\zf \colon M \rightarrow M$ with
a differentiable inverse.  The mapping
		$$\xD\zf(x,\cdot) \colon V \rightarrow V \colon v \mapsto \xD\zf(x,v)
																										\tag \label{For81}$$
	is a linear automorphism at each $x \in M$.  It is an element of the group $\sG(V)$.  The derivative of a
diffeomorphism $\zf$ and the derivative of its inverse are related by
		$$\xD\zf^{-1}(\zf(x),\xD\zf(x,v)) = v.
																										\tag \label{For82}$$

	A diffeomorphism $\zf$ applied to a $q$\,-\,form $A$ results in the {\it pull back}
		$$\align
		(\zf^{-1})^{\*}A\, &\colon M \times (\times^q V) \times \sO(V) \rightarrow \R \\
			&\colon (x,v_1, \ldots ,v_q,o) \mapsto A(\zf^{-1}(x),\xD\zf^{-1}(x,v_1), \ldots ,
\xD\zf^{-1}(x,v_q),[\xD\zf(x,\cdot)](o)),
																										\tag \label{For83}\endalign$$
	where $[\xD\zf(x,\cdot)]$ is the class of $\xD\zf(x,\cdot)$ in the quotient group $\sH(V) = \sG(V) \big/ \sG^E(V)$.  This class is
constant on $M$ due to continuity.  The pull back of the differential of a form is the differential of the pull back:
		$$\zf^{\*}\rd A = \rd\zf^{\*}A.
																										\tag \label{For84}$$
	The pull back of the exterior product of forms is the exterior product of pull backs:
		$$\zf^{\*}(A_1 \wedge A_2) = \zf^{\*}A_1 \wedge \zf^{\*}A_2.
																										\tag \label{For85}$$

	The value of an even form $A$ does not depend on the orientation.  Hence,
		$$\align
			A(\zf^{-1}(x),\xD\zf^{-1}(x,v_1), & \ldots , \xD\zf^{-1}(x,v_q),[\xD\zf(x,\cdot)](o))\\
					&= A(\zf^{-1}(x),\xD\zf^{-1}(x,v_1), \ldots , \xD\zf^{-1}(x,v_q),o).
																										\tag \label{For86}\endalign$$
	For an odd form we have
		$$\align
			A(\zf^{-1}(x),\xD\zf^{-1}(x,v_1), & \ldots , \xD\zf^{-1}(x,v_q),[\xD\zf(x,\cdot)](o))\\
					&= \frac{\zD}{|\zD|}A(\zf^{-1}(x),\xD\zf^{-1}(x,v_1), \ldots , \xD\zf^{-1}(x,v_q),o),
																										\tag \label{For87}\endalign$$
	with
		$$\zD = \det(\xD\zf(x,\cdot)).
																										\tag \label{For88}$$
	Based on the transformation properties an even $q$\,-\,form should be considered a {\it covariant $q$\,-\,vector} and
an odd $q$\,-\,form should be classified as a {\it covariant $W$-$\,q$\,-\,vector}.  This terminology is taken from
Schouten [1].

	The pull back $(\zf^{-1})^{\*}A$ of a $q$\,-\,form $A$ represented by the mapping $\wA$ is represented by
		$$(\zf^{-1})^\*\wA \colon M \rightarrow \wedge_p^qV^{\*} \colon x \mapsto \xD\zf^{-1}(x,\cdot)^\*\wA(\zf^{-1}(x)).
																										\tag \label{For89}$$

	The pull back of an odd $m$\,-\,form or a {\it scalar density} $E$ is represented by
		$$(\zf^{-1})^\*\wE \colon M \rightarrow \wedge_o^mV^{\*} \colon x \mapsto
|\det(\xD\zf^{-1}(x,\cdot))|^{-1}\wE(\zf^{-1}(x)).
																										\tag \label{For90}$$
	A $q$\,-\,vector field
		$$W \colon M \rightarrow \wedge_e^qV
																										\tag \label{For91}$$
	transforms according to the rule
		$$\zf_\*W \colon M \rightarrow \wedge_e^qV \colon x \mapsto \xD\zf(x,\cdot)_\*W(\zf^{-1}(x)).
																										\tag \label{For92}$$
	A $q$\,-\,vector density $\wA$ can be composed as a tensor product
		$$W \otimes \wE \colon M \rightarrow \wedge_e^qV \otimes \wedge_o^m V^\* \colon x \mapsto W(x) \otimes \wE(x)
																										\tag \label{For93}$$
	of a $q$\,-\,vector field $W$ and a scalar density $\wE$.  This tensor product follows the transformation rule
		$$\zf_\*(W \otimes \wE) = \zf_\*W \otimes (\zf^{-1})^{\*}\wE
																										\tag \label{For94}$$
	or
		$$\zf_\*(W \otimes \wE) \colon M \rightarrow \wedge_e^qV \otimes \wedge_o^m V^\* \colon x \mapsto
|\det(\xD\zf^{-1}(x,\cdot))|^{-1}\xD\zf(x,\cdot)_\*W(\zf^{-1}(x)) \otimes \wE(\zf^{-1}(x)).
																										\tag \label{For95}$$
	In Schouten's terminology a scalar density is a {\it covariant $W$-$\,m$\,-\,vector} or  a {\it scalar density of weight} 1.  A
$q$\,-\,vector density is a {\it contravariant $q$\,-\,vector density of weight} 1.

	The invariance property
		$$\overline{(\zf^{-1})^{\*}A} = \zf_{\*}\ovA.
																										\tag \label{For96}$$
	of Weyl duality is a direct consequence of the invariance of the Weyl isomorphism.

		\sect{Integral relations of electrodynamics.}
	Let $M$ be the affine Minkowski space-time of special relativity with the model space $V$ and the metric tensor $g
\colon V \rightarrow V^\*$ of signature $(1,3)$.

	Electromagnetic phenomena are described by the {\it electromagnetic field}
		$$F \in \zF_e^2(M),
																										\tag \label{For97}$$
	the {\it electromagnetic induction}
		$$G \in \zF_o^2(M),
																										\tag \label{For98}$$
	and the {\it four-current}
		$$J \in \zF_o^3(M).
																										\tag \label{For99}$$
	The integral equalities
		$$\int_{\partial\bc} F = 0
																										\tag \label{For100}$$
	and
		$$\int_{\partial\bc} G  = - \frac{4\zp}{c}\int_\bc J
																										\tag \label{For101}$$
	hold for each chain $\bc$ of dimension 3.  The integral equations hold for currents as well as chains.

		\sect{Maxwell's equations.}
	The integral relations \Ref{For100} and \Ref{For101} are equivalent to the differential {\it Maxwell's equations}
		$$\rd F = 0
																										\tag \label{For102}$$
	and
		$$\rd G = - \frac{4\zp}{c}J.
																										\tag \label{For103}$$

	The objects $F$, $G$ and $J$ can be interpreted as mappings
		$$\wF \colon M \rightarrow \wedge_e^2 V^\*,
																										\tag \label{For104}$$
		$$\wG \colon M \rightarrow \wedge_o^2 V^\*,
																										\tag \label{For105}$$
	and
		$$\wJ \colon M \rightarrow \wedge_o^3 V^\*
																										\tag \label{For106}$$
	satisfying Maxwell's equations
		$$\rd \wF = 0
																										\tag \label{For107}$$
	and
		$$\rd \wG = - \frac{4\zp}{c}\wJ.
																										\tag \label{For108}$$

	In addition to Maxwell's equations we have the {\it constitutive relation} between $\wF$ and $\ovG$ expressed by
		$$\ovG = \left(\WE^2_e\, g^{-1} \circ \wF\right) \otimes \sqrt{|g|}.
																										\tag \label{For109}$$
	This is the constitutive relation for electromagnetic fields in vacuum.  We are using the mapping
		$$\WE^2_e\, g^{-1} \colon \WE^2_e V^\* \rightarrow \WE^2_e V
																										\tag \label{For110}$$
	characterized by the equality
		$$\WE^2_e\, g^{-1}(a_1 \wedge a_2) = g^{-1}(a_1) \wedge g^{-1}(a_2)
																										\tag \label{For111}$$
	for simple even bicovectors.

	If the electromagnetic field $F$ is defined on all the affine space $M$, then the equation \Ref{For107} implies the
existence of a {\it potential}
		$$A \in \zF_e^1(M)
																										\tag \label{For112}$$
	such that
		$$F = \rd A.
																										\tag \label{For113}$$
	This is a consequence of the Poincar\'e lemma.

		\sect{Invariance of Maxwell's equations.}
	If objects $F$, $G$, and $J$ satisfy Maxwell's equations \Ref{For102} and \Ref{For103}, then the pullbacks of these
objects by a diffeomorphism again satisfy the equations.  This not in general true of the constitutive relation
\Ref{For109}.

	Let $M$ be an affine space modelled on a vector space $V$.  A mapping $\zf \colon M \rightarrow M$ is said to be {\it
affine} if there is a linear mapping $\chi \colon V \rightarrow V$ such that
	$$\zf(x') - \zf(x) = \chi(x' - x).
																										\tag \label{For114}$$
	If the mapping $\chi \colon V \rightarrow V$ satisfying the above condition exists, then it is unique.  It is called
the {\it linear part} of the affine mapping $\zf$ and is denoted by $\ozf$.

	An affine mapping $\zf$ is invertible if and only if its linear part $\ozf$ is invertible.  If $\zf$ is
invertible then $\zf^{-1}$ is an affine mapping and $\overline{\zf^{-1}} = (\overline{\zf})^{-1}$.

	An affine mapping $\zf$ is differentiable.  Its differential and its linear part are related by
		$$\xD\zf(x,v) = \ozf(v).
																										\tag \label{For115}$$

	Let $V$ be a vector space of dimension $4$ with a Minkowski metric $g \colon V \rightarrow V^\*$ of signature $(1,3)$.
The {\it Lorentz group} for this space is the group of linear automorphisms
		$$\sG(V,g) = \left\{\zr \in \sG(V) ;\; \zr^\* \circ g \circ \zr = g \right\}.
																										\tag \label{For116}$$
	Let $M$ be the Minkowski space-time of special relativity.  A {\it Poincar\'e transformation} is an affine mapping $\zf
\colon M \rightarrow M$ such that $\ozf$ is a Lorentz transformation.  Poincar\'e transformations form a group denoted by
$\sP(M,g)$.

	The mapping $\WE^2_e g^{-1}$ and the scalar density $\sqrt{|g^{-1}|}$ are both Poincar\'e invariant.  It follows that
the constitutive relation for electromagnetic fields in vacuum is Poincar\'e invariant.

		\sect{Inertial reference frames.}
	An inertial observer establishes a {\it reference frame} in the Minkowski space-time $M$.  The world line of an inertial
observer is the image of a mapping
		$$\zg \colon \R \rightarrow M \colon s \mapsto x_0 + s u,
																										\tag \label{For117}$$
	where $u \in V$ is the time-like {\it velocity four-vector} and $x_0$ is the {\it initial event}.  The parameter $s$ is
the {\it proper time} of the observer.  The set
 		$$N = \left\{y \in M ;\; \langle g(u), y - x_0\rangle = 0 \right\}
																										\tag \label{For118}$$
	is the set of events considered by the observer simultaneous with the initial event $x_0$.  It is an affine space and
the space
		$$W = \left\{w \in V ;\; \langle g(u), w\rangle = 0 \right\}
																										\tag \label{For119}$$
	of vectors orthogonal to $u$ is its model space.  The quotient $V^\*\big/W^\polar$ of the dual space $V^\*$ by the polar
		$$W^\polar = \left\{a \in V^\* ;\; \langle a, w\rangle = 0\;\text{ for each }\; w \in W \right\}
																										\tag \label{For120}$$
	is the natural choice of dual space $W^\*$.  A Euclidean metric tensor in $W$ is defined by
		$$h \colon W \rightarrow W^\* \colon w \mapsto -\zp(g(w)),
																										\tag \label{For121}$$
	where $\zp \colon V^\* \rightarrow W^\*$ is the canonical projection.

	The pair
		$$(t,y) = (\zt(x),\zy(x)) = (c^{-1}\langle g(u), x - x_0\rangle, x - \langle g(u), x - x_0\rangle u) \in \R \times N
																										\tag \label{For122}$$
	is assigned to an event $x \in M$.  The inverse assignment is expressed by
		$$x = \zn(t,y) = y + ctu.
																										\tag \label{For123}$$

	Proper time is measured in units of length and the observer time $t$ is measured in units of time.

		\sect{Time-dependent forms and time-dependent currents.}
	Let $N$ be an affine space with a Euclidean model space $W$.  Let $h \colon W \rightarrow W^\*$ be the metric tensor.  A
{\it time dependent-differential $q$\,-\,form} is a differentiable mapping
		$$A \colon \R \times N \times W^q \times \sO(W) \rightarrow \R
																										\tag \label{For124}$$
	multilinear and totally antisymmetric in its vector arguments.  A time-dependent differential form $A$ is said to be
{\it even}, if
		$$A(t,y,w_1, w_2, \ldots ,w_q,Po) = A(t,y,w_1, w_2, \ldots ,w_q,o).
																										\tag \label{For125}$$
	It is said to be {\it odd}, if
		$$A(t,y,w_1, w_2, \ldots ,w_q,Po) = - A(t,y,w_1, w_2, \ldots ,w_q,o).
																										\tag \label{For126}$$

	The vector space of even time-dependent differential $q$\,-\,forms will be denoted by $\zF^q_e(\R,N)$ and space of odd
time-dependent differential $q$\,-\,forms will be denoted by $\zF^q_o(\R,N)$.  Even 0-forms are functions on $\R \times N$.

	If $A$ is a time-dependent differential $q$\,-\,form, then for each $t \in \R$ there is a differential $q$\,-\,form on $N$
defined as the mapping
		$$A_t \colon N \times W^q \times \sO(W) \rightarrow \R \colon (y,w_1, w_2, \ldots ,w_q,o) \mapsto
A(t,y,w_1, w_2, \ldots ,w_q,o).
																										\tag \label{For127}$$

	A {\it time dependent $q$-chain} is a mapping
		$$\bd \colon \R \rightarrow \zX_q^o(N).
																										\tag \label{For128}$$
	The space of time-dependent $q$-chains will be denoted by $\zX_q^p(\R,N)$

	The integral of a time-dependent form $A \in \zF^q_p(\R,N)$ on a time dependent chain $\bd \in \zX^p_q(\R,N)$ is the
function
		$$\int_\bd A \colon \R \rightarrow \R \colon t \mapsto \int_{\bd(t)} A_t.
																										\tag \label{For129}$$

	Constructions such as the exterior product and the exterior differential are easily extended to time-dependent forms.
The {\it exterior product} of a time-dependent $q$\,-\,form $A$ with a time-dependent $q'$\,-\,form $A'$ is the
$(q\+q')$\,-\,form
		$$\align
	A \wedge A' &\colon \R \times N \times W^{q+q'} \times \sO(W) \rightarrow \R \colon (t,y,w_1,\ldots,w_{q+q'},o) \\
		&\hskip6mm\mapsto  {\sum_{\zs \in S(q+q')}} \dsize{\frac{\sgn(\zs)}{q!q'!}}
A(t,y,w_{\zs(1)},\ldots,w_{\zs(q)},o) A'(t,y,w_{\zs(q+1)},\ldots,w_{\zs(q+q')},o).
																										\tag \label{For130}\endalign$$
	The {\it exterior differential} of a $q$\,-\,form $A$ is the $(q\+1)$\,-\,form
		$$\align
	\rd A \colon \R \times N \times &W^{q+1} \times \sO(W) \rightarrow \R \colon (t,y,w_1,w_2,\ldots,w_{q+1},o) \\
		&\mapsto - {\sum_{i=1}^{q+1}(-1)^i} \dsize{\frac{\rd}{\rd s}} A(t,y +
sw_i,w_1,w_2,\ldots,\widehat{w_i},\ldots,w_{q+1},o)\big|_{s=0}.
																										\tag \label{For131}\endalign$$
	Properties of these constructions established for ordinary forms have obvious generalizations to the case of
time-dependent forms.

	The {\it time derivative} of a time-dependent differential $q$\,-\,form $A$ is the time-dependent $q$\,-\,form
		$$\partial_t A \colon M \times W^q \times \sO(W) \rightarrow \R \colon (t,y,w_1,w_2,\ldots,w_q,o) \mapsto
\frac{\partial}{\partial t} A(t,y,w_1,w_2,\ldots,w_q,o).
																										\tag \label{For132}$$

	Given a reference frame we introduce the mapping
		$$\hat\zn\, \colon \R \times N \times \sO(W) \rightarrow M \times \sO(V) \colon (t,y,o) \mapsto (\zn(t,y),[u,o]) =
(y + ctu,[u,o]).
																										\tag \label{For133}$$
	The symbol $[u,o]$ is used to denote the orientation of $V$ represented by the base $(u,e_1,e_2,e_3)$, where
$(e_1,e_2,e_3)$ is the base of $W$ associated with the orientation $o$.  The mapping $\hat\zn$, the vector $u$, and the
constant vector field
		$$U \colon M \rightarrow V \colon x \mapsto u
																										\tag \label{For134}$$
	 are then used to associate with a $q$\,-\,form $A$ on $M$ a time-dependent $q$\,-\,form
		$$\hat\zn^{\*}A \colon \R \times N \times W^q \times \sO(W) \rightarrow \R \colon (t,y,w_1, w_2, \ldots
,w_q,o) \mapsto A(y + ctu,w_1, w_2, \ldots ,w_q,[u,o])
																										\tag \label{For135}$$
	and a time-dependent $(q\-1)$\,-\,form
		$$\hat\zn^{\*}\xi_U A \colon \R \times N \times W^q \times \sO(W) \rightarrow \R \colon (t,y,w_1, w_2, \ldots
,w_{q-1},o) \mapsto A(y + ctu,u,w_1, w_2, \ldots ,w_{q-1},[u,o]).
																										\tag \label{For136}$$
	The two time-dependent forms together fully represent the original form $A$.  The symbol $\hat\zn^\*$ is used because
the operation it denotes is similar to the pull back.  The image by $\hat\zn$ of a time-dependent chain $\bd \in
\zX^p_q(\R,N)$ is a {\it time-dependent chain}
		$$\hat\zn(\bd) \colon \R \rightarrow \zX^p_q(M) \colon t \mapsto \hat\zn(t,\cdot)(\bd(t)).
																										\tag \label{For137}$$
	The integral of a form $A \in \zF_p^q(M)$ on this chain is the function
		$$\int_{\hat\zn(\bd)} A = \int_\bd \hat\zn^{\*}A \colon \R \rightarrow \R \colon t \mapsto \int_{\bd(t)}
(\hat\zn^\*A)_t.
																										\tag \label{For138}$$
	From a time-dependent chain $\bd \in \zX^p_{q\-1}(\R,N)$ and a field $X \colon M \rightarrow V$ we construct the {\it
time-dependent current}
		$$X \wedge \hat\zn(\bd) \colon \zF_p^q(M) \rightarrow \R \colon A \mapsto \int_{X \wedge \hat\zn(\bd)} A =
\int_{\hat\zn(\bd)} \xi_X A.
																										\tag \label{For139}$$

		\sect{Reference frames and electromagnetic objects.}
	Electromagnetic phenomena are represented in relation to a reference frame by the following time-dependent forms on $N$.
			\list
	\item The {\it electric field} --- a time-dependent even 1-form
		$$\iE = \hat\zn^{\*}\xi_U F.
																										\tag \label{For140}$$
	\item The {\it magnetic induction} --- a time-dependent even 2-form
		$$\iB = \hat\zn^{\*} F.
																										\tag \label{For141}$$
	\item The {\it electric induction} --- a time-dependent odd 2-form
		$$\iD = \hat\zn^{\*} G.
																										\tag \label{For142}$$
	\item The {\it magnetic field} --- a time-dependent odd 1-form
		$$\iH = \hat\zn^{\*}\xi_U G.
																										\tag \label{For143}$$
	\item The {\it charge} --- a time-dependent odd 3-form
		$$\iQ = \hat\zn^{\*} J.
																										\tag \label{For144}$$
	\item The {\it current} --- a time-dependent odd 2-form
		$$\iJ = \hat\zn^{\*}\xi_U J.
																										\tag \label{For145}$$
	\item The {\it electric potential} --- a time-dependent function

		$$\iU = \hat\zn^{\*}\xi_U A.
																										\tag \label{For146}$$
	\item The {\it vector potential} --- a time-dependent even 1-form
		$$\iA = - \hat\zn^{\*} A.
																										\tag \label{For147}$$
			\endlist

	These symbolic definitions are expressed by the following more explicit relations:
		$$\iE \colon \R \times N \times W \times \sO(W) \rightarrow \R \colon (t,y,w,o) \mapsto F(y + ctu,u,w,[u,o]),
																										\tag \label{For148}$$
		$$\iB \colon \R \times N \times W \times W \times \sO(W) \rightarrow \R \colon (t,y,w_1,w_2,o) \mapsto F(y +
ctu,w_1,w_2,[u,o]),
																										\tag \label{For149}$$
		$$\iD \colon \R \times N \times W \times W \times \sO(W) \rightarrow \R \colon (t,y,w_1,w_2,o) \mapsto G(y +
ctu,w_1,w_2,[u,o]),
																										\tag \label{For150}$$
		$$\iH \colon \R \times N \times W \times \sO(W) \rightarrow \R \colon (t,y,w,o) \mapsto G(y + ctu,u,w,[u,o]),
																										\tag \label{For151}$$
		$$\iQ \colon \R \times N \times W \times W \times W \times \sO(W) \rightarrow \R \colon (t,y,w_1,w_2,w_3,o) \mapsto
J(y + ctu,w_1,w_2,w_3,[u,o]),
																										\tag \label{For152}$$
		$$\iJ \colon \R \times N \times W \times W \times \sO(W) \rightarrow \R \colon (t,y,w_1,w_2,o) \mapsto J(y +
ctu,u,w_1,w_2,[u,o]),
																										\tag \label{For153}$$
		$$\iU \colon \R \times N \times \sO(W) \rightarrow \R \colon (t,y,o) \mapsto A(y + ctu,u,[u,o]),
																										\tag \label{For154}$$
	and
		$$\iA \colon \R \times N \times W \times \sO(W) \rightarrow \R \colon (t,y,w,o) \mapsto -A(y + ctu,w,[u,o]).
																										\tag \label{For155}$$

	Even objects $\iE$, $\iB$, $\iU$, and $\iA$ can be replaced by mappings
		$$\wiE \colon \R \times N \rightarrow \wedge_e^1 W^\*,
																										\tag \label{For156}$$
		$$\wiB \colon \R \times N \rightarrow \wedge_e^2 W^\*,
																										\tag \label{For157}$$
		$$\wiU \colon \R \times N \rightarrow \R,
																										\tag \label{For158}$$
	and
		$$\wiA \colon \R \times N \rightarrow \wedge_e^1 W^\*.
																										\tag \label{For159}$$

		\sect{Reference frames and integral relations.}
	We list the well known integral relations relative to a reference frame.
			\list
		\item  Faraday--Lenz:
		$$\frac{1}{c} \,\xD\!\!\int_\bd \iB + \int_{\partial\bd}\left(\iE + \frac{1}{c}\,\xi_Y \iB\right) = 0.
																										\tag \label{For160}$$
	The symbol $\bd$ denotes a moving even 2-chain in $N$ and $Y \colon \R \times N \rightarrow W$ is the velocity field.
At each point of the moving boundary $\partial\bd$ the combination $\iE + \frac{1}{c}\,\xi_Y \iB$ is the electric field in the
rest frame of the point.
	\vskip2mm

	\item  No magnetic poles:
		$$\int_{\partial\bd} \iB = 0.
																										\tag \label{For161}$$
	Here, $\bd$ denotes a moving even 3-chain in $N$.
	\vskip2mm

	\item  Amp\`ere:
		$$\frac{1}{c}\xD\int_{\bd}\iD - \int_{\partial\bd}\left(\iH + \frac{1}{c}\,\xi_Y \iD\right) = -
\frac{4\zp}{c}\int_{\bd}\left(\iJ + \frac{1}{c}\,\xi_Y \iQ\right).
																										\tag \label{For162}$$
	The symbol $\bd$ denotes a moving odd 2-chain in $N$ and $Y \colon \R \times N \rightarrow W$ is the velocity field.
The better known version
		$$\frac{1}{c}\xD\int_{\bd}\iD - \int_{\partial\bd}\iH = - \frac{4\zp}{c}\int_{\bd}\iJ
																										\tag \label{For163}$$
	applies to stationary chains with $Y = 0$.
  	\vskip2mm

	\item  Gauss:
		$$\int_{\partial\bd}\iD = 4\zp\int_{\bd}\iQ.
																										\tag \label{For164}$$
	The symbol $\bd$ denotes a moving odd 3-chain in $N$.
			\endlist

	The first two integral relations are derived from
		$$\int_{\partial\bc} F = 0.
																										\tag \label{For165}$$

	Starting with a constant chain $\bd_0 \in \zX_2^e(N)$ and a flow $\zh \colon \R \times N \rightarrow N$ we introduce a
time-dependent chain
		$$\bd \colon \R \rightarrow \zX_2^e(N) \colon t \mapsto \hat\zh(t,\bd_0)
																										\tag \label{For166}$$
	denoted by $\hat\zh(\bd_0)$.  The mapping
		$$\hat\zh \colon \R \times N \times \sO(W)\rightarrow N \times \sO(W) \colon (t,y,o) \mapsto (\zh(t,y),o)
																										\tag \label{For167}$$
	is used.  The velocity field in $N$ associated with the flow $\zh$ is the time-dependent vector field
		$$Y \colon \R \times N \rightarrow W \colon (t,y) \mapsto \left(\frac{\partial}{\partial t}
\zh(t,\cdot)\right)\left(\zh(t,\cdot)^{-1}(y)\right).
																										\tag \label{For168}$$
	The mapping $\zh(t,\cdot)$ is assumed to be a diffeomorphism for each $t$.  The vector field
		$$X \colon M \rightarrow V \colon x \mapsto u + c^{-1}Y(\zt(x),\zy(x))
																										\tag \label{For169}$$
	is the velocity field in $M$.  The mapping
		$$\zx \colon \R \times M \rightarrow M \colon (s,x) \mapsto \zh\left(\zt(x) +c^{-1}s,\zh(\zt(x),\cdot)^{-1}(\zy(x))\right)
																										\tag \label{For170}$$
	is the flow in $M$ for this velocity field.  The field is related to the flow by
		$$X(x) = \frac{\partial}{\partial s} \zx(s,x)\big|_{s=0}.
																										\tag \label{For171}$$
	Setting $\bc = X \wedge \hat\zn(\bd)$ we arrive at
		$$\int_{\partial\bc} F = \int_{\partial(X \wedge \hat\zn(\bd))} F = \int_{X \wedge \hat\zn(\bd)} \rd F =
\int_{\hat\zn(\bd)} \xi_X\rd F = \int_{\hat\zn(\bd)}\rd_X F - \int_{\partial\hat\zn(\bd)}\xi_X F.
																										\tag \label{For172}$$

	The first integral involves the Lie derivative
		$$\rd_X F = \frac{\rd}{\rd s} \hat\zx(s,\cdot)^{\*}F\big|_{s=0}.
																										\tag \label{For173}$$
	This integral is a function of time since $\hat\zn(\bd)$ is a time dependent chain.  The following precise expression
		$$\align
			\hskip-5mm\int_{\hat\zn(\bd)}\rd_X F &\colon \R \rightarrow \R \\
				&\colon t \mapsto \int_{\hat\zn(t,\bd(t))}\rd_X F = \int_{\hat\zn(t,\bd(t))} \frac{\rd}{\rd
s}\hat\zx(s,\cdot)^\* F\big|_{s=0} = \frac{\partial}{\partial s} \int_{\bd_0} \left(\hat\zx(s,\cdot) \circ \hat\zn(t,\cdot)
\circ \hat\zh(t,\cdot)\right)^\* F\big|_{s=0}
																										\tag \label{For174}\endalign$$
	is obtained for this function.  The mapping
		$$\hat\zx \colon \R \times M \times \sO(V)\rightarrow M \times \sO(V) \colon (s,x,o) \mapsto (\zx(s,x),o)
																										\tag \label{For175}$$
	is used.  The derivation of the final expression
		$$\align
			\frac{\partial}{\partial s} \int_{\bd_0} \left(\hat\zx(s,\cdot) \circ \hat\zn(t,\cdot) \circ
\hat\zh(t,\cdot)\right)^\* F\big|_{s=0} &= \frac{1}{c} \frac{\rd}{\rd t} \int_{\bd_0} \left(\hat\zn(t,\cdot) \circ
\hat\zh(t,\cdot)\right)^\* F \\
				&= \frac{1}{c} \frac{\rd}{\rd t} \int_{\hat\zh(t,\bd_0)} \hat\zn(t,\cdot)^\* F \\
				&= \frac{1}{c} \frac{\rd}{\rd t} \int_{\bd(t)} B_t
																										\tag \label{For176}\endalign$$
	for the value of the integral at $t$ is based on
		$$\left(\zx(s,\cdot) \circ \zn(t,\cdot) \circ \zh(t,\cdot)\right)(y) = \zx(s,\zn(t,\zh(t,y))) = \zx(s,\zh(t,y) +
ctu) = \zh\left(t + c^{-1}s,y\right) + (ct + s)u.
																										\tag \label{For177}$$

	For the expression $\hat\zn^{\*}\rd\xi_X F$ we have
		$$\hat\zn^{\*}\rd\xi_X F = \rd\hat\zn^{\*}\xi_X F = \rd\left(\iE + c^{-1}\,\xi_Y \iB\right).
																										\tag \label{For178}$$
	Hence,
		$$\int_{\partial(X \wedge \hat\zn(\bd))} F \colon t \mapsto  \frac{1}{c} \frac{\rd}{\rd t} \int_{\bd(t)} \iB_t +
\int_{\partial\bd(t)}\left(\iE_t + c^{-1}\,\xi_{Y(t)} \iB_t\right).
																										\tag \label{For179}$$
	The Faraday-Lenz relation follows from \Ref{For165}.

	Starting with a time-dependent chain $\bd \in \zX_3^e(\R,N)$ and $\bc = \hat\zn(\bd)$ we derive the equality
		$$\int_{\partial\bc} F = \int_{\partial\hat\zn(\bd)} F = \int_{\hat\zn(\bd)}\rd F = \int_\bd
\hat\zn^{\*}\rd F = \int_\bd \rd\zn^{\*}F = \int_\bd \rd\iB = \int_{\partial\bd} \iB.
																										\tag \label{For180}$$
	The equality \Ref{For161} follows from \Ref{For165}.

	The Amp\`ere law \Ref{For162} is derived from
		$$\int_{\partial\bc} G  = - \frac{4\zp}{c}\int_\bc J.
																										\tag \label{For181}$$
	Its derivation is analogous to the derivation of the Faraday-Lenz relation.  Starting with a chain $\bd_0 \in \zX_2^o(N)$
and a flow $\zh \colon \R \times N \rightarrow N$ we obtain the expressions
		$$\int_{\partial(X \wedge \hat\zn(\bd))} G \colon t \mapsto  \frac{1}{c} \frac{\rd}{\rd t} \int_{\bd(t)} \iD_t +
\int_{\partial\bd(t)}\left(\iH_t + c^{-1}\,\xi_{Y(t)} D_t\right)
																										\tag \label{For182}$$
	and
		$$\int_{X \wedge \hat\zn(\bd)} J = \int_{\bd}\hat\zn^\*\xi_X J \colon t \mapsto  \int_{\bd(t)}\left(\iJ_t +
c^{-1}\xi_{Y(t)}\iQ_t\right).
																										\tag \label{For183}$$
	The Amp\`ere law follows from \Ref{For181}.

	Choosing a chain $\bd \in \zX_3^o(N)$ and setting $\bc = \hat\zn(\bd)$ we obtain
		$$\int_{\partial\bc} G = \int_{\partial\hat\zn(\bd)} G = \int_{\partial\bd} \iD
																										\tag \label{For184}$$
	and
		$$\int_\bc J = \int_{\hat\zn(\bd)} J = \int_\bd \hat\zn^{\*}J = \int_\bd \iQ.
																										\tag \label{For185}$$
	The Gauss law follows from from \Ref{For181}.

		\sect{Reference frames and Maxwell's equations.}
	Maxwell's equations
		$$\frac{1}{c} \partial_t \iB + \rd \iE = 0,
																										\tag \label{For186}$$
		$$\rd \iB = 0,
																										\tag \label{For187}$$
		$$\frac{1}{c} \partial_t\oviD - \Div\oviH = - \frac{4\zp}{c} \oviJ,
																										\tag \label{For188}$$
	and
		$$\Div\oviD = 4\zp\oviQ
																										\tag \label{For189}$$
	will be derived from Maxwell's equations in $M$.

	Constitutive relations
		$$\oviD = \left(\WE^1_e\, h^{-1} \circ \wiE\right) \otimes \sqrt{|h|}
																										\tag \label{For190}$$
and
		$$\oviH = \left(\WE^2_e\, h^{-1} \circ \wiB\right) \otimes \sqrt{|h|}
																										\tag \label{For191}$$
	and also equations
		$$\iE = \frac{1}{c} \partial_t\iA - \rd\iU
																										\tag \label{For192}$$
	and
		$$\iB = \rd\iA
																										\tag \label{For193}$$
	will be derived.

		\sect{Derivation of Maxwell's equations.}
	Equations
		$$\frac{1}{c} \partial_t \iB + \rd \iE = 0
																										\tag \label{For194}$$
	and
		$$\rd\iB = 0
																										\tag \label{For195}$$
	follow from
		$$\rd F = 0,
																										\tag \label{For196}$$
		$$\align
			\rd F(y + ctu,u,w_1,w_2,[u,o]) &= \frac{\rd}{\rd s} F(y + ctu + su,w_1,w_2,[u,o])\big|_{s=0} - \frac{\rd}{\rd s}
F(y + ctu + sw_1,u,w_2,[u,o])\big|_{s=0}\\
				&\hskip25mm + \frac{\rd}{\rd s} F(y + ctu + sw_2,u,w_1,[u,o])\big|_{s=0} \\
					&= \frac{1}{c}\frac{\partial}{\partial t} \iB(t,y,w_1,w_2,o) - \frac{\rd}{\rd s} \iE(t,y +
sw_1,w_2,o)\big|_{s=0}\\
  				&\hskip25mm + \frac{\rd}{\rd s} \iE(t,y + sw_2,w_1,o)\big|_{s=0} \\
					&= \frac{1}{c} \partial_t \iB(t,y,w_1,w_2,o) - \rd \iE(t,y,w_1,w_2,o),
																										\tag \label{For197}\endalign$$
	and
		$$\align
			\rd F(y + ctu,w_1,w_2,w_3,[u,o]) &= \frac{\rd}{\rd s} F(y + ctu + sw_1,w_2,w_3,[u,o])\big|_{s=0} - \frac{\rd}{\rd
s} F(y + ctu + sw_2,w_1,w_3,[u,o])\big|_{s=0} \\
					&\hskip30mm + \frac{\rd}{\rd s} F(y + ctu + sw_3,w_1,w_2,[u,o])\big|_{s=0} \\
					&= \frac{\rd}{\rd s} \iB(t,y + sw_1,w_2,w_3,o) - \frac{\rd}{\rd s} \iB(t,y + sw_2,w_1,w_3,o)\big|_{s=0} \\
					&\hskip30mm + \frac{\rd}{\rd s} \iB(t,y + sw_3,w_1,w_2,o)\big|_{s=0} \\
					&= \rd \iB(t,y,w_1,w_2,w_3,o).
																										\tag \label{For198}\endalign$$

	Equations
		$$\frac{1}{c} \partial_t\iD - \rd\iH = -\frac{4\zp}{c}\iJ
																										\tag \label{For199}$$
	and
		$$\rd\iD = 4\zp \iQ
																										\tag \label{For200}$$
	follow from
		$$\rd G = - \frac{4\zp}{c} J,
																										\tag \label{For201}$$
		$$\rd G(y + ctu,u,w_1,w_2,[u,o]) = \partial_t \iD(t,y,w_1,w_2,o) - \rd \iH(t,y,w_1,w_2,o),
																										\tag \label{For202}$$
	and
		$$\rd G(y + ctu,w_1,w_2,w_3,[u,o]) = \rd \iD(t,y,w_1,w_2,w_3,o).
																										\tag \label{For203}$$
	These equations are converted to
		$$\frac{1}{c}\partial_t\oviD - \Div\oviH = - \frac{4\zp}{c}\oviJ
																										\tag \label{For204}$$
	and
		$$\Div\oviD = 4\zp\oviQ.
																										\tag \label{For205}$$

	Equations
		$$\iE = \frac{1}{c} \partial_t\iA - \rd\iU
																										\tag \label{For206}$$
	and
		$$\iB = \rd\iA
																										\tag \label{For207}$$
	follow from
		$$F = \rd A,
																										\tag \label{For208}$$
		$$\align
			\rd A(y + ctu,u,w,[u,o]) &= \frac{\rd}{\rd s} A(y + ctu + su,w,[u,o])\big|_{s=0} - \frac{\rd}{\rd s} A(y + ctu +
sw,u,[u,o])\big|_{s=0} \\
					&= \frac{1}{c} \frac{\partial}{\partial t} \iA(t,y,w,o) - \frac{\rd}{\rd s} \iU(t,y + sw,o)\big|_{s=0}\\
					&= \frac{1}{c} \partial_t \iA(t,y,w,o) - \rd \iU(t,y,w,o),
																										\tag \label{For209}\endalign$$
	and
		$$\align
			\rd A(y + ctu,w_1,w_2,[u,o]) &= \frac{\rd}{\rd s} A(y + ctu + sw_1,w_2,[u,o])\big|_{s=0} - \frac{\rd}{\rd s} A(y
+ ctu + sw_2,w_1,[u,o])\big|_{s=0} \\
					&= \frac{\rd}{\rd s} \iA(t,y + sw_1,w_2,o) - \frac{\rd}{\rd s} \iA(t,y + sw_2,w_1,o)\big|_{s=0} \\
					&= \rd \iA(t,y,w_1,w_2,o).
																										\tag \label{For210}\endalign$$

	We start with the constitutive relation in the form
		$$\ovG(y + ctu) = \left(\WE^2_e\, g^{-1}(\wF(y + ctu))\right) \otimes \sqrt{|g|}.
																										\tag \label{For211}$$
	If
		$$\WE^2_e\, g^{-1}(\wF(y + ctu)) = \left[(u,w,[u,o]) + \tsize\sum_{i=1}^n \zl_i(w_1^i,w_2^i,[u,o^i])\right],
																										\tag \label{For212}$$
	then
		$$\oviD(t,y) = \left[(w,o)\right] \otimes \sqrt{|h|},
																										\tag \label{For213}$$
		$$\oviH(t,y) = \left[\tsize\sum_{i=1}^n \zl_i(w_1^i,w_2^i,o^i)\right] \otimes \sqrt{|h|},
																										\tag \label{For214}$$
	and
		$$\wF(y + ctu) = \WE^2_e\, g\left(\left[(u,w,[u,o]) + \tsize\sum_{i=1}^n \zl_i(w_1^i,w_2^i,[u,o^i])\right]\right).
																										\tag \label{For215}$$

	For each vector $w'$ in $W$ and each orientation $o'$ of $W$ we have
		$$\wF(y + ctu)(u,w',[u,o']) = F(y + ctu,u,w',[u,o']) = \iE(t,y,w',o') = \wiE(t,u)(w',o')
																										\tag \label{For216}$$
	and
		$$\align
				\WE^2_e\, g\left(\left[(u,w,[u,o]) + \tsize\sum_{i=1}^n
\zl_i(w_1^i,w_2^i,[u,o^i])\right]\right)(u,w',[u,o']) &= \langle g(w), w'\rangle \\
			&= -\langle h(w), w'\rangle \\
			&= - \WE^1_e h([(w,o)])(w',o').
																										\tag \label{For217}\endalign$$
	Hence,
		$$\wiE(t,u)(w',o') = - \WE^1_e h([(w,o)])(w',o'),
																										\tag \label{For218}$$
		$$\wiE(t,u) = - \WE^1_e h([(w,o)]),
																										\tag \label{For219}$$
		$$\WE^1_e h^{-1}(\wiE(t,u)) = - [(w,o)],
																										\tag \label{For220}$$
	and
		$$\WE^1_e h^{-1}(\wiE(t,u)) \otimes \sqrt{|h|} = - [(w,o)] \otimes \sqrt{|h|} = \oviD(t,y).
																										\tag \label{For221}$$
	We have obtained the constitutive relation
		$$\oviD = \left(\WE^1_e\, h^{-1} \circ \wiE\right) \otimes \sqrt{|h|}.
																										\tag \label{For222}$$

	For each pair of vectors $w_1$ and $w_2$ in $W$ and each orientation $o'$ of $W$ we have
		$$\wF(y + ctu)(w_1,w_2,[u,o']) = F(y + ctu,w_1,w_2,[u,o']) = \iB(t,y,w_1,w_2,o') = \wiB(t,y)(w_1,w_2,o')
																										\tag \label{For223}$$
	and
		$$\align
				&\WE^2_e\, g\left(\left[(u,w,[u,o]) + \tsize\sum_{i=1}^n
\zl_i(w_1^i,w_2^i,[u,o^i])\right]\right)(w_1,w_2,[u,o']) \\
			&\hskip40mm = \tsize\sum_{i=1}^n \zl_i\left(\langle g(w_1^i),
w_1\rangle\langle g(w_2^i), w_2\rangle - \langle g(w_1^i), w_2\rangle\langle g(w_2^i), w_1\rangle\right) \\
			&\hskip40mm = \tsize\sum_{i=1}^n \zl_i\langle h(w_1^i), w_1\rangle\langle h(w_2^i), w_2\rangle - \langle
h(w_1^i), w_2\rangle\langle h(w_2^i), w_1\rangle \\
			&\hskip40mm = \tsize\sum_{i=1}^n \zl_i\WE^2_e h([(w_1^i,w_2^i,o^i)])(w_1,w_2,o').
																										\tag \label{For224}\endalign$$
	Hence,
		$$\wiB(t,u)(w_1,w_2,o') = \tsize\sum_{i=1}^n \zl_i\WE^2_e h([(w_1^i,w_2^i,o^i)])(w_1,w_2,o'),
																										\tag \label{For225}$$
		$$\wiB(t,u) = \tsize\sum_{i=1}^n \zl_i\WE^2_e h([(w_1^i,w_2^i,o^i)]),
																										\tag \label{For226}$$
		$$\WE^2_e h^{-1}(\wiB(t,u)) = \tsize\sum_{i=1}^n \zl_i([(w_1^i,w_2^i,o^i)]),
																										\tag \label{For227}$$
	and
		$$\WE^2_e h^{-1}(\wiB(t,u)) \otimes \sqrt{|h|} = \tsize\sum_{i=1}^n \zl_i([(w_1^i,w_2^i,o^i)]) \otimes \sqrt{|h|} =
\oviH(t,y).
																										\tag \label{For228}$$
	We have derived the constitutive relation
		$$\oviH = \left(\WE^2_e\, h^{-1} \circ \wiB\right) \otimes \sqrt{|h|}.
																										\tag \label{For229}$$

\newpage

		\sect{Traditional formulation of Maxwell's equations.}

	By using the Weyl isomorphism, the metric $h$, and the metric volume form $\sqrt{|h|}$ physical objects are usually
identified as scalars, pseudoscalars, vectors, and pseudovectors.  The scalar product and the vector product are used.  The
identifications used in traditional formulations of electrodynamics are listed in the following table.  The {\it temporal
parity} is also listed.

\vskip4mm

\setbox80=\hbox{\vrule height-2.5pt width130mm depth3pt}
\setbox81=\hbox{even}
\setbox82=\hbox{odd}

\setbox1=\hbox{The object}
\setbox2=\hbox{is identified as}
\setbox85=\hbox{temporal parity}

\setbox3=\hbox{The even 1-form}
\setbox4=\hbox{$\iE$}
\setbox5=\hbox{a vector field}
\setbox6=\hbox{$\vec E$}

\setbox7=\hbox{The even 2-form}
\setbox8=\hbox{$\iB$}
\setbox9=\hbox{a pseudovector field}
\setbox10=\hbox{$\vec B$}

\setbox11=\hbox{The odd\hskip6.5pt 2-form}
\setbox12=\hbox{$\iD$}
\setbox13=\hbox{a vector field}
\setbox14=\hbox{$\vec D$}

\setbox15=\hbox{The odd\hskip6.5pt 1-form}
\setbox36=\hbox{$\iH$}
\setbox18=\hbox{a pseudovector field}
\setbox19=\hbox{$\vec H$}

\setbox20=\hbox{The odd\hskip6.5pt 3-form}
\setbox21=\hbox{$\iQ$}
\setbox22=\hbox{a scalar field}
\setbox23=\hbox{$Q$}

\setbox24=\hbox{The odd\hskip6.5pt 2-form}
\setbox25=\hbox{$\iJ$}
\setbox26=\hbox{a vector field}
\setbox27=\hbox{$\vec J$}

\setbox28=\hbox{The even 0-form}
\setbox29=\hbox{$\iU$}
\setbox30=\hbox{a scalar field}
\setbox31=\hbox{$U$}

\setbox32=\hbox{The even 1-form}
\setbox33=\hbox{$\iA$}
\setbox34=\hbox{a vector field}
\setbox35=\hbox{$\vec A$}

\setbox0=\hbox{
\hskip010mm\lower-01mm\copy01\hskip-\wd01\hskip-010mm
\hskip063mm\lower-01mm\copy02\hskip-\wd02\hskip-063mm
\hskip115mm\lower-01mm\copy85\hskip-\wd85\hskip-115mm

\hskip10mm \copy80 \hskip-140mm

\hskip010mm\lower06mm\copy03\hskip-\wd03\hskip-010mm
\hskip040mm\lower06mm\copy04\hskip-\wd04\hskip-040mm
\hskip063mm\lower06mm\copy05\hskip-\wd05\hskip-063mm
\hskip098mm\lower06mm\copy06\hskip-\wd06\hskip-098mm
\hskip125mm\lower06mm\copy82\hskip-\wd82\hskip-125mm

\hskip010mm\lower11mm\copy07\hskip-\wd07\hskip-010mm
\hskip040mm\lower11mm\copy08\hskip-\wd08\hskip-040mm
\hskip063mm\lower11mm\copy09\hskip-\wd09\hskip-063mm
\hskip098mm\lower11mm\copy10\hskip-\wd10\hskip-098mm
\hskip125mm\lower11mm\copy81\hskip-\wd81\hskip-125mm

\hskip010mm\lower16mm\copy11\hskip-\wd11\hskip-010mm
\hskip040mm\lower16mm\copy12\hskip-\wd12\hskip-040mm
\hskip063mm\lower16mm\copy13\hskip-\wd13\hskip-063mm
\hskip098mm\lower16mm\copy14\hskip-\wd14\hskip-098mm
\hskip125mm\lower16mm\copy82\hskip-\wd82\hskip-125mm

\hskip010mm\lower21mm\copy15\hskip-\wd15\hskip-010mm
\hskip040mm\lower21mm\copy36\hskip-\wd36\hskip-040mm
\hskip063mm\lower21mm\copy18\hskip-\wd18\hskip-063mm
\hskip098mm\lower21mm\copy19\hskip-\wd19\hskip-098mm
\hskip125mm\lower21mm\copy81\hskip-\wd81\hskip-125mm

\hskip010mm\lower26mm\copy20\hskip-\wd20\hskip-010mm
\hskip040mm\lower26mm\copy21\hskip-\wd21\hskip-040mm
\hskip063mm\lower26mm\copy22\hskip-\wd22\hskip-063mm
\hskip098mm\lower26mm\copy23\hskip-\wd23\hskip-098mm
\hskip125mm\lower26mm\copy82\hskip-\wd82\hskip-125mm

\hskip010mm\lower31mm\copy24\hskip-\wd24\hskip-010mm
\hskip040mm\lower31mm\copy25\hskip-\wd25\hskip-040mm
\hskip063mm\lower31mm\copy26\hskip-\wd26\hskip-063mm
\hskip098mm\lower31mm\copy27\hskip-\wd27\hskip-098mm
\hskip125mm\lower31mm\copy81\hskip-\wd81\hskip-125mm

\hskip010mm\lower36mm\copy28\hskip-\wd28\hskip-010mm
\hskip040mm\lower36mm\copy29\hskip-\wd29\hskip-040mm
\hskip063mm\lower36mm\copy30\hskip-\wd30\hskip-063mm
\hskip098mm\lower36mm\copy31\hskip-\wd31\hskip-098mm
\hskip125mm\lower36mm\copy82\hskip-\wd82\hskip-125mm

\hskip010mm\lower41mm\copy32\hskip-\wd32\hskip-010mm
\hskip040mm\lower41mm\copy33\hskip-\wd33\hskip-040mm
\hskip063mm\lower41mm\copy34\hskip-\wd34\hskip-063mm
\hskip098mm\lower41mm\copy35\hskip-\wd35\hskip-098mm
\hskip125mm\lower41mm\copy81\hskip-\wd81\hskip-125mm

}\box0
\vskip4mm

	Temporal parity of objects is determined by inspecting the response to time reflection.  The {\it time reflection}
relative to an inertial frame is represented in $M$ and in $\R \times N$ by the transformations
		$$\zf \colon M \rightarrow M \colon x \mapsto x - 2\langle g(u), x - x_0\rangle u
																										\tag \label{For230}$$
	and
		$$\zc \colon \R \times N \rightarrow \R \times  N \colon (t,y) \mapsto (-t,y)
																										\tag \label{For231}$$
	with derivatives
		$$\xD\zf \colon M \times V \rightarrow V \colon (x,v) \mapsto v - 2\langle g(u), v\rangle u
																										\tag \label{For232}$$
	and
		$$\xD\zc \colon \R \times N \times \R \times W \rightarrow \R \times  W \colon (t,y,t',w) \mapsto (- t',w).
																										\tag \label{For233}$$
	Note that these mappings are involutions and that $\det(\xD\zf(x,\cdot)) = -1$.

	Here are the effects of the time reflection applied to the forms $F$, $G$, $J$, and $A$:
		$$\align
			\zf^{\*}F(x,v_1,v_2,o) &= F(\zf(x),\xD\zf(x,v_1),\xD\zf(x,v_2),[\xD\zf(x,\cdot)](o)) \\
					&= F(x - 2\langle g(u), x - x_0\rangle u,v_1 - 2\langle g(u), v_1\rangle u,v_2 - 2\langle g(u),
v_2\rangle u,o),
																										\tag \label{For234}\endalign$$
		$$\align
			\zf^{\*}G(x,v_1,v_2,o) &= G(\zf(x),\xD\zf(x,v_1),\xD\zf(x,v_2),[\xD\zf(x,\cdot)](o)) \\
					&= -G(x - 2\langle g(u), x - x_0\rangle u,v_1 - 2\langle g(u), v_1\rangle u,v_2 - 2\langle g(u),
v_2\rangle u,o),
																										\tag \label{For235}\endalign$$
		$$\align
			\zf^{\*}&J(x,v_1,v_2,v_3,o) = J(\zf(x),\xD\zf(x,v_1),\xD\zf(x,v_2),\xD\zf(x,v_3),[\xD\zf(x,\cdot)](o)) \\
					&= -J(x - 2\langle g(u), x - x_0\rangle u,v_1 - 2\langle g(u), v_1\rangle u,v_2 - 2\langle g(u),
v_2\rangle u,v_3 - 2\langle g(u), v_3\rangle u,o),
																										\tag \label{For236}\endalign$$
	and
		$$\align
			\zf^{\*}A(x,v,o) &= A(\zf(x),\xD\zf(x,v),[\xD\zf(x,\cdot)](o)) \\
					&= A(x - 2\langle g(u), x - x_0\rangle u,v - 2\langle g(u), v\rangle u,o).
																										\tag \label{For237}\endalign$$

	The time reflection transformation rules for $\iE$, $\iB$, $\iD$, $\iH$, $\iQ$, $\iJ$, $\iU$, and $\iA$  are obtained
from
		$$\align
			\zc^{\*}\iE(t,y,w,o) &= \hat\zn^{\*}\xi_U \zf^{\*}F(t,y,w,o) \\
					&= \zf^{\*}F(y + ctu,u,w,[u,o]) \\
					&= F(y - ctu,-u,w,[u,o]) \\
					&= -\iE(-t,y,w.o),
																										\tag \label{For238}\endalign$$
		$$\align
			\zc^{\*}\iB(t,y,w_1,w_2,o) &= \hat\zn^{\*}\zf^{\*}F(t,y,w_1,w_2,o) \\
					&= \zf^{\*}F(y + ctu,w_1,w_2,[u,o]) \\
					&= F(y - ctu,w_1,w_2,[u,o]) \\
					&= \iB(-t,y,w_1,w_2.o),
																										\tag \label{For239}\endalign$$
		$$\align
			\zc^{\*}\iD(t,y,w_1,w_2,o) &= \hat\zn^{\*}\zf^{\*}G(t,y,w_1,w_2,o) \\
					&= \zf^{\*}G(y + ctu,w_1,w_2,[u,o]) \\
					&= -G(y - ctu,w_1,w_2,[u,o]) \\
					&= -\iD(-t,y,w_1,w_2.o),
																										\tag \label{For240}\endalign$$
		$$\align
			\zc^{\*}\iH(t,y,w,o) &= \hat\zn^{\*}\xi_U \zf^{\*}G(t,y,w,o) \\
					&= \zf^{\*}G(y + ctu,u,w,[u,o]) \\
					&= -G(y - ctu,-u,w,[u,o]) \\
					&= \iH(-t,y,w.o),
																										\tag \label{For241}\endalign$$
		$$\align
			\zc^{\*}\iQ(t,y,w_1,w_2,w_3,o) &= \hat\zn^{\*}\zf^{\*}J(t,y,w_1,w_2,w_3,o) \\
					&= \zf^{\*}J(y + ctu,w_1,w_2,w_3,[u,o]) \\
					&= -J(y - ctu,w_1,w_2,w_3,[u,o]) \\
					&= -\iQ(-t,y,w_1,w_2,w_3.o),
																										\tag \label{For242}\endalign$$
		$$\align
			\zc^{\*}\iJ(t,y,w_1,w_2,o) &= \hat\zn^{\*}\xi_U\zf^{\*}J(t,y,w_1,w_2,o) \\
					&= \zf^{\*}J(y + ctu,u,w_1,w_2,[u,o]) \\
					&= -J(y - ctu,-u,w_1,w_2,[u,o]) \\
					&= \iJ(-t,y,w_1,w_2.o),
																										\tag \label{For243}\endalign$$
		$$\align
			\zc^{\*}\iU(t,y,o) &= \hat\zn^{\*}\xi_U \zf^{\*}A(t,y,o) \\
					&= \zf^{\*}A(y + ctu,u,[u,o]) \\
					&= A(y - ctu,-u,[u,o]) \\
					&= -\iU(-t,y.o),
																										\tag \label{For244}\endalign$$
	and
		$$\align
			\zc^{\*}\iA(t,y,w,o) &= \hat\zn^{\*}\zf^{\*}A(t,y,w,o) \\
					&= \zf^{\*}A(y + ctu,w,[u,o]) \\
					&= A(y - ctu,w,[u,o]) \\
					&= \iA(-t,y,w.o).
																										\tag \label{For245}\endalign$$

	The equations
		$$\frac{1}{c} \partial_t \vec B + \vec\nabla \times \vec E = 0,
																										\tag \label{For246}$$
		$$\vec\nabla \cdot \vec B = 0,
																										\tag \label{For247}$$
		$$\partial_t \vec D - \vec\nabla \times \vec H = - \frac{4\zp}{c} \vec J,
																										\tag \label{For248}$$
		$$\vec\nabla \cdot \vec D = 4\zp Q,
																										\tag \label{For249}$$
		$$\vec E = \frac{1}{c} \partial_t\vec A - \vec\nabla U,
																										\tag \label{For250}$$
		$$\vec B = \vec\nabla \times \vec A,
																										\tag \label{For251}$$
	are the Maxwell's equations in the traditional form.  The vector operator $\vec\nabla$ replaces the exterior
differential $\rd$.  These are the equations found in standard texts of electrodynamics such as {\it Classical
Electrodynamics} by John David Jackson [10] and {\it Field Theory} by L. D. Landau and E. M. Lifshitz [11]. Below is the
table of parity determinations extracted from the two texts.

\vskip4mm

\setbox0=\hbox{
\hskip010mm\lower-01mm\copy01\hskip-\wd01\hskip-010mm
\hskip063mm\lower-01mm\copy02\hskip-\wd02\hskip-063mm
\hskip115mm\lower-01mm\copy85\hskip-\wd85\hskip-115mm

\hskip10mm \copy80 \hskip-140mm

\hskip010mm\lower06mm\copy03\hskip-\wd03\hskip-010mm
\hskip040mm\lower06mm\copy04\hskip-\wd04\hskip-040mm
\hskip063mm\lower06mm\copy05\hskip-\wd05\hskip-063mm
\hskip098mm\lower06mm\copy06\hskip-\wd06\hskip-098mm
\hskip125mm\lower06mm\copy81\hskip-\wd81\hskip-125mm

\hskip010mm\lower11mm\copy07\hskip-\wd07\hskip-010mm
\hskip040mm\lower11mm\copy08\hskip-\wd08\hskip-040mm
\hskip063mm\lower11mm\copy09\hskip-\wd09\hskip-063mm
\hskip098mm\lower11mm\copy10\hskip-\wd10\hskip-098mm
\hskip125mm\lower11mm\copy82\hskip-\wd82\hskip-125mm

\hskip010mm\lower16mm\copy11\hskip-\wd11\hskip-010mm
\hskip040mm\lower16mm\copy12\hskip-\wd12\hskip-040mm
\hskip063mm\lower16mm\copy13\hskip-\wd13\hskip-063mm
\hskip098mm\lower16mm\copy14\hskip-\wd14\hskip-098mm
\hskip125mm\lower16mm\copy81\hskip-\wd81\hskip-125mm

\hskip010mm\lower21mm\copy15\hskip-\wd15\hskip-010mm
\hskip040mm\lower21mm\copy36\hskip-\wd36\hskip-040mm
\hskip063mm\lower21mm\copy18\hskip-\wd18\hskip-063mm
\hskip098mm\lower21mm\copy19\hskip-\wd19\hskip-098mm
\hskip125mm\lower21mm\copy82\hskip-\wd82\hskip-125mm

\hskip010mm\lower26mm\copy20\hskip-\wd20\hskip-010mm
\hskip040mm\lower26mm\copy21\hskip-\wd21\hskip-040mm
\hskip063mm\lower26mm\copy22\hskip-\wd22\hskip-063mm
\hskip098mm\lower26mm\copy23\hskip-\wd23\hskip-098mm
\hskip125mm\lower26mm\copy81\hskip-\wd81\hskip-125mm

\hskip010mm\lower31mm\copy24\hskip-\wd24\hskip-010mm
\hskip040mm\lower31mm\copy25\hskip-\wd25\hskip-040mm
\hskip063mm\lower31mm\copy26\hskip-\wd26\hskip-063mm
\hskip098mm\lower31mm\copy27\hskip-\wd27\hskip-098mm
\hskip125mm\lower31mm\copy82\hskip-\wd82\hskip-125mm

\hskip010mm\lower36mm\copy28\hskip-\wd28\hskip-010mm
\hskip040mm\lower36mm\copy29\hskip-\wd29\hskip-040mm
\hskip063mm\lower36mm\copy30\hskip-\wd30\hskip-063mm
\hskip098mm\lower36mm\copy31\hskip-\wd31\hskip-098mm
\hskip125mm\lower36mm\copy81\hskip-\wd81\hskip-125mm

\hskip010mm\lower41mm\copy32\hskip-\wd32\hskip-010mm
\hskip040mm\lower41mm\copy33\hskip-\wd33\hskip-040mm
\hskip063mm\lower41mm\copy34\hskip-\wd34\hskip-063mm
\hskip098mm\lower41mm\copy35\hskip-\wd35\hskip-098mm
\hskip125mm\lower41mm\copy82\hskip-\wd82\hskip-125mm

}\box0
\vskip4mm

	The temporal parities of all objects in this table are opposite to the ones in our list based on the assumption that
the electromagnetic field $F$ is an even form and the electromagnetic induction $G$ is an odd form.

		\Title{B. Electrodynamics with relativistic orientations.}

		\sect{Orientation in the Minkowski space-time.}
	Let $V$ be a vector space of dimension $4$ with a Minkowski metric $g \colon V \rightarrow V^\*$ of signature $(1,3)$.
The {\it Lorentz group} for this space is the group of linear automorphisms
		$$\sG(V,g) = \left\{\zr \in \sG(V) ;\; \zr^\* \circ g \circ \zr = g \right\}.
																										\tag \label{For252}$$
	A linear automorphism
		$$\zh \colon V \rightarrow \R^4
																										\tag \label{For253}$$
	is called a {\it Lorentz frame} if
		$$(\zh^\* \circ g \circ \zh^{-1})\pmatrix\mathstrut v^0 \\ \mathstrut v^1 \\ \mathstrut v^2 \\ \mathstrut v^3
\endpmatrix = (v^0, - v^1, - v^2, - v^3)
																										\tag \label{For254}$$
for each vector
		$$\pmatrix\mathstrut v^0 \\ \mathstrut v^1 \\ \mathstrut v^2 \\ \mathstrut v^3 \endpmatrix \in \R^4.
																										\tag \label{For255}$$
	We denote by $\sF(V,g)$ the space of Lorentz frames.  This space is a homogeneous space for the group $\sG(V,g)$ with
the natural group action
		$$\sG(V,g) \times \sF(V,g) \rightarrow \sF(V,g) \colon (\zr,\zx) \mapsto \zx \circ \zr^{-1}.
																										\tag \label{For256}$$

	The {\it light cone}
		$$C = \left\{v \in  V;\; \langle g(v), v\rangle = 0 \right\}
																										\tag \label{For257}$$
	divides the space $V$ in three disjoint connected regions.  There is the  region
		$$S = \left\{v \in  V;\; \langle g(v), v\rangle < 0 \right\}
																										\tag \label{For258}$$
	of {\it space-like} vectors and the region
		$$T = \left\{v \in  V;\; \langle g(v), v\rangle > 0 \right\}
																										\tag \label{For259}$$
	of {\it time-like} vectors.  This region is the union of two disjoint regions $T_1$ and $T_2$.

	The group $\sG(V,g)$ has four connected components:
		$$\sC^E(V,g) = \left\{\zr \in \sG(V,g) ;\; \det(\zr) = 1,\; \zr(T_1) = T_1 \right\},
																										\tag \label{For260}$$
		$$\!\quad\sC^T(V,g) = \left\{\zr \in \sG(V,g) ;\; \det(\zr) = - 1,\; \zr(T_1) = T_2 \right\},
																										\tag \label{For261}$$
		$$\kern-1pt\quad\sC^S(V,g) = \left\{\zr \in \sG(V,g) ;\; \det(\zr) = - 1,\; \zr(T_1) = T_1 \right\},
																										\tag \label{For262}$$
	and
		$$\!\quad\sC^{TS}(V,g) = \left\{\zr \in \sG(V,g) ;\; \det(\zr) = 1,\; \zr(T_1) = T_2 \right\}.
																										\tag \label{For263}$$
	The component of the unit element $\sC^E(V,g)$ is a normal subgroup denoted by $\sG^E(V,g)$.

	The set of {\it orientations}
		$$\sO(V,g) = \sF(V,g) \big/ \sG^E(V,g)
																										\tag \label{For264}$$
	 has four elements.  This set is a homogeneous
space for the quotient group $\sH(V,g) = \sG(V,g) \big/ \sG^E(V,g)$.  The quotient group is commutative.  Its elements are
the four components $\sC^E(V,g)$, $\sC^T(V,g)$, $\sC^S(V,g)$, and $\sC^{TS}(V,g)$ denoted simply by $E$, $T$, $S$. and $TS$
respectively.  The element $E = \sC^E(V,g)$ is the unit and all elements are involutions.  The composition rule of $T$ with
$S$ is incorporated in the notation used.

	There is an ordered base $(u_0,u_1,u_2,u_3)$ of $V$ associated with each frame $\zh$.  If
		$$\zh(v) = \pmatrix\mathstrut v^0 \\ \mathstrut v^1 \\ \mathstrut v^2 \\ \mathstrut v^3 \endpmatrix,
																										\tag \label{For265}$$
	then $v = u_\zk v^\zk$.  For each $\zr \in \sG(V,g)$ the base $(\zr(u_0),\zr(u_1),\zr(u_2),\zr(u_3))$ is associated
with the frame $\zh \circ \zr^{-1}$ if $(u_0,u_1,u_2,u_3)$ is the base associated with $\zh$.  Orthonormality relations
		$$\align
	 \langle g(u_\zk), u_\zl\rangle = \cases \hphantom{-}1 ,&\text{if }\; \zl = \zk = 0 \\
											-1 ,&\text{if } \;\zl = \zk \neq 0 \\
							  				 \hphantom{-}0 ,&\text{if } \;\zl \neq \zk \endcases
																										\tag \label{For266}\endalign$$
	follow from \Ref{For254}.

		\sect{Multicovectors in the Minkowski space.}
	A {\it relativistic $q$\,-\,covector} in a the Minkowski vector space $V$ is a mapping
		$$a\, \colon V^q \times \sO(V,g) \rightarrow \R.
																										\tag \label{For267}$$
	This mapping is $q$\,-\,linear and totally antisymmetric in its vector arguments.  A $q$\,-\,covector $a$ is said to have {\it
even temporal parity} if
		$$a(v_0,v_1,v_2,v_3,To) = a(v_0,v_1,v_2,v_3,o).
																										\tag \label{For268}$$
	It is said to have {\it odd temporal parity}, if
		$$a(v_0,v_1,v_2,v_3,To) = - a(v_0,v_1,v_2,v_3,o).
																										\tag \label{For269}$$
	It is said to have {\it even spatial parity}, if
		$$a(v_0,v_1,v_2,v_3,So) = a(v_0,v_1,v_2,v_3,o).
																										\tag \label{For270}$$
	It is said to have {\it odd spatial parity}, if
		$$a(v_0,v_1,v_2,v_3,So) = - a(v_0,v_1,v_2,v_3,o).
																										\tag \label{For271}$$

	Relativistic $q$\,-\,covectors with different parities form four distinct vector spaces $\WE^q_{e,e}V^\*$,
$\WE^q_{o,e}V^\*$, $\WE^q_{e,o}V^\*$, $\WE^q_{o,o}V^\*$.  The first of the two subscripts identifies the temporal
parity and the second identifies the spatial parity of the forms.  The {\it exterior product} of relativistic
multicovectors is defined as in \Ref{For10} and has the usual properties.  The parities of the exterior products of
multicovectors are indicated in the following table.

\vskip8mm

\setbox80=\hbox{\vrule height-2.5pt width098mm depth3pt}
\setbox81=\hbox{\vrule height-2.5pt width.7pt depth41mm}

\setbox01=\hbox{$\WE^{q_1}_{e,e}(V)$}
\setbox02=\hbox{$\WE^{q_1}_{o,e}(V)$}
\setbox03=\hbox{$\WE^{q_1}_{e,o}(V)$}
\setbox04=\hbox{$\WE^{q_1}_{o,o}(V)$}
\setbox05=\hbox{$\WE^{q_2}_{e,e}(V)$}
\setbox06=\hbox{$\WE^{q_2}_{o,e}(V)$}
\setbox07=\hbox{$\WE^{q_2}_{e,o}(V)$}
\setbox08=\hbox{$\WE^{q_2}_{o,o}(V)$}
\setbox09=\hbox{$\WE^{q_1+q_2}_{e,e}(V)$}
\setbox10=\hbox{$\WE^{q_1+q_2}_{o,e}(V)$}
\setbox11=\hbox{$\WE^{q_1+q_2}_{e,o}(V)$}
\setbox12=\hbox{$\WE^{q_1+q_2}_{o,o}(V)$}

\setbox0=\hbox{
\hskip055mm\lower-01mm\copy05\hskip-\wd05\hskip-055mm
\hskip075mm\lower-01mm\copy06\hskip-\wd06\hskip-075mm
\hskip095mm\lower-01mm\copy07\hskip-\wd07\hskip-095mm
\hskip115mm\lower-01mm\copy08\hskip-\wd08\hskip-115mm

\hskip35mm \lower01mm\copy80 \hskip-133mm

\hskip050mm\lower-05mm\copy81\hskip-\wd81\hskip-050mm

\hskip035mm\lower09mm\copy01\hskip-\wd01\hskip-035mm
\hskip055mm\lower09mm\copy09\hskip-\wd09\hskip-055mm
\hskip075mm\lower09mm\copy10\hskip-\wd10\hskip-075mm
\hskip095mm\lower09mm\copy11\hskip-\wd11\hskip-095mm
\hskip115mm\lower09mm\copy12\hskip-\wd12\hskip-115mm

\hskip035mm\lower17mm\copy02\hskip-\wd02\hskip-035mm
\hskip055mm\lower17mm\copy10\hskip-\wd10\hskip-055mm
\hskip075mm\lower17mm\copy09\hskip-\wd09\hskip-075mm
\hskip095mm\lower17mm\copy12\hskip-\wd12\hskip-095mm
\hskip115mm\lower17mm\copy11\hskip-\wd11\hskip-115mm

\hskip035mm\lower25mm\copy03\hskip-\wd03\hskip-035mm
\hskip055mm\lower25mm\copy11\hskip-\wd11\hskip-055mm
\hskip075mm\lower25mm\copy12\hskip-\wd12\hskip-075mm
\hskip095mm\lower25mm\copy09\hskip-\wd09\hskip-095mm
\hskip115mm\lower25mm\copy10\hskip-\wd10\hskip-115mm

\hskip035mm\lower33mm\copy04\hskip-\wd04\hskip-035mm
\hskip055mm\lower33mm\copy12\hskip-\wd12\hskip-055mm
\hskip075mm\lower33mm\copy11\hskip-\wd11\hskip-075mm
\hskip095mm\lower33mm\copy10\hskip-\wd10\hskip-095mm
\hskip115mm\lower33mm\copy09\hskip-\wd09\hskip-115mm

}\box0
\vskip10mm

	Elements of a base $\{e^0,e^1,e^2,e^3\}$ dual to a base $\{e_0,e_1,e_2,e_3\}$ of $V$ define covectors
		$$e_{e,e}^\zk \colon V \times \sO(V,g) \rightarrow \R \colon (v,o) \mapsto \langle e^\zk, v\rangle
																										\tag \label{For272}$$
	of parity $e,e$.  The base is not necessarily orthonormal.  The use of orthonormal bases offers no simplification.  In
addition to this base we choose an orientation $o \in \sO(V,g)$ and introduce 0\,-\,covectors $e_{p,p}$ defined by
		$$e_{e,e}(o) = 1,\;\;\;e_{e,e}(To) = \hphantom{-}1,\;\;\;e_{e,e}(So) = \hphantom{-}1,\;\;\;\text{ and
}\;\;\;e_{e,e}(TSo) = \hphantom{-}1,
																										\tag \label{For273}$$
		$$e_{o,e}(o) = 1,\;\;\;e_{o,e}(To) = -1,\;\;\;e_{o,e}(So) = \hphantom{-}1,\;\;\;\text{ and }\;\;\;e_{o,e}(TSo) = -1,
																										\tag \label{For274}$$
		$$e_{e,o}(o) = 1,\;\;\;e_{e,o}(To) = \hphantom{-}1,\;\;\;e_{e,o}(So) = -1,\;\;\;\text{ and }\;\;\;e_{e,o}(TSo) = -1,
																										\tag \label{For275}$$
		$$e_{o,o}(o) = 1,\;\;\;e_{o,o}(To) = -1,\;\;\;e_{o,o}(So) = -1,\;\;\;\text{ and }\;\;\;e_{o,o}(TSo) = \hphantom{-}1.
																										\tag \label{For276}$$
	The covectors $e_{p,p}$ could be characterized as the 0\,-\,covectors of parity $p,p$ such that
		$$e_{p,p}(o) = 1.
																										\tag \label{For277}$$

	Each $q$\,-\,covector of parity $p,p$ has a representation as the combination
		$$a = \frac{1}{q!}\; a_{\zk_1 \zk_2 \ldots \zk_q} e_{p,p}^{\zk_1 \zk_2 \ldots \zk_q}
																										\tag \label{For278}$$
	of $q$\,-\,covectors
		$$e_{p,p}^{\zk_1 \zk_2 \ldots \zk_q} = e_{p,p}^{\zk_1} \wedge e_{e,e}^{\zk_2} \wedge \cdots \wedge
e_{e,e}^{\zk_q}.
																										\tag \label{For279}$$
	Antisymmetric coefficients in this combination are unique.  A base is extracted from each of the systems by selecting
elements with superscripts forming ascending sequences
		$$\zk_1 < \zk_2 < \ldots < \zk_q.
																										\tag \label{For280}$$
	The spaces of relativistic multicovectors have dimensions
		$$\dim(\wedge_{p,p}^q V^\*) = \binom 4 q.
																										\tag \label{For281}$$

	Each of the relativistic orientations in $\sO(V,g)$ is included in one of the elements of the set $\sO(V)$.  It follows
that it makes sense to assign to each multicovector defined on orientations in $\sO(V)$ a multicovector defined on
relativistic orientations.  The inclusions $\sC^T(V,g) \subset \sC^P(V)$ and $\sC^S(V,g) \subset \sC^P(V)$ imply that an
even $q$\,-\,covector will be assigned an element in $\WE^q_{e,e}(V)$ and an odd $q$\,-\,covector will be assigned an
element in $\WE^q_{o,o}(V)$.

	The metric volume $\sqrt{|g|}$ defined in Section 5 is an odd 4-covector.  It is interpreted as an element of the space
$\wedge_{o,o}^4$.

	The group $\sG(V,g)$ has natural representations in the spaces $\wedge_{p,p}^q V^\*$.  A Lorentz transformation $\zr
\in \sG(V,g)$  applied to a $q$\,-\,covector $a$ produces the $q$\,-\,covector
		$$\align
		(\zr^{-1})^{\*}a\, &\colon V^q \times \sO(V,g) \rightarrow \R \\
			&\colon (v_1, v_2, \ldots ,v_q,o) \mapsto a(\zr^{-1}(v_1), \zr^{-1}(v_2), \ldots ,
\zr^{-1}(v_q),[\zr](o)),
																										\tag \label{For282}\endalign$$
	where $[\zr]$ is the class of $\zr$ in the quotient group $\sH(V,g) = \sG(V,g) \big/ \sG^E(V,g)$.  We introduce the
{\it index} $\idx_{p,p}(\zr)$ of a Lorentz transformation $\zr$ in terms of the result
		$$(\zr^{-1})^{\*}a \colon \sO(V,g) \rightarrow \R \colon o \mapsto  a([\zr](o)) = \idx_{p,p}a(o)
																										\tag \label{For283}$$
	of the transformation applied to a 0\,-\,covector
		$$a \colon \sO(V,g) \rightarrow \R
																										\tag \label{For284}$$
	of parity $p,p$.  The values of the index are listed in the following table.

\setbox2=\hbox{\vrule height-2.5pt width97mm depth3pt}

\setbox4=\hbox{\vrule height-2.5pt width.7pt depth32mm}

\setbox11=\hbox{$\hphantom{-}1$}
\setbox12=\hbox{$-1$}

\setbox41=\hbox{$\zr \in E$}
\setbox42=\hbox{$\zr \in T$}
\setbox43=\hbox{$\zr \in S$}
\setbox44=\hbox{$\zr \in TS$}

\setbox51=\hbox{$\idx_{e,e}(\zr)$}
\setbox52=\hbox{$\idx_{o,e}(\zr)$}
\setbox53=\hbox{$\idx_{e,o}(\zr)$}
\setbox54=\hbox{$\idx_{o,o}(\zr)$}

\vskip5mm
\setbox0=\hbox{
\hskip055mm\lower-01mm\copy41\hskip-\wd41\hskip-055mm
\hskip075mm\lower-01mm\copy42\hskip-\wd42\hskip-075mm
\hskip095mm\lower-01mm\copy43\hskip-\wd43\hskip-095mm
\hskip115mm\lower-01mm\copy44\hskip-\wd44\hskip-115mm

\hskip033mm \copy02 \hskip-130mm
\hskip50mm\lower-05mm\copy04\hskip-\wd04\hskip-50mm

\hskip035mm\lower06mm\copy51\hskip-\wd51\hskip-035mm
\hskip056mm\lower06mm\copy11\hskip-\wd11\hskip-056mm
\hskip076mm\lower06mm\copy11\hskip-\wd11\hskip-076mm
\hskip096mm\lower06mm\copy11\hskip-\wd11\hskip-096mm
\hskip116mm\lower06mm\copy11\hskip-\wd11\hskip-116mm

\hskip035mm\lower12mm\copy52\hskip-\wd52\hskip-035mm
\hskip056mm\lower12mm\copy11\hskip-\wd11\hskip-056mm
\hskip076mm\lower12mm\copy12\hskip-\wd12\hskip-076mm
\hskip096mm\lower12mm\copy11\hskip-\wd11\hskip-096mm
\hskip116mm\lower12mm\copy11\hskip-\wd12\hskip-116mm

\hskip035mm\lower18mm\copy53\hskip-\wd53\hskip-035mm
\hskip056mm\lower18mm\copy11\hskip-\wd11\hskip-056mm
\hskip076mm\lower18mm\copy11\hskip-\wd11\hskip-076mm
\hskip096mm\lower18mm\copy12\hskip-\wd12\hskip-096mm
\hskip116mm\lower18mm\copy12\hskip-\wd12\hskip-116mm

\hskip035mm\lower24mm\copy54\hskip-\wd54\hskip-035mm
\hskip056mm\lower24mm\copy11\hskip-\wd11\hskip-056mm
\hskip076mm\lower24mm\copy12\hskip-\wd12\hskip-076mm
\hskip096mm\lower24mm\copy12\hskip-\wd12\hskip-096mm
\hskip116mm\lower24mm\copy11\hskip-\wd11\hskip-116mm

}\box0
\vskip9mm

	In terms of the index the result of the application of a Lorentz transformation $\zr$ to a $q$\,-\,covector a of parity
$p,p$ is the $q$\,-\,covector
		$$\align
		(\zr^{-1})^{\*}a\, &\colon V^q \times \sO(V,g) \rightarrow \R \\
			&\colon (v_1, v_2, \ldots ,v_q,o) \mapsto \idx_{p,p}a(\zr^{-1}(v_1), \zr^{-1}(v_2), \ldots , \zr^{-1}(v_q),o).
																										\tag \label{For285}\endalign$$

\newpage

		\sect{Multivectors in the Minkowski space-time.}
	We consider again the space $\sK(V^q \times \sO(V,g))$ of formal linear combinations of sequences
		$$(v_1, v_2, \ldots ,v_q,o) \in V^q \times \sO(V,g).
																										\tag \label{For286}$$
	In the space $\sK(V^q \times \sO(V,g))$ we introduce subspaces
		$$\align
		\sA^{\,p,p}_q(V) &= \left\{\tsize{\sum_{\,i=1}^{\,n}}\, \zl_i(v^i_1, v^i_2, \ldots ,v^i_q,o^i) \in \sK(V^q \times
\sO(V)) ;\; \tsize{\sum_{\,i=1}^{\,n}}\, \zl_i a(v^i_1, v^i_2, \ldots ,v^i_q,o^i) = 0 \right. \\
	&\hskip50mm \left. \vphantom{\tsize{\sum_{\,i=1}}}\text{ for each }\;\; a \in \WE^q_{p,p} V^\* \right\}.
																										\tag \label{For287}\endalign$$
	We then define quotient spaces
		$$\WE^q_{p,p} V = \sK(V^q \times \sO(V)) \big/ \sA^{\,p,p}_q(V).
																										\tag \label{For288}$$
	Elements of spaces $\WE^q_{p,p} V$ {\it $q$\,-\,vectors of parity $p,p$}.

	Evaluation of $q$\,-\,covectors on sequences $(v_1, v_2, \ldots ,v_q,o) \in V^q \times \sO(V,g)$ extends to linear
combinations and their equivalence classes.  If $w$ is a $q$\,-\,vector represented by the linear combination
		$$\tsize{\sum_{\,i=1}^{\,n}}\, \zl_i(v^i_1, v^i_2, \ldots ,v^i_q,o^i)
																										\tag \label{For289}$$
	and $a$ is a $q$\,-\,covector of the same parity as $w$, then
		$$\langle a, w\rangle = \tsize{\sum_{\,i=1}^{\,n}}\, \zl_i a(v^i_1, v^i_2, \ldots ,v^i_q,o^i)
																										\tag \label{For290}$$
	is the evaluation of $a$ on $w$.  We have constructed pairings
		$$\langle \;,\,\rangle \colon \WE^q_{p,p} V^\* \times \WE^q_{p,p} V \rightarrow \R.
																										\tag \label{For291}$$

	Let $\{e_0,e_1,e_2,e_3\}$ be a base of $V$ and let $o$ be an orientation of $(V,g)$.  Each $q$\,-\,vector
$w$ has a representation as the combination
		$$w = \frac{1}{q!}\; w^{\zk_1 \zk_2 \ldots \zk_q} e^{p,p}_{\zk_1 \zk_2 \ldots \zk_q}
																										\tag \label{For292}$$
	with unique antisymmetric coefficients.  The $q$\,-\,vectors
		$$e^e_{\zk_1 \zk_2 \ldots \zk_q}
																										\tag \label{For293}$$
	are equivalence classes in $\WE^q_{p,p} V$ of the sequences
		$$(e_{\zk_1},e_{\zk_2},\ldots,e_{\zk_q},o).
																										\tag \label{For294}$$
	of elements of the base $\{e_0,e_1,e_2,e_3\}$ and the orientation $o$.  The coefficients in the formula
\Ref{For292} are obtained from
		$$w^{\zk_1 \zk_2 \ldots \zk_q} = \langle e_{p,p}^{\zk_1 \zk_2 \ldots \zk_q}, w\rangle.
																										\tag \label{For295}$$

	The spaces $\WE^q_{p,p} V$ have the dimension
		$$\dim(\wedge_{p,p}^q V) = \binom 4 q.
																										\tag \label{For296}$$
	The sets
		$$\{e^{p,p}_{\zk_1 \zk_2 \ldots \zk_q}\}_{\zk_1 < \zk_2 < \ldots < \zk_q}
																										\tag \label{For297}$$
	of independent multivectors form bases of these spaces.

	Let $w$ be a $q$\,-\,vector represented by the linear combination
		$$\tsize{\sum_{\,i=1}^{\,n}}\, \zl_i(v^i_1, v^i_2, \ldots ,v^i_q,o^i).
																										\tag \label{For298}$$
	The image of this multivector by an automorphism $\zr \in \sG(V)$ is the multivector $\zr_\* w$ represented by the
combination
		$$\tsize{\sum_{\,i=1}^{\,n}}\, \zl_i(\zr(v^i_1), \zr(v^i_2), \ldots ,\zr(v^i_q),[\zr](o^i)),
																										\tag \label{For299}$$
	where $[\zr]$ is again the class of $\zr$ in the quotient group $\sH(V) = \sG(V) \big/ \sG^E(V)$.  If $w \in
\wedge_{p,p}^q V$, then the multivector $\zr_\*w$ is represented by
		$$\idx_{p,p}\tsize{\sum_{\,i=1}^{\,n}}\, \zl_i(\zr(v^i_1), \zr(v^i_2), \ldots ,\zr(v^i_q),o^i).
																										\tag \label{For300}$$

	It follows from the definition \Ref{For290} that the pairings \Ref{For291} are invariant in the sense that
		$$\langle (\zr^{-1})^{\*}a, \zr_\* w\rangle = \langle a, w\rangle.
																										\tag \label{For301}$$

		\sect{Differential forms in the Minkowski space-time.}
	A {\it relativistic differential $q$\,-\,form} on the Minkowski space-time $M$ is a differentiable function
		$$A \colon M \times V^4 \times \sO(V,g) \rightarrow \R
																										\tag \label{For302}$$
	multilinear and totally antisymmetric in its vector arguments.  A differential form $A$ is said to have {\it even
temporal parity} if
		$$A(x,v_0,v_1,v_2,v_3,To) = A(x,v_0,v_1,v_2,v_3,o).
																										\tag \label{For303}$$
	It is said to have {\it odd temporal parity}, if
		$$A(x,v_0,v_1,v_2,v_3,To) = - A(x,v_0,v_1,v_2,v_3,o).
																										\tag \label{For304}$$
	It is said to have {\it even spatial parity}, if
		$$A(x,v_0,v_1,v_2,v_3,So) = A(x,v_0,v_1,v_2,v_3,o).
																										\tag \label{For305}$$
	It is said to have {\it odd spatial parity}, if
		$$A(x,v_0,v_1,v_2,v_3,So) = - A(x,v_0,v_1,v_2,v_3,o).
																										\tag \label{For306}$$

	For each degree $q$ there are four spaces $\zF^q_{e,e}$\,, $\zF^q_{o,e}$\,, $\zF^q_{e,o}$\,, and $\zF^q_{o,o}$\,\, of
forms of different parities.  As in the case of $q$\,-\,covectors the first of the two subscripts identifies the temporal
parity and the second identifies the spatial parity of the forms.  The parities of the exterior product of forms are listed
in the following table.

\vskip8mm

\setbox80=\hbox{\vrule height-2.5pt width098mm depth3pt}
\setbox81=\hbox{\vrule height-2.5pt width.7pt depth41mm}

\setbox01=\hbox{$\zF^{q_1}_{e,e}(M)$}
\setbox02=\hbox{$\zF^{q_1}_{o,e}(M)$}
\setbox03=\hbox{$\zF^{q_1}_{e,o}(M)$}
\setbox04=\hbox{$\zF^{q_1}_{o,o}(M)$}
\setbox05=\hbox{$\zF^{q_2}_{e,e}(M)$}
\setbox06=\hbox{$\zF^{q_2}_{o,e}(M)$}
\setbox07=\hbox{$\zF^{q_2}_{e,o}(M)$}
\setbox08=\hbox{$\zF^{q_2}_{o,o}(M)$}
\setbox09=\hbox{$\zF^{q_1+q_2}_{e,e}(M)$}
\setbox10=\hbox{$\zF^{q_1+q_2}_{o,e}(M)$}
\setbox11=\hbox{$\zF^{q_1+q_2}_{e,o}(M)$}
\setbox12=\hbox{$\zF^{q_1+q_2}_{o,o}(M)$}

\setbox0=\hbox{
\hskip055mm\lower-01mm\copy05\hskip-\wd05\hskip-055mm
\hskip075mm\lower-01mm\copy06\hskip-\wd06\hskip-075mm
\hskip095mm\lower-01mm\copy07\hskip-\wd07\hskip-095mm
\hskip115mm\lower-01mm\copy08\hskip-\wd08\hskip-115mm

\hskip35mm \lower01mm\copy80 \hskip-133mm

\hskip050mm\lower-05mm\copy81\hskip-\wd81\hskip-050mm

\hskip035mm\lower09mm\copy01\hskip-\wd01\hskip-035mm
\hskip055mm\lower09mm\copy09\hskip-\wd09\hskip-055mm
\hskip075mm\lower09mm\copy10\hskip-\wd10\hskip-075mm
\hskip095mm\lower09mm\copy11\hskip-\wd11\hskip-095mm
\hskip115mm\lower09mm\copy12\hskip-\wd12\hskip-115mm

\hskip035mm\lower17mm\copy02\hskip-\wd02\hskip-035mm
\hskip055mm\lower17mm\copy10\hskip-\wd10\hskip-055mm
\hskip075mm\lower17mm\copy09\hskip-\wd09\hskip-075mm
\hskip095mm\lower17mm\copy12\hskip-\wd12\hskip-095mm
\hskip115mm\lower17mm\copy11\hskip-\wd11\hskip-115mm

\hskip035mm\lower25mm\copy03\hskip-\wd03\hskip-035mm
\hskip055mm\lower25mm\copy11\hskip-\wd11\hskip-055mm
\hskip075mm\lower25mm\copy12\hskip-\wd12\hskip-075mm
\hskip095mm\lower25mm\copy09\hskip-\wd09\hskip-095mm
\hskip115mm\lower25mm\copy10\hskip-\wd10\hskip-115mm

\hskip035mm\lower33mm\copy04\hskip-\wd04\hskip-035mm
\hskip055mm\lower33mm\copy12\hskip-\wd12\hskip-055mm
\hskip075mm\lower33mm\copy11\hskip-\wd11\hskip-075mm
\hskip095mm\lower33mm\copy10\hskip-\wd10\hskip-095mm
\hskip115mm\lower33mm\copy09\hskip-\wd09\hskip-115mm

}\box0
\vskip10mm

	The exterior differential of a form preserves the parity of the original form.

		\sect{Vector densities and Weyl duality.}
	Differentiable mappings
		$$\ovA \colon M \rightarrow \wedge_{o,e}^qV \otimes \wedge_{o,o}^4 V^\*
																										\tag \label{For307}$$
	play the role of {\it $q$\,-\,vector densities}.  A 0-vector density is called a {\it scalar density}.

	The Weyl isomorphism is used to associate with an $(4\- q)$\,-\,form $A$ a {\it $q$\,-\,vector density}
		$$\ovA \colon M \rightarrow \wedge_e^qV \otimes \wedge_o^4 V^\*
																										\tag \label{For308}$$
	defined by
		$$\ovA = \We^{-1}_q \circ \wA
																										\tag \label{For309}$$
	and called the {\it relativistic Weyl dual} of $A$.  If $\ovA$ is the relativistic Weyl dual of an $(m\- q)$\,-\,form
$A$, then the relativistic Weyl dual of the $(m\- q\+ 1)$\,-\,form $\rd A$ is a $(q\- 1)$\,-\,vector density denoted by
$\Div\ovA$.  Thus,
		$$\Div\ovA = \We^{-1}_{q+1} \circ \rd(\We_q \circ \ovA).
																										\tag \label{For310}$$

		\sect{Transformation properties of forms and densities.}
	Let $M$ be an affine space modelled on a vector space $V$.  A mapping $\zf \colon M \rightarrow M$ is said to be {\it
affine} if there is a linear mapping $\chi \colon V \rightarrow V$ such that
	$$\zf(x') - \zf(x) = \chi(x' - x).
																										\tag \label{For311}$$
	If the mapping $\chi \colon V \rightarrow V$ satisfying the above condition exists, then it is unique.  It is called
the {\it linear part} of the affine mapping $\zf$ and is denoted by $\ozf$.

	An affine mapping $\zf$ is invertible if and only if its linear part $\ozf$ is invertible.  If $\zf$ is
invertible then $\zf^{-1}$ is an affine mapping and $\overline{\zf^{-1}} = (\overline{\zf})^{-1}$.

	An affine mapping $\zf$ is differentiable.  Its differential and its linear part are related by
		$$\xD\zf(x,v) = \ozf(v).
																										\tag \label{For312}$$

	Let $M$ be the Minkowski space-time of special relativity.  A {\it Poincar\'e transformation} is an affine mapping $\zf
\colon M \rightarrow M$ such that $\ozf$ is a Lorentz transformation.  Poincar\'e transformations form a group denoted by
$\sP(M,g)$.

	A Poincar\'e transformation $\zf$ applied to a $q$\,-\,form $A$ produces the $q$\,-\,form $A$
		$$\align
		(\zf^{-1})^{\*}A\, &\colon M \times V^q \times \sO(V,g) \rightarrow \R \\
			&\colon (x,v_1, \ldots ,v_q,o) \mapsto A(\zf^{-1}(x),\ozf^{-1}(x,v_1), \ldots ,
\ozf^{-1}(x,v_q),[\ozf](o)),
																										\tag \label{For313}\endalign$$
	where $[\ozf]$ is the class of $\ozf$ in the quotient group $\sH(V,g) = \sG(V,g) \big/ \sG^E(V,g)$.

	The pull back $(\zf^{-1})^{\*}A$ of a $q$\,-\,form $A$ represented by the mapping $\wA$ is represented by
		$$(\zf^{-1})^\*\wA \colon M \rightarrow \wedge_p^qV^{\*} \colon x \mapsto (\ozf^{-1})^\*\wA(\zf^{-1}(x)).
																										\tag \label{For314}$$

	The pull back of a $4$\,-\,form $E$ of parity $o,o$ is the form $E$ itself.  A $q$\,-\,vector field
		$$W \colon M \rightarrow \wedge_{o,e}^qV
																										\tag \label{For315}$$
	transforms according to the rule
		$$\zf_\*W \colon M \rightarrow \wedge_{o,e}^qV \colon x \mapsto \ozf_\*W(\zf^{-1}(x)).
																										\tag \label{For316}$$
	A $q$\,-\,vector density $\wA$ can be composed as a tensor product
		$$W \otimes \wE \colon M \rightarrow \wedge_{o,e}^qV \otimes \wedge_{o,o}^4 V^\* \colon x \mapsto W(x) \otimes
\wE(x)
																										\tag \label{For317}$$
	of a $q$\,-\,vector field $W$ and a scalar density $\wE$.  This tensor product follows the transformation rule
		$$\zf_\*(W \otimes \wE) = \zf_\*W \otimes (\zf^{-1})^{\*}\wE
																										\tag \label{For318}$$
	or
		$$\zf_\*(W \otimes \wE) \colon M \rightarrow \wedge_e^qV \otimes \wedge_o^m V^\* \colon x \mapsto
|\det(\xD\zf^{-1}(x,\cdot))|^{-1}\xD\zf(x,\cdot)_\*W(\zf^{-1}(x)) \otimes \wE(\zf^{-1}(x)).
																										\tag \label{For319}$$
	In Schouten's terminology a scalar density is a {\it covariant $W$-$\,m$\,-\,vector} or  a {\it scalar density of weight} 1.  A
$q$\,-\,vector density is a {\it contravariant $q$\,-\,vector density of weight} 1.

	The invariance property
		$$\overline{(\zf^{-1})^{\*}A} = \zf_{\*}\ovA.
																										\tag \label{For320}$$
	of Weyl duality is a direct consequence of the invariance of the Weyl isomorphism.

		\sect{Space-time parity of electromagnetic objects.}
	We will see that the assignments
		$$F \in \zF^2_{o,e}\,, \hskip10mm G \in \zF^2_{e,0}\,, \hskip10mm J \in \zF^3_{e,o}\,, \hskip10mm A \in
\zF^1_{o,e}\,. \hskip15mm
																										\tag \label{For321}$$
	of parities of electromagnetic fields result in parities postulated in the literature.

		\sect{Time reflection and electromagnetic objects.}
	Temporal parity of objects is determined by inspecting the response to time reflection.  The {\it time reflection}
relative to an inertial frame is represented in $M$ and in $\R \times N$ by the transformations
		$$\zf \colon M \rightarrow M \colon x \mapsto x - 2\langle g(u), x - x_0\rangle u
																										\tag \label{For322}$$
	and
		$$\zc \colon \R \times N \rightarrow \R \times  N \colon (t,y) \mapsto (-t,y)
																										\tag \label{For323}$$
	wth linear parts
		$$\ozf \colon M \times V \rightarrow V \colon (x,v) \mapsto v - 2\langle g(u), v\rangle u
																										\tag \label{For324}$$
	and
		$$\overline{\zc} \colon \R \times N \times \R \times W \rightarrow \R \times  W \colon (t,y,\zd t,w) \mapsto (-\zd
t,w).
																										\tag \label{For325}$$
	Note that these mappings are involutions and that $\ozf \in T$.  Hence,
		$$\idx_{e,e} = 1,\;\;\;\; \idx_{o,e} = - 1,\;\;\;\; \idx_{e,o} = 1,\;\;\;\; \idx_{o,o} = - 1,\;\;\;\; 
																										\tag \label{For326}$$

	Here are the effects of the time reflection applied to the forms $F$, $G$, $J$, and $A$:
		$$\align
			\zf^{\*}F(x,v_1,v_2,o) &= F(\zf(x),\ozf(v_1),\ozf(v_2),[\ozf](o)) \\
					&= \idx_{0,e}(\ozf)F(x - 2\langle g(u), x - x_0\rangle u,v_1 - 2\langle g(u), v_1\rangle u,v_2 - 2\langle g(u),
v_2\rangle u,o) \\
					&= - F(x - 2\langle g(u), x - x_0\rangle u,v_1 - 2\langle g(u), v_1\rangle u,v_2 - 2\langle g(u),
v_2\rangle u,o),
																										\tag \label{For327}\endalign$$
		$$\align
			\zf^{\*}G(x,v_1,v_2,o) &= G(\zf(x),\ozf(v_1),\ozf(v_2),[\ozf](o)) \\
					&= \idx_{e,0}(\ozf)G(x - 2\langle g(u), x - x_0\rangle u,v_1 - 2\langle g(u), v_1\rangle u,v_2 - 2\langle g(u),
v_2\rangle u,o) \\
					&= G(x - 2\langle g(u), x - x_0\rangle u,v_1 - 2\langle g(u), v_1\rangle u,v_2 - 2\langle g(u),
v_2\rangle u,o),
																										\tag \label{For328}\endalign$$
		$$\align
			\zf^{\*}&J(x,v_1,v_2,v_3,o) = J(\zf(x),\ozf(v_1),\ozf(v_2),\ozf(v_3),[\ozf](o)) \\
					&= \idx_{e,0}(\ozf)J(x - 2\langle g(u), x - x_0\rangle u,v_1 - 2\langle g(u), v_1\rangle u,v_2 - 2\langle g(u),
v_2\rangle u,v_3 - 2\langle g(u), v_3\rangle u,o) \\
					&= J(x - 2\langle g(u), x - x_0\rangle u,v_1 - 2\langle g(u), v_1\rangle u,v_2 - 2\langle g(u),
v_2\rangle u,v_3 - 2\langle g(u), v_3\rangle u,o),
																										\tag \label{For329}\endalign$$
	and
		$$\align
			\zf^{\*}A(x,v,o) &= A(\zf(x),\ozf(v),[\ozf](o)) \\
					&= \idx_{0,e}(\ozf)A(x - 2\langle g(u), x - x_0\rangle u,v - 2\langle g(u), v\rangle u,o) \\
					&= - A(x - 2\langle g(u), x - x_0\rangle u,v - 2\langle g(u), v\rangle u,o).
																										\tag \label{For330}\endalign$$

	The time reflection transformation rules for $\iE$, $\iB$, $\iD$, $\iH$, $\iQ$, $\iJ$, $\iU$, and $\iA$  are obtained
from
		$$\align
			\zc^{\*}\iE(t,y,w,o) &= \hat\zn^{\*}\xi_U \zf^{\*}F(t,y,w,o) \\
					&= \zf^{\*}F(y + ctu,u,w,[u,o]) \\
					&= - F(y - ctu,-u,w,[u,o]) \\
					&= \iE(-t,y,w.o),
																										\tag \label{For331}\endalign$$
		$$\align
			\zc^{\*}\iB(t,y,w_1,w_2,o) &= \hat\zn^{\*}\zf^{\*}F(t,y,w_1,w_2,o) \\
					&= \zf^{\*}F(y + ctu,w_1,w_2,[u,o]) \\
					&= - F(y - ctu,w_1,w_2,[u,o]) \\
					&= - \iB(-t,y,w_1,w_2.o),
																										\tag \label{For332}\endalign$$
		$$\align
			\zc^{\*}\iD(t,y,w_1,w_2,o) &= \hat\zn^{\*}\zf^{\*}G(t,y,w_1,w_2,o) \\
					&= \zf^{\*}G(y + ctu,w_1,w_2,[u,o]) \\
					&= G(y - ctu,w_1,w_2,[u,o]) \\
					&= \iD(-t,y,w_1,w_2.o),
																										\tag \label{For333}\endalign$$
		$$\align
			\zc^{\*}\iH(t,y,w,o) &= \hat\zn^{\*}\xi_U \zf^{\*}G(t,y,w,o) \\
					&= \zf^{\*}G(y + ctu,u,w,[u,o]) \\
					&= G(y - ctu,-u,w,[u,o]) \\
					&= - \iH(-t,y,w.o),
																										\tag \label{For334}\endalign$$
		$$\align
			\zc^{\*}\iQ(t,y,w_1,w_2,w_3,o) &= \hat\zn^{\*}\zf^{\*}J(t,y,w_1,w_2,w_3,o) \\
					&= \zf^{\*}J(y + ctu,w_1,w_2,w_3,[u,o]) \\
					&= J(y - ctu,w_1,w_2,w_3,[u,o]) \\
					&= \iQ(-t,y,w_1,w_2,w_3.o),
																										\tag \label{For335}\endalign$$
		$$\align
			\zc^{\*}\iJ(t,y,w_1,w_2,o) &= \hat\zn^{\*}\xi_U\zf^{\*}J(t,y,w_1,w_2,o) \\
					&= \zf^{\*}J(y + ctu,u,w_1,w_2,[u,o]) \\
					&= J(y - ctu,-u,w_1,w_2,[u,o]) \\
					&= - \iJ(-t,y,w_1,w_2.o),
																										\tag \label{For336}\endalign$$
		$$\align
			\zc^{\*}\iU(t,y,o) &= \hat\zn^{\*}\xi_U \zf^{\*}A(t,y,o) \\
					&= \zf^{\*}A(y + ctu,u,[u,o]) \\
					&= - A(y - ctu,-u,[u,o]) \\
					&= \iU(-t,y.o),
																										\tag \label{For337}\endalign$$
	and
		$$\align
			\zc^{\*}\iA(t,y,w,o) &= \hat\zn^{\*}\zf^{\*}A(t,y,w,o) \\
					&= \zf^{\*}A(y + ctu,w,[u,o]) \\
					&= - A(y - ctu,w,[u,o]) \\
					&= - \iA(-t,y,w.o).
																										\tag \label{For338}\endalign$$

	The derived time reflection transformation rules are exactly the ones postulated by Jackson and Landau and Lifshitz.

\newpage
		\leftline{\bf References.}
\vskip2mm
\noindent [1] J. A. Schouten, {\it Ricci calculus}, Springer, Berlin, 1955.

\noindent [2] J. A. Schouten, {\it Tensor Analysis for Physicists}, Oxford University Press, London, 1951.

\noindent [3] G. de Rham, {\it Vari\'et\'es Differentiables}, Hermann, Paris, 1955.

\noindent [4] E. Cartan, Ann. Ec. Norm. {\bf 41} (1924).

\noindent [5] R. S. Ingarden and A. Jamio\l kowski, {\it Classical Electrodynamics}, PWN, Warszawa, 1985.

\noindent [6] E. J. Post, {\it Formal Structure of Electromagnetics}, North-Holland, Amsterdam, 1962.

\noindent [7] William L. Burke, {\it Applied differential geometry}, CUP, Cambridge, 1985.

\noindent [8] S. Parrott, {\it Relativistic Electrodynamics and Differential geometry}, Springer-Verlag, New York, 1987.

\noindent [9] C. W. Misner, K. S. Thorn, and J. A. Wheeler, {\it Gravitation}, W. H. Freeman, San Francisco, 1973.

\noindent [10] J. D. Jackson, {\it Classical Electrodynamics}, John Wiley, N. Y., 1975.

\noindent [11] L. D. Landau and E. M. Lifshitz, {\it The Classical Theory of Fields} Addison-Wesley, Reading, MA, 1970.

\noindent [12] T. Frankel, {\it The Geometry of Physics: An Introduction} Cambridge University Press, Cambridge, 1997.

\noindent [13] F. W. Hehl and Yu. N. Obukhov, {\it Foundations of Classical Electrodynamics, Charge, Flux, and Metric}

\hskip12mm Birkh\"auser, Boston, MA, 2003.

\noindent [14] I. V. Lindell, {\it Differential Forms in Electromagnetics} IEEE Press, Piscataway, NJ, and
Wiley-Interscience, 2004.

\end